\documentclass[sigconf]{acmart}

\AtBeginDocument{%
  }

\copyrightyear{2025}
\acmYear{2025}
\setcopyright{cc}
\setcctype{by}
\acmConference[UIST '25]{The 38th Annual ACM Symposium on User Interface

Software and Technology}{September 28-October 1, 2025}{Busan, Republic of
Korea}
\acmBooktitle{The 38th Annual ACM Symposium on User Interface Software and
Technology (UIST '25), September 28-October 1, 2025, Busan, Republic of
Korea}\acmDOI{10.1145/3746059.3747610}
\acmISBN{979-8-4007-2037-6/2025/09}
 
\usepackage[utf8]{inputenc}   
\usepackage[T1]{fontenc}
\usepackage{listings}

\lstdefinelanguage{text}{
  morecomment=[l]{\%},     % comments start with %
  morestring=[b]",         % strings in double quotes
}

\usepackage{array}
\usepackage{multirow}
\usepackage{ragged2e}
\usepackage{tabularray}
\usepackage{comment}
\usepackage{color,soul}
\usepackage{enumitem}

\usepackage{xcolor}

\usepackage{soul}    
\sethlcolor{yellow} 

\DeclareRobustCommand{\hl}[1]{#1}

\begin{document}


\title{StepWrite: Adaptive Planning for Speech-Driven Text Generation}

\author{Hamza El Alaoui}  
\email{helalaou@cs.cmu.edu}
\affiliation{ 
  \institution{School of Computer Science \\ Carnegie Mellon University}
  %\institution{Carnegie Mellon University}
   \city{Pittsburgh}
   \state{PA}
  \country{USA}
}

\author{Atieh Taheri}  
\email{ataheri@cs.cmu.edu}
\affiliation{ 
  \institution{School of Computer Science \\ Carnegie Mellon University}
  %\institution{Carnegie Mellon University}
  \city{Pittsburgh}
  \state{PA}
  \country{USA}
}
 
\author{Yi-Hao Peng}  
\email{yihaop@cs.cmu.edu}
\affiliation{ 
  \institution{School of Computer Science \\ Carnegie Mellon University}
  %\institution{Carnegie Mellon University}
  \city{Pittsburgh}
  \state{PA}
  \country{USA}
}

\author{Jeffrey P. Bigham}  
\email{jbigham@cs.cmu.edu}
\affiliation{ 
  \institution{School of Computer Science \\ Carnegie Mellon University}
  %\institution{Carnegie Mellon University}
  \city{Pittsburgh}
  \state{PA}
  \country{USA}
}

%   \institution{School of Computer Science \\ Carnegie Mellon University}

\renewcommand{\shortauthors}{El Alaoui et al.}

\begin{abstract}
People frequently use speech-to-text systems to compose short texts with voice. However, current voice-based interfaces struggle to support composing more detailed, contextually complex texts, especially in scenarios where users are on the move and cannot visually track progress. Longer-form communication, such as composing structured emails or thoughtful responses, requires persistent context tracking, structured guidance, and adaptability to evolving user intentions---capabilities that conventional dictation tools and voice assistants do not support. We introduce StepWrite, a large language model-driven voice-based interaction system that augments human writing ability by enabling structured, hands-free and eyes-free composition of longer-form texts while on the move. StepWrite decomposes the writing process into manageable subtasks and sequentially guides users with contextually-aware non-visual audio prompts. StepWrite reduces cognitive load by offloading the context-tracking and adaptive planning tasks to the models. Unlike baseline methods like standard dictation features (e.g., Microsoft Word) and conversational voice assistants (e.g., ChatGPT Advanced Voice Mode), StepWrite dynamically adapts its prompts based on the evolving context and user intent, and provides coherent guidance without compromising user autonomy. An empirical evaluation with 25 participants engaging in mobile or stationary hands-occupied activities demonstrated that StepWrite significantly reduces cognitive load, improves usability and user satisfaction compared to baseline methods. Technical evaluations further confirmed StepWrite's capability in dynamic contextual prompt generation, accurate tone alignment, and effective fact checking. This work highlights the potential of structured, context-aware voice interactions in enhancing hands-free and eye-free communication in everyday multitasking scenarios.

\end{abstract}

\ccsdesc[500]{Human-centered computing~Natural language interfaces}
\ccsdesc[500]{Computing methodologies~Natural language processing}
\ccsdesc[500]{Human-centered computing~Interactive systems and tools}
\ccsdesc[100]{Human-centered computing~Interaction techniques}
\ccsdesc[100]{Human-centered computing~Sound-based input / output}
\ccsdesc[100]{Human-centered computing~Empirical studies in HCI}
%%

\keywords{voice interfaces, adaptive planning, task decomposition, speech-to-text, text composition, hands-free writing, eyes-free writing, large language models}

%    \centering
%    \includegraphics[width=\textwidth, height=4.835in, keepaspectratio]{figures/teaser_3.png}
%    \caption{StepWrite enables hands-free, eyes-free writing through an adaptive Q\&A dialogue that incrementally elicits task-relevant context and automatically synthesizes a coherent draft, allowing users to compose text while simultaneously performing everyday activities. Each row illustrates a three-step interaction: (A) the user engages in a hands-busy task, (B) responds to contextual prompts from the AI agent via voice, and (C) receives a coherent, personalized draft. The scenarios shown—walking with a smartwatch and cooking at home—are two example use cases among many where StepWrite supports nonvisual, multitasking-friendly text composition.}
%    \label{fig:teaser}

\begin{teaserfigure}
   \centering
   \includegraphics[width=\textwidth, height=4.6in, keepaspectratio]{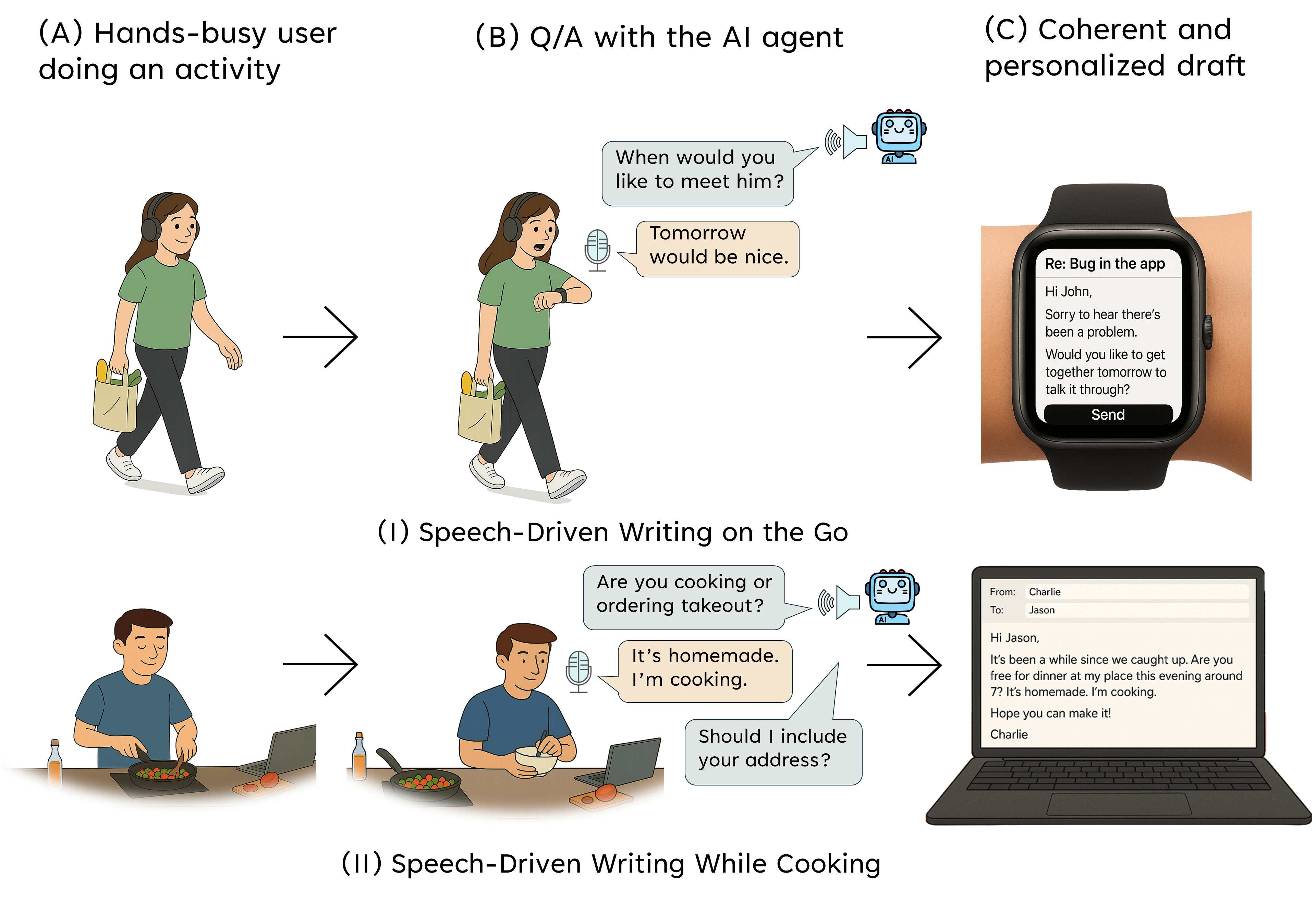}
   \caption{\hl{StepWrite enables hands-free, eyes-free writing through an adaptive Q\&A dialogue that incrementally elicits task-relevant context and synthesizes coherent drafts. Each row illustrates a three-step interaction: (A) the user performs a hands-busy activity, (B) engages in voice-based Q\&A with the AI agent, and (C) receives a personalized draft. The examples shown: (I) walking and (II) cooking highlight two of many multitasking contexts where StepWrite supports hands-free nonvisual composition.}}
   \label{fig:teaser}
\end{teaserfigure}


\maketitle

\section{Introduction}
% Task 

Voice-based interfaces have become increasingly prevalent in everyday computing, offering an appealing alternative to traditional keyboard input in hands-busy or eyes-free scenarios. Users routinely employ speech-to-text (STT) tools and voice assistants to compose short texts, set reminders, or issue commands in contexts such as driving, cooking, or walking. As speech recognition technology has matured, its speed and accuracy have made it a viable modality for many casual communication tasks~\cite{ruan2016speech, mehra2023gist}. However, despite these advances, current voice input systems remain ill-equipped for cognitively demanding activities like composing structured, long-form text. 

% SOTA 
Composing a complete email, persuasive message, or narrative explanation involves more than converting speech into text — it requires planning, organizing content, tracking structure, and revising based on evolving goals. Conventional STT systems offer limited support beyond basic transcription and rudimentary editing commands. Voice assistants, while interactive, operate in a command-response paradigm and lack memory, persistence, or the ability to manage document structure across multiple turns. As a result, users attempting longer-form writing via voice often experience high cognitive load, fragmented outputs, and substantial post-editing demands~\cite{schmidt2020classifying, mehra2023gist}.

% Gap in SOTA 
Recent work in AI-assisted writing tools has shown that large language models (LLMs) can enhance the writing process through completion, summarization, and rewriting capabilities~\cite{lee2022coauthor, shen2023beyond, coenen2021wordcraft}. These systems typically assume visual interfaces and manual input, such as co-authoring with keyboard prompts in a web-based editor. While generative voice assistants like ChatGPT’s Advanced Voice Mode (AVM) offer conversational writing support, they often lack scaffolding and structure, leaving users to navigate the planning and composition process on their own. Furthermore, these tools may not accommodate non-visual workflows and are not optimized for hands-free interaction in multitasking settings. As prior work has shown, users engaging with AI systems in such contexts may struggle to maintain intent, track content, or feel a sense of authorship~\cite{ippolito2022creative}.

% Idea 
This paper introduces StepWrite, a novel voice-based writing system designed to support structured, hands-free composition of long-form text. StepWrite transforms the writing process into a contextual, spoken dialogue, guiding users through a stepwise sequence of high-utility prompts that scaffold idea generation, elaboration, and narrative development. Powered by LLMs, StepWrite dynamically adapts its guidance based on prior user responses and document state, helping users maintain coherence and structure without visual feedback. The system is designed explicitly for multitasking and eyes-free contexts, enabling users to compose meaningful messages through voice alone, even while walking or engaging in other manual tasks.

Unlike conventional dictation or generic conversational assistants, StepWrite offers persistent context, tailored writing scaffolds, and adaptive prompting strategies that encourage user reflection and agency. Drawing from principles of pedagogical scaffolding~\cite{bereiter1982conversation, graham2007meta}, the system aims to reduce cognitive load by offloading planning and structure management to the model, allowing users to focus on content.

To investigate the effectiveness of this structured, voice-first approach to hands-free writing, we address the following research questions:
\begin{itemize}[leftmargin=*]
    \item \textbf{RQ1:} How does a structured, voice-guided writing assistant like StepWrite affect the cognitive effort and revision burden compared to conventional dictation and generative AI voice tools?
    \item \textbf{RQ2:} How does StepWrite influence the quality, readability, and semantic alignment of user-generated texts relative to other voice-based writing approaches?
    \item \textbf{RQ3:} How do users perceive the usability, emotional engagement, and overall writing experience of StepWrite during multitasking, hands-busy scenarios?
    \item \textbf{RQ4:} What is the utility of dynamically generated scaffolding prompts in StepWrite, and how often do these prompts contribute meaningfully to users' final written outputs?
\end{itemize}

We evaluated StepWrite in a within-subjects user study with 25 participants, who completed two writing tasks (email composition and reply) using three tools: (1) StepWrite, (2) Microsoft Word’s built-in dictation feature, and (3) ChatGPT AVM. Participants engaged in both stationary and mobile (walking) conditions to simulate real-world multitasking scenarios. We measured revision effort, text quality, temporal efficiency, and question necessity, between spoken drafts and revised outputs. Subjective feedback was collected using NASA TLX for workload, SUS for usability, a custom Emotional Engagement Questionnaire (EEQ), and a custom Hands-Free Writing Tools Assessment instrument.

Our findings show that StepWrite significantly reduced revision effort and cognitive workload compared to both baselines, while producing outputs more closely aligned with users’ intended meaning. Participants reported high usability, low frustration, and strong emotional engagement with the system’s guided process. Moreover, technical analyses of StepWrite’s prompt quality and semantic contribution revealed that over 77\% of system-generated prompts were essential to final outputs.

This work makes the following contributions:
\begin{itemize}[leftmargin=*]
    \item The design and implementation of StepWrite, an LLM-powered system that supports hands-free, structured writing through contextual, voice-based scaffolding;
    \item A controlled user study comparing StepWrite to both conventional dictation and conversational LLM voice interaction across mobile and stationary use cases;
    \item Empirical evidence demonstrating that structured voice interaction, through adaptive planning, reduces cognitive load, improves semantic alignment, and enhances user satisfaction in hands-free, writing tasks.
\end{itemize}

\section{Related Work}
\subsection{Voice-Based Text Composition}

Speech-to-text (STT) technologies have become increasingly integrated into daily communication workflows, particularly for short-form content like messages~\cite{ruan2018comparing, xu2007empirical, haas2020they}, reminders~\cite{liu2023older, esquivel2024voice, pradhan2020use}, and voice memos~\cite{thomas2024me, hohne2024automatic}. STT systems offer convenience in hands-busy or eyes-free contexts, such as driving or exercising, but their use for long-form writing remains limited~\cite{ruan2016speech}. This is largely due to their linear transcription approach, which lacks support for revision, structural planning, or content reorganization.

Prior work on voice-driven text entry systems, such as VoicePen~\cite{harada2007voicepen}, explored multimodal interactions where users combined voice and gestures for document annotation and creation. Other systems attempted to enhance dictation with speech commands for editing. Yet, these systems often required visual attention, struggled with maintaining document coherence, and offered limited support for context persistence or complex branching structures.

Conversational voice agents, including Apple Siri~\footnote{Apple Siri: \url{https://www.apple.com/siri}}, Google Assistant~\footnote{Google Assistant: \url{https://assistant.google.com}}, and ChatGPT Voice Mode~\footnote{ChatGPT: \url{https://chatgpt.com}}, have introduced more dynamic turn-taking and basic task completion. However, these systems typically operate with shallow task representations and lack a high-level understanding of writing goals or structure~\cite{volkel2021eliciting}. Research has shown that multi-turn conversations with an assistant (where the system can ask follow-up questions or confirmations) rather than the traditional single-turn Question and Answer (Q\&A) format has shown to be strongly preferred by the users~\cite{burggraf2022preferences}. Recent innovations like Rambler~\cite{lin2024rambler} seek to address this gap by supporting iterative editing of spoken content through gist extraction, summarization, and semantic manipulation, but remain dependent on visual interfaces. Our work builds on this space to enable a structured, non-visual writing process through contextual dialogue.

\subsection{Intelligent Writing Assistance}

AI-powered writing tools have made notable progress in supporting users during the composition process. Tools such as Grammarly~\footnote{Grammarly: \url{https://www.grammarly.com}} and QuillBot~\footnote{QuillBot: \url{https://quillbot.com}} offer real-time grammar correction, paraphrasing, and tone adjustments. More advanced systems now integrate large language models (LLMs) to provide content suggestions, sentence completions, and summarization capabilities~\cite{shi2023effidit}.

Research in this space has focused on co-authoring interfaces~\cite{lee2022coauthor}, idea generation~\cite{wan2024felt, huang2023conveying} and narrative scaffolding~\cite{coenen2021wordcraft, mirowski2023co}. Ippolito et al.\cite{ippolito2022creative} evaluated Wordcraft, an LLM-powered editor used by professional writers, revealing the importance of workflows that support elaboration and preserve authorial control. Similarly, Shen et al.~\cite{shen2023beyond} emphasized modular support for complex expository writing, highlighting the need for AI systems that assist reading, synthesis, and composition across stages of the writing pipeline. Recent work ~\cite{el2023building} provides a comprehensive survey of AI-powered dialogue systems, emphasizing both the strengths and current limitations of pipeline and end-to-end architectures. Their findings underscore the challenges of adaptivity, error propagation, and interpretability, concerns also central to StepWrite’s goal of providing structure and transparency through voice-based interaction. \citet{shao2024assisting} introduced STORM, a system that enhances long-form article generation by simulating multi-perspective, question-driven research during the pre-writing stage. Their work highlights the value of guided question-asking and outline planning, aligning closely with StepWrite’s emphasis on structured guidance and dynamic scaffolding in composition. \citet{goodman2022lampost} developed LaMPost, an AI-assisted email writing tool for adults with dyslexia, which incorporated features such as outlining, rewriting, and suggestion generation. Their evaluation highlighted both the promise and limitations of LLM-based writing assistance, especially the need for greater accuracy, personalization, and user control—principles that also guide StepWrite’s interactive scaffolding approach.

While these tools have demonstrated creative potential, they often rely on visual interfaces and assume constant user attention. StepWrite departs from passive suggestion-based models by engaging the user in an active dialogue, prompting reflection, elaboration, and decision-making through voice. Drawing inspiration from pedagogical scaffolding~\cite{graham2007meta, bereiter1982conversation, hu2025enhancing}, the system prompts users to elaborate, reflect, and refine their ideas step-by-step, functioning more like a conversational writing coach. This method encourages progressive development of content while alleviating the cognitive overhead associated with planning and organization.

\subsection{Hands-Free and Multitasking Interfaces}

The design of hands-free and eyes-free interfaces has been studied extensively in wearable computing, automotive interaction, and mobile computing contexts~\cite{larsen2020hands, oakley2007designing, taheri2021design, hendrik2006designing}. Systems like WatchIt~\cite{perrault2013watchit} and wearable AR displays~\cite{lee2024gazepointar, ch2022gesture, khurana2024just, korkiakoski2024preference} explored gesture and voice inputs to support lightweight task execution. In in-vehicle environments, voice interfaces are used to reduce driver distraction by enabling eyes-free navigation and communication~\cite{strayer2016talking, tsimhoni2004address}.

However, these systems often focus on command-and-control tasks, such as launching applications or issuing navigation instructions, rather than open-ended cognitive tasks like writing. Multitasking research has emphasized reducing cognitive load and improving task switching~\cite{nith2024splitbody}, but less attention has been paid to how users can engage in complex content creation while physically or visually occupied. For example,~\citet{edwards2019multitasking} found that using a voice assistant while writing can significantly disrupt the writing process when the task involves content generation as opposed to simple transcription.

Recent systems such as GlassMail~\cite{zhou2024glassmail} demonstrate the potential of LLMs on wearable devices to support hands-free composition, but are constrained by visual attention demands. WearWrite~\cite{nebeling2016wearwrite} explored crowd-powered writing from smartwatches, highlighting the feasibility and value of distributed collaboration in mobile writing. StepWrite builds on this trajectory by enabling cognitively rich writing tasks in multitasking environments through structured audio prompting. By minimizing visual demands and adapting its prompts to the user's evolving input, it supports a fluid writing flow in mobile, everyday contexts.

\subsection{Adaptive Planning and LLM Interaction}

Large language models have demonstrated remarkable capabilities in reasoning, planning, and language generation. Recent techniques such as chain-of-thought prompting~\cite{wei2022chain}, tool use via prompting~\cite{schick2023toolformer}, and conversational memory~\cite{xu2023rewoo} have explored how LLMs can assist users in decomposing tasks and managing complex information flows.

Adaptive planning strategies with LLMs function as a form of scaffolding by dynamically adjusting support to match a user's evolving needs during the writing process. These strategies include techniques such as dynamic prompting, reflective planning, and user-guided questioning~\cite{feng2025reflective}, all of which help writers navigate complex tasks by breaking them into manageable steps. For example, Miura et al.~\cite{miura2025understanding} introduced a QA-based email writing system that provides iterative support through interactive question-answering, exemplifying how adaptive scaffolding can facilitate progressive idea development. This approach informed StepWrite’s design for enabling dynamic and discovery-oriented writing workflows. Similarly, Dhillon et al.~\cite{dhillon2024shaping} examined how different levels of LLM assistance influence writing productivity and ownership, finding that well-calibrated scaffolding can significantly improve outcomes by maintaining a balance between guidance and user autonomy.

These approaches are primarily deployed in screen-based tools where users can visualize and modify generated content. Few systems have explored the potential of real-time, non-visual scaffolding via audio. StepWrite contributes a novel interaction modality: adaptive, voice-guided writing scaffolding. It incrementally builds a shared context with the user, asking clarifying and elaborative questions one at a time rather than presenting static, pre-scripted prompts. This method not only enhances contextual relevance but also supports user agency and personalization during composition.

\section{StepWrite}

\hl{StepWrite is an intelligent, voice-first writing system designed to provide \textit{context-aware scaffolding} for hands-free composition with minimal visual attention. It guides users through a stepwise, adaptive dialogue that collects essential content, infers appropriate tone, and produces a coherent draft—entirely through speech. Built as a responsive React-based web application, StepWrite runs in the browser and requires only a modern internet-connected device. All heavy computation is offloaded to cloud APIs, enabling responsive performance even on low-tier laptops, desktops, wearables, and mobile devices. We tested StepWrite across heterogeneous form factors and operating systems, and observed consistent performance, enabling real-world use in multitasking scenarios such as composing emails, messages, or notes while cooking or commuting. StepWrite is available as open-source software.}\footnote{\url{https://github.com/helalaou/StepWrite}}

\subsection{Interaction Model and Scaffolding}

StepWrite is grounded in the pedagogical notion of \textit{scaffolding}—a method of support that provides structure while preserving user autonomy. Rather than expecting users to dictate a message in a single stream of thought, the system breaks writing into manageable subtasks, asking focused, relevant questions one at a time. Each question is adaptively generated by an LLM based on prior responses, user intent, and inferred genre. These questions help users elaborate on their goals, clarify context, and make decisions about tone, audience, or timing—all without needing to plan the entire message upfront.

This interaction model serves dual purposes: it reduces cognitive load by externalizing planning, and it provides semantic structure that enables downstream modules to compose high-quality drafts. The process is fully hands-free and eyes-free, navigable via voice commands, and reversible: users can skip questions, revise earlier answers, and re-enter the editor at any point. The system also supports keyboard-only and hybrid (text + speech) modes, allowing users to benefit from the adaptive planning and structured prompts even without using the hands-free features.

\subsection{Audio Input and Voice Command Handling}

StepWrite continuously listens for user input using an efficient audio pipeline designed to balance responsiveness with accuracy. Incoming microphone data is first passed through a noise-filtering layer, tuned to discard ambient environmental sounds (e.g., keyboard taps, wind, or appliance hum). Users may customize this filtering behavior based on their environment. Next, a voice activity detection (VAD) algorithm identifies the onset and end of speech. The VAD model distinguishes natural pauses from utterance completions using a \textit{thinking window} that can either be manually tuned or adaptively adjusted over time, allowing the system to accommodate user-specific pacing and pause duration. Only speech identified as human voice is processed further.

Before any transcription occurs, StepWrite performs client-side macro command recognition. Each transcribed segment is first checked for voice commands such as \textit{“skip question,” “repeat that,” “go back,” “pause output,” “go to editor”}, or user-defined equivalents. \hl{To avoid accidental triggers from natural speech, all built-in commands were two-word phrases (e.g., “skip question”) rather than single words like “skip.” Commands are matched using a token-level fuzzy matching algorithm that supports both exact and partial phrase recognition, with a minimum cosine similarity threshold of 0.85. This enabled recognition of phrasings such as “please skip this question” without interfering with typical user responses. A complete list of supported voice commands is provided in Appendix}~\ref{voice_commands}.

\hl {During the experiment, no speech segments were mistakenly classified as commands. The system is configurable to support custom macros, but for consistency, we used only the standard command set during the study. In future deployments—particularly on wearable devices—gesture-based shortcuts (e.g., shaking a smartwatch to navigate between next and previous questions) could serve as an alternative hands-free control modality.}

If no command is detected, the speech segment is sent via API call to OpenAI’s Whisper model for transcription. The resulting transcript is then passed to the question engine if the user is answering a prompt, or to the text editor engine if the user is executing a command within the editor (e.g., “go back to questions,” “play output again”). Speech recognition outputs are streamed in real time, with visual confirmation and audio playback available for each response.

\subsection{Modular Prompt Pipeline: Q\&A to Final Output}

Once transcribed, the user’s answer enters the Q\&A pipeline. The system maintains a linear conversation graph consisting of prompt–
response pairs, which are evaluated after each turn. An LLM receives the full Q\&A history—along with context flags and optional memory cues—and generates the next question. Questions are crafted to elicit actionable, factual details without suggesting formatting, wording, or stylistic preferences. Memory is used selectively to enhance prompt quality—for example, by recalling preferred task types, common recipients, user preferences, or frequently used phrases from previous sessions. With each new prompt, the full Q\&A history is appended and the LLM is asked to determine whether sufficient context has been collected. The model returns a Boolean flag, \textit{followup\_needed}, \hl{which serves as a functional end-of-sequence signal. When this flag is false, the system triggers text generation.}

Throughout the interaction, users may navigate forward or backward through the Q\&A flow. After each question is answered, the updated conversational state is stored. Users can issue commands such as “modify answer” after navigating to a given question. Upon detecting such edits, StepWrite removes all subsequent question–answer pairs. This prevents later content from becoming irrelevant or inconsistent with the updated response. To avoid repetition, the system is instructed to skip over any questions that were previously marked as skipped by the user, preventing their reintroduction in subsequent generations.

After the dialogue concludes, StepWrite invokes the \textit{tone classification module} which analyzes the full Q\&A history to determine an appropriate voice for the output. The classifier draws on lexical choice, sentiment, and inferred context to assign a tone label (e.g., \textit{empathetic}, \textit{assertive}, \textit{apologetic}), which is passed along with the \textit{text generation module}. The \textit{text generation module} produces a draft and then passes it to the \textit{fact checking module}.

\subsection{Text Generation and Fact Checking}

Text generation begins when all question–answer pairs and tone metadata are passed to an the \textit{text generation module}. This module synthesizes a complete draft, constrained to reference only user-provided facts and context. Memory is optionally passed in this stage as well—allowing the model to align with the user’s prior conventions or use domain-specific content that was previously surfaced. 

The resulting draft is then routed to the \textit{fact-checking module}, which performs a multi-step verification process to identify factual inconsistencies, omissions, or contradictions relative to the Q\&A input. For each issue detected, the fact checker returns a structured report that includes a type ("missing", "inconsistent", "inaccurate", or "unsupported"), a detail describing the issue, and a QA reference pointing to the relevant excerpt from the question–answer history. This metadata is then passed to the \textit{text adjustment module}, which selectively rewrites only the problematic segments while preserving tone, structure, and fluency. The revised text is returned to the fact checker for re-evaluation, and this loop continues until no further issues are found or a user-defined maximum number of passes (default: 10) is reached. Memory, when available, is used to verify persistent facts across sessions. In testing, this verification-adjustment loop added no more than 7 seconds of latency for short-form drafts; we expect this duration to scale with document length and complexity. System prompts used in these modules are referenced in Appendix \ref{appendix:prompts}.

Importantly, the fact checker enforces information integrity while remaining agnostic to superficial preferences. It avoids enforcing rigid template structures or unnecessary rephrasings, focusing solely on semantic alignment.

\subsection{Session Control and Editor Transition}

StepWrite supports both linear progression and cyclical revision. Users can move freely between dialogue and editor views, repeat system prompts, replay prior answers, or issue corrective commands via voice. All interaction state is persistently stored in a session object indexed by UUID and maintained throughout the session, ensuring continuity even across temporary disconnects.

Auditory and visual feedback are integrated to ensure clarity: system prompts are accompanied by synthesized text-to-speech (TTS) playback, speech detection is visualized via waveform animation, and navigation events—such as modifications, skips, and command triggers—are signaled through both auditory chimes, audio cues stating the commands being executed, and subtle interface animations.

\subsection{Scenario: Composing While Occupied}

Consider a user who wishes to respond to a time-sensitive email while unpacking groceries. With both hands full, they initiate StepWrite via a voice trigger: "Hey StepWrite! Help me with this email." The system reads the original email aloud, then asks: ``What would you like to say in response?'' As the user continues unloading bags, StepWrite proceeds through a set of targeted questions—``Do you want to accept the offer?'', ``Is there a specific time that works best?'', ``Should we confirm any details with them now?''—each spoken aloud and answered conversationally. At any point, the user may say ``go back,'' ``change my answer,'' or ``that’s enough,'' without breaking flow. A complete, fact-checked draft is presented once the user issues the “finish” command, ready for final review or automatic sending.

In summary, StepWrite provides a modular, cloud-backed, and device-agnostic scaffolding architecture for hands-free writing. By chaining audio processing, adaptive Q\&A planning, tone selection, structured drafting, and iterative fact-checking, it forms an integrated and responsive authoring system. Its support for non-linear interaction, fuzzy macro invocation, robust correction logic, and cross-platform deployment enables rich composition in scenarios where traditional writing tools fall short.

\subsection{System Architecture Overview}

System-level, speech-to-text, and text-to-speech diagrams are provided in Appendix~\ref{appendix:diagrams} (Figures~\ref{fig:stepwrite_system_diagram}–\ref{fig:tts_pipeline}).

The following table summarizes the core models used across each stage of the StepWrite pipeline. These models were chosen based on a balance between speed, accuracy, and compatibility with their assigned tasks.

\begin{table}[t]
  \centering
  \caption{Core components and models in the StepWrite pipeline.}
  \label{tab:models}
  \begin{tabular}{ll}
    \toprule
    \textbf{Module} & \textbf{Implementation} \\
    \midrule
    Speech-to-Text & \texttt{whisper-1} \\
    Text-to-Speech & \texttt{tts-1} + caching \\
    Question Generation & \texttt{4.1-mini} \\
    Text Generation & \texttt{4.1-mini} \\
    Fact Checking & \texttt{4o-mini} \\
    Tone Classification & \texttt{gpt-4.1} \\
    %Dependency Analysis & \texttt{o3} \\ this was added later
    Memory Management & JSON-based store \\
    Audio Processing & VAD + Custom filters\\
   
    Command Recognition & Custom fuzzy engine \\
    Voice Activity Detection (VAD)  & Silero VAD \\
    \bottomrule
  \end{tabular}
\end{table}

\section{Pilot Study}
After implementing a minimally functional prototype of StepWrite, we conducted a formative study to evaluate how users interacted with the system in realistic hands-free scenarios. Rather than beginning with speculative design probes or mockups, we prioritized testing a live version of the system early in development to observe actual usage and gather actionable feedback. The goal was to uncover usability challenges, surface desired features, and inform iterative refinements before formal evaluation.

The initial implementation of StepWrite already supported dynamic question generation, voice-based navigation, and full draft composition. Users could proceed through system-generated prompts using voice commands like ``next question,'' ``previous question,'' and ``modify answer,'' and switch between answering questions and editing text. The system determined when to stop asking questions based on inferred completeness, and it automatically compiled the responses into a polished draft. A basic visual interface displayed ongoing progress and draft content, though with limited animations and minimal visual cues. While this version provided end-to-end functionality, it lacked personalization, non-visual feedback, and fine-grained control over flow—issues we targeted during iterative testing.

We worked with 4 university students (2 male, 2 female; ages 22–24), all native English speakers with varied experience using speech-based systems. Each participant selected a realistic writing scenario of their choice, including email replies, scheduling messages, casual updates, and short announcements. They completed the writing task using the prototype while either walking slowly or performing stationary activities (e.g., handling small objects, light snacking). These sessions helped us evaluate how the system performed in real-world multitasking conditions and uncover early usability concerns.

Participants responded positively to the system’s incremental guidance. Several appreciated being asked focused, specific questions instead of needing to dictate full messages at once. One participant noted, “It felt more manageable than just talking into a void — I didn’t have to plan everything in my head.” Another described StepWrite as more of a “thinking partner” than a writing assistant—something that could prompt ideas, support brainstorming, and then generate a polished draft based on the interactive exchange.

However, the study surfaced key areas for improvement. Participants struggled when unexpected interruptions occurred—such as being approached mid-task or needing to attend to their surroundings—often losing track of where they left off. We added a dedicated pause/resume command pair that allowed users to momentarily disengage and return to the same point in the writing process without resetting context or content. This feature was especially valued by those who envisioned using StepWrite in semi-public settings like gyms, cafes, or while commuting.

Users also expressed the need for a “finish” command that could end the interaction early. While StepWrite was designed to scaffold writing through several focused prompts, participants noted that in some cases—such as quick notes or casual replies—they only wanted to provide a small amount of context and have the system generate a complete draft without going through the entire Q\&A flow. In response, we added functionality allowing users to signal that they had said enough and wished to stop receiving questions, at which point StepWrite would generate an output based on the information gathered so far.

While the original prototype offered visual cues on the display, participants noted that this was insufficient in eyes-free scenarios. We added both audio feedback (e.g., light clicks and confirmation tones) and more detailed visual indicators such as “moving to next question,” “returning to previous,” and “generating final output” to increase transparency and reduce uncertainty during use.

Participants also expressed interest in a system that could learn from them over time. In response, we implemented a lightweight memory layer that stores user preferences and basic recurring information. Users can optionally predefine facts about themselves (e.g., name, role, writing style, calendar patterns), or allow the system to build up this memory gradually across sessions. This personalization layer enables StepWrite to tailor prompts, omit redundant questions, and improve drafting relevance based on prior interactions.

Additionally, participants noted a desire to use their own phrasing for voice commands. To support this, we added a customizable command-mapping interface, allowing users to rename any core command and define multiple trigger phrases (macros) per function. For example, a user might set both “stop now” and “that’s enough” as equivalents for the finish command. These small adjustments gave participants greater control and comfort in tailoring the interaction to their own speaking habits.

During testing in ambient environments such as student lounges, we observed additional issues with noise interference and unintentional activations. We integrated noise filtering, voice activity detection, and automatic sensitivity adjustment to improve robustness across varied environments.

This formative work took place after an initial working version of StepWrite had been implemented, and directly informed the feature set, robustness improvements, and user experience enhancements evaluated in the main study.

\section{Methods}

\subsection{Participants}

We recruited 25 participants (\textit{M}\textsubscript{age} = 24.80, \textit{SD} = 4.39; range = 18–35) from the university community and the broader metropolitan area via campus mailing lists, online postings, and local outreach. Participants included students, researchers, educators, engineers, and healthcare professionals.

12 participants identified as male, 11 as female, and 2 as non-binary or preferred not to disclose. Most were highly confident with digital tools (\textit{M} = 4.80/5 on confidence scale) and regularly engaged in diverse writing tasks, including messaging, note-taking, and document authoring.

While the majority reported no writing-related impairments (84.0\%), a small number identified physical or learning disabilities (8.0\% each). Most participants (96.0\%) expressed potential interest in using a guided, hands-free writing assistant, with many describing scenarios—such as commuting, cooking, or experiencing fatigue—where multitasking made traditional writing inconvenient or inaccessible. Participants had varied levels of exposure to AI-powered writing tools and speech-to-text interfaces. They were compensated at a rate of \$20 per hour for their time.

\begin{table}[]
    \caption{Participant Demographics and Writing Background}
    \label{tab:participant_summary}
    \setlength{\tabcolsep}{8pt}
    \renewcommand{\arraystretch}{1.3}
    \newcolumntype{C}[1]{>{\centering\arraybackslash}m{#1}}    
    \centering
    \begin{tabular}
    {C{.15\linewidth}|C{.2\linewidth}C{.5\linewidth}}
        \toprule
         & \textbf{Measure} & \textbf{Summary} \\ 
         \midrule
        \multirow{3}{=}{\textbf{Basic Demographic}} & Age & Mean = 24.8, SD = 4.39, range 18–35 \\ \cline{2-3}
         & Gender & 48\% Male, 44\% Female, 8\% Other \\ \cline{2-3}
         & Primary Language & 60\% English only, 40\% multilingual \\ \hline
        \textbf{Education} & Level & 60\% College, 32\% Graduate, 8\% Other \\ \hline
        \multirow{4}{=}{\textbf{Tech \& Writing}} & Tech Confidence & Mean = 4.8 (1–5 scale) \\ \cline{2-3}
         & STT Use & 72\% used it before, 28\% never used \\ \cline{2-3}
         & Writing Challenges & Common issues: phrasing, structure, editing \\ \cline{2-3}
         & Guided Tool Interest & 96\% expressed interest (answered "yes" or "maybe") \\ \bottomrule
    \end{tabular}
\end{table}

\subsection{Study Design}

We employed a within-subjects design in which participants completed two writing tasks—one \textit{write} and one \textit{reply}—using each of three hands-free writing tools: \textbf{StepWrite} (structured AI guidance), \textbf{ChatGPT Advanced Voice Mode} (conversational generative AI), and a \textbf{Dictation Tool} (Microsoft Word’s built-in speech-to-text).

Each participant completed all six tool-task combinations, counterbalanced using a Latin square design. To simulate real-world constraints on attention and mobility, we introduced two activity contexts: \textit{stationary} and \textit{movement}. Participants completed one task under each condition, with the assignment of activities randomized per task. Stationary activities included solving a Rubik’s cube, folding origami, snacking, or selecting items from a snack table; movement activities involved walking at a comfortable pace within a defined 4~$\times$~2.5 meter area. Two participants with mobility disabilities completed both tasks under stationary conditions using adapted hands-occupied activities. The study session lasted approximately 60--70 minutes per participant.

\subsection{Apparatus}

All tools ran on a MacBook Pro (M3 Pro, 36GB RAM) connected to a small external display positioned approximately 2 meters from participants. To simulate conditions where visual attention may be intermittently disrupted (e.g., mobile or multitasking settings), the external display was intermittently disabled for brief periods during task execution. This was done to encourage participants to rely primarily on auditory cues—closer to real-world scenarios involving wearables or heads-free interaction.

Participants wore a Poly Voyager 4320 headset with an integrated noise-canceling microphone throughout the study. A wireless keyboard and mouse were provided solely for text editing during the revision stage. All equipment (headset, keyboard, mouse) was sanitized between sessions.

\textbf{StepWrite} was implemented as a client-server web app (described earlier), offering structured, AI-guided writing assistance through interactive Q\&A-based dialogue. For the purposes of this study, memory and personalization functionalities were disabled.

\textbf{ChatGPT Advanced Voice Mode (AVM)} was accessed via OpenAI’s official ChatGPT interface (Plus subscription tier), enabling conversational, open-ended voice interaction.

\textbf{Dictation Tool} employed Microsoft Word’s built-in speech-to-text feature, selected for its widespread familiarity and command support.

\subsection{Procedure}

\hl{This study was approved by the IRB office at Carnegie Mellon University (protocol number: \textit{STUDY2024\textunderscore00000515}). }
Participants first provided informed consent and received a study overview. They were then introduced to two tasks: \emph{\textbf{Write Task}} and \emph{\textbf{Reply Task}}. Each task were explained by the researcher and printed on a reference sheet. The Write Task involved composing a casual email inviting friends or family to a weekend outing. The Reply Task required responding to a professor’s inquiry about a missing final project submission. Participants were given time to review the task sheet and were allowed to keep it with them throughout the study.

Immediately before using each tool, the researcher provided a live demonstration. For both StepWrite and the Dictation Tool, participants also received a reference sheet listing available voice commands. They were given as much time as they needed to review these commands, and most participants indicated readiness to begin the study within two minutes. For ChatGPT AVM, the researcher demonstrated typical interaction patterns and explained the kinds of conversational prompts and voice instructions participants could use to request modifications, clarifications, or readbacks.

Each task consisted of two phases: \emph{\textbf{Drafting}} and \emph{\textbf{Revision}}. In the Drafting phase, participants composed initial drafts entirely via voice—engaging interactively with StepWrite and ChatGPT AVM, or dictating directly into Microsoft Word. 

In the Revision phase, participants were instructed to revise their initial draft using the keyboard and mouse only, with the explicit objective of producing a message that they would personally send. Revision criteria included content, tone, formatting, and clarity. \hl{This manual revision step was applied uniformly across all tool to ensure a consistent evaluation of final output quality. While StepWrite fully supports hands-free drafting and in-flow voice-based revision (e.g., via “modify answer” command), voice-based editing during the final revision stage is a planned feature for future versions.}

\hl{Drafting Time was measured from the start of the Q\&A interaction until the draft was shown, including any voice-based edits made during this phase. Revision Time began once the draft appeared and included only keyboard-and-mouse edits. If users navigated back into the Q\&A flow after viewing the draft (e.g., to modify a previous answer by voice), Revision Time was paused and Drafting Time resumed until they returned to the editor.}

Participants completed tasks under alternating stationary and movement conditions, randomized to control for order effects and minimize potential bias. After using each tool to complete the writing and replying tasks, participants filled out a corresponding post-task questionnaire. This process was repeated for each tool. At the end of the session, participants were debriefed by the researcher.
% Upon completing all tasks, participants filled out post-study questionnaires and were debriefed by the researcher.

\subsection{Measures}

We collected both objective and subjective measures to evaluate participants' interaction with each hands-free writing tool. Our measures spanned \hl{seven} key dimensions: \textit{revision effort}, \textit{text quality}, \textit{temporal efficiency}, \textit{question necessity}, \hl{\textit{tone assessment}}, \textit{user experience}, \hl{and \textit{order effect} }.

\subsubsection{Revision Effort}

To quantify the extent of post-generation editing required, we calculated the number of word-level insertions, deletions, and replacements between each participant’s original drafts (speech-generated) and final drafts (manually revised). This was operationalized using the Ratcliff and Obershelp sequence-matching algorithm~\cite{ratcliff1988pattern, wikipediaGestaltPatternMatching}, which identifies meaningful changes at the phrase and sentence level. This metric reflects both the quantity and nature of participant effort required to bring AI-generated or dictated drafts to a personally acceptable standard.

\subsubsection{Text Quality and Structure}

We analyzed four aspects of textual output to assess writing quality and linguistic coherence:

\begin{itemize}[leftmargin=*]
    \item \textbf{Readability:} Measured with the Flesch Reading Ease (FRE)~\cite{flesch1948new}, where higher scores (max = 100) indicate simpler, more accessible language; and the Flesch–Kincaid Grade Level, estimating the U.S. school grade level required for comprehension.
    
    \item \textbf{Sentence Structure:} Average sentence length was computed for both original and revised drafts to assess syntactic coherence and structural clarity. Large reductions signal the correction of run‑on or disfluent output.
    
    \item \textbf{Lexical Diversity:} Type–token ratio (TTR)—the ratio of unique words to total words—captured vocabulary richness. Higher TTR indicates more varied and expressive language.
    
    \item \textbf{Semantic Diversity:} To quantify meaning‑level changes, we embedded the original and revised texts with \textit{gte-\allowbreak Qwen1.5-\allowbreak 7B-\allowbreak instruct}\footnote{\url{https://huggingface.co/spaces/mteb/leaderboard}}. Cosine similarity between the two embeddings was converted to diversity as $(1-\text{similarity})$. Lower scores therefore signify that a tool’s first draft already matched the user’s intended meaning, whereas higher scores reflect substantial semantic edits.
    
    \item \hl{\textbf{Final Draft Length:} We computed the total word count of each submitted draft to evaluate how writing tool and task type influenced overall verbosity. These statistics help assess whether structured scaffolding led to longer or more detailed outputs.}

\end{itemize}

All text analyses were performed separately on original and revised outputs to examine how much participants altered their language during the editing phase.

\subsubsection{Temporal Metrics}

We logged time spent on each phase of the writing task:

\begin{itemize}[leftmargin=*]
    \item \textbf{Drafting Time:} Measured from the start of voice input until the user indicated they had completed drafting.
    \item \textbf{Revision Time:} Measured from the moment the participant began editing until they confirmed completion.
    \item \textbf{Total Task Time:} Combined duration of the drafting and revision phases.
\end{itemize}

These metrics helped assess the time-efficiency tradeoffs between the three tools.

\subsubsection{Question–Necessity Analysis (StepWrite Only)}

Because StepWrite’s workflow is driven by AI‑generated follow‑up questions, we conducted an additional corpus study to gauge the \emph{necessity} of those questions.  
For every StepWrite session we logged each system question, the participant’s spoken answer, and the participant’s \emph{final} text after revisions.  
Two independent annotators classified every question into one of three mutually exclusive categories:

\begin{description}[leftmargin=*]
  \item[\textit{necessary}] answering the question contributed information that appears in the participant’s final text; omitting the answer would have prevented reproduction of that final output;
  \item[\textit{unnecessary}] the answer was ultimately unused or contradicted in the final text;
  \item[\textit{skipped}] the participant explicitly invoked the ``skip'' command or otherwise declined to answer.
\end{description}

We report raw counts, percentages, and an \textbf{Essential Question Fraction (EQF)} defined as

\[
\text{EQF}= \frac{\text{necessary}}{\text{necessary}+\text{unnecessary}+\text{skipped}}
\]

that is, the proportion of \emph{all} StepWrite questions that proved indispensable for the revised text. \hl{In addition to annotating question necessity, we also analyzed the average number of questions received, answered, and skipped per task, as well as the average word counts of StepWrite-generated questions and user responses.}

\subsubsection{\hl{Evaluation of Tone Classification (StepWrite Only)}}

\hl{Because no widely accepted taxonomy of written tones exists in the scholarly literature, we developed a 14-category schema informed by two tone-of-voice web-based sources}\footnote{{NNgroup: \url{https://www.nngroup.com/articles/tone-of-voice-dimensions}}}\,\footnote{{Grammarly: \url{https://www.grammarly.com/blog/writing-techniques/types-of-tone}}} \hl{and linguistic works on tone and voice}~\cite{stoehr1968tone}.

\hl{The resulting categories are:} \emph{formal}, \emph{informal}, \emph{friendly}, \emph{diplomatic}, \emph{urgent}, \emph{concerned}, \emph{optimistic}, \emph{curious}, \emph{encouraging}, \emph{surprised}, \emph{cooperative}, \emph{empathetic}, \emph{apologetic}, and \emph{assertive}. \hl{These tones reflect the range of stylistic intentions commonly encountered in everyday email communication; full definitions appear in Appendix}~\ref{tones}.

\hl{For evaluation, we first sampled 200 messages from the university-email corpus of Singh et al.}~\cite{singh2018email}. \hl{Three trained raters labeled each message with our tone schema. The vast majority fell into five tones—}\emph{formal}, \emph{cooperative}, \emph{friendly}, \emph{empathetic}, and \emph{urgent}—\hl{while the remaining categories were sparsely represented or absent. To obtain balanced coverage across all 14 tones, we authored and annotated an additional 150 messages, yielding a comprehensive 350-message benchmark for tone-classification evaluation. Although a single message can contain several tonal nuances, annotators were instructed to assign the one} \emph{dominant} \hl{tone that best captured the overall intent. This curated dataset will be released publicly.}

\subsubsection{Subjective Metrics / User Experience}

Participants completed a series of standardized and custom post-condition questionnaires:

\begin{itemize}[leftmargin=*]
    \item \textbf{NASA Task Load Index (TLX)}~\cite{NASATLX}: Assessed perceived workload across six dimensions—mental demand, physical demand, temporal demand, effort, performance, and frustration. We used the raw TLX (unweighted) for analysis.
    
    \item \textbf{System Usability Scale (SUS)}~\cite{SUS}: A 10-item measure of perceived system usability, scored from 0–100. A score above 68 is generally interpreted as above-average usability.
    
    \item \textbf{Emotional Experience Questionnaire (EEQ)}: Assessed participants' emotional engagement with the writing tool using a customized survey adapted from the User Engagement Scale by~\citet{o2010development}. Our adapted scale included five items specifically targeting users' \emph{enjoyment}, \emph{motivation}, \emph{stress reduction}, \emph{creativity}, and \emph{overall engagement}, aligning with O'Brien and Toms' framework that emphasizes emotional, cognitive, and behavioral components. This measure provided insights into the tool's effectiveness in improving emotional engagement during creative writing. \hl{The EEQ instrument can be found in Appendix} \ref{eeq}.

    \item \textbf{Hands-Free Writing Tools Assessment (HFWTA)}: We developed a 30-item questionnaire to evaluate seven dimensions of hands-free writing experience: \emph{Guided Writing Process}, \emph{Hands-Free Interaction}, \emph{Adaptability \& Contextual Awareness}, \emph{Multitasking Capability}, \emph{Content Refinement}, \emph{Output Quality \& Satisfaction}, and \emph{Ownership \& Agency}. Each dimension contained 3–5 items rated on a 7-point Likert scale (1 = Strongly Disagree, 7 = Strongly Agree). We designed this instrument to capture domain-specific aspects of hands-free writing tools that standard usability measures don't address. \hl{The full instrument is included in Appendix}~\ref{sec:hfwta}.

\end{itemize}

Together, these instruments provided a multidimensional understanding of cognitive effort, usability, affective response, perceived creativity, guidance effectiveness, multitasking capability, and user agency in hands-free writing contexts.

\subsubsection{Order–Effect Analysis}

\hl{Because each participant used all three tools in a counter-balanced sequence, we tested whether the \textit{first tool encountered} influenced performance on \textit{later} tools.}

\begin{itemize}[leftmargin=*]
  \item \textbf{Grouping.} Participants were assigned to three groups according to their initial tool: \textit{StepWrite First} ($n{=}9$),
        \textit{ChatGPT AVM First} ($n{=}8$), and \textit{Dictation Tool First}
        ($n{=}8$).
  \item \textbf{Data filtering.} For every participant we discarded the
        first-tool session and retained only their second and third sessions.
  \item \textbf{Metrics examined.} Raw NASA-TLX, revision effort, revision time, and semantic diversity.
  \item \textbf{Statistical test.} A Kruskal–Wallis $H$ test compared the three groups for each metric (\(\alpha = .05\)).
\end{itemize}

\hl{This analysis serves as a manipulation-check on the Latin-square design: if the test is non-significant the main results can be interpreted without sequence concerns; if significant, the size and direction of any order effect are reported in the Results section.}

\section{Results}
We report our findings across six main dimensions introduced in the \emph{Measures} section: \emph{revision effort}, \emph{text quality and structure}, \emph{temporal efficiency},\emph{questions necessity}, \emph{evaluation of tone classification}, and \emph{user experience}. Within user experience, we specifically examine workload (NASA TLX), usability (SUS), and emotional responses (EEQ). All statistical analyses used repeated-measures ANOVAs unless otherwise noted, with appropriate post-hoc pairwise tests (Bonferroni-corrected).

\subsection{Revision Effort}
\label{sec:revision-effort}

\paragraph{Operationalization.} We quantified how much participants had to manually edit their automatically generated drafts by counting word-level insertions, deletions, and replacements between each participant’s original (speech-generated) draft and final (revised) draft. A Ratcliff-Obershelp sequence-matching algorithm~\cite{ratcliff1988pattern} identified these edits at the phrase and sentence levels, capturing the extent of user modifications required to achieve an acceptable final version. To focus on substantive revisions, we removed all extraneous whitespace, line breaks, and formatting artifacts prior to comparison. This approach slightly favored tools like Dictation, which tended to produce disorganized formatting; had we included formatting issues, its revision count would likely have been even higher.

\begin{figure}[t]
    \centering
    \includegraphics[width=1\linewidth]{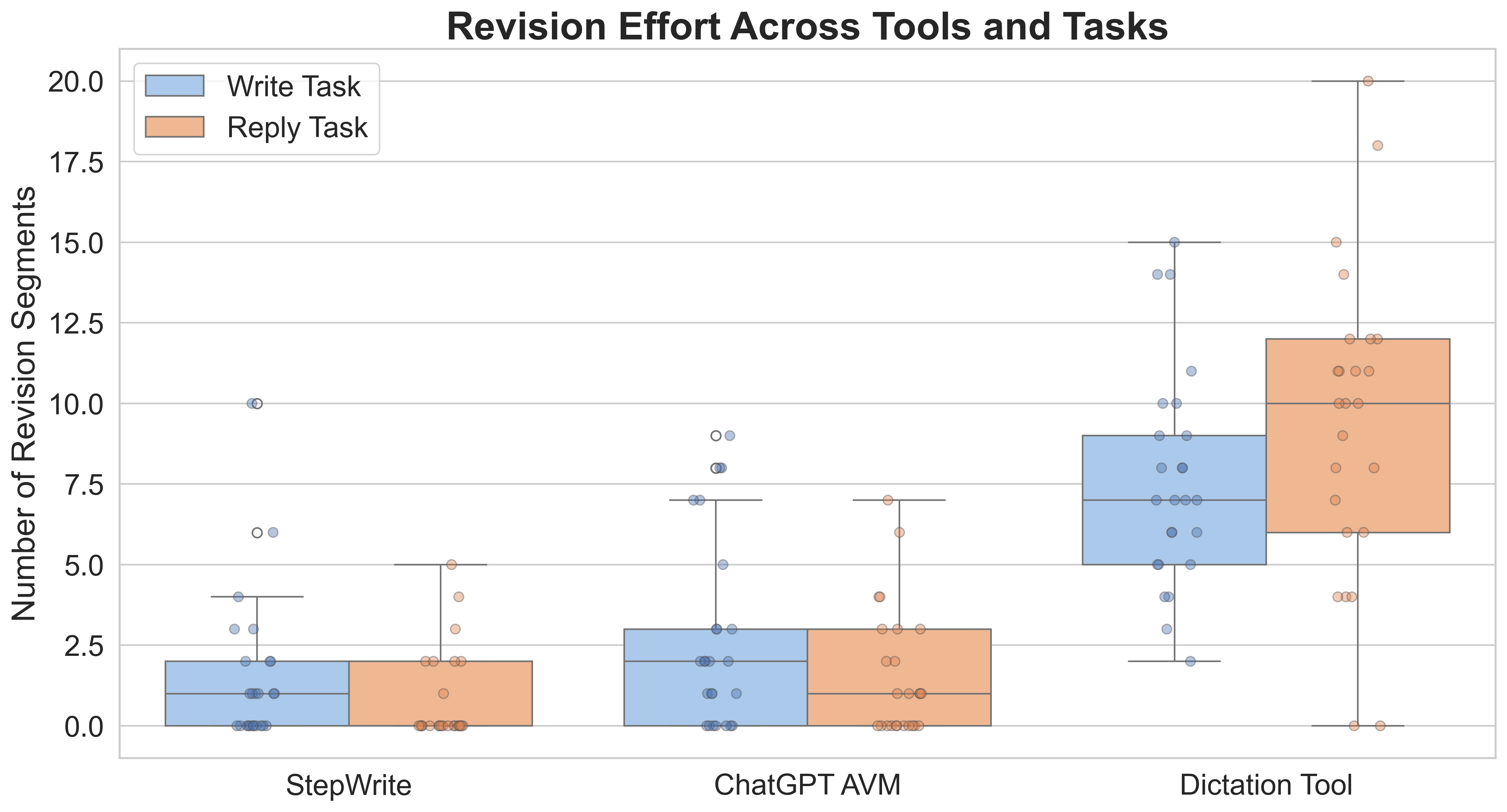}
    \caption{Revision effort (number of word-level edits) for the \emph{write} and \emph{reply} tasks under each tool. StepWrite drafts required the fewest edits, while Dictation required substantially more.}
    \label{fig:revision_effort}
\end{figure}

\paragraph{Findings.}
\autoref{fig:revision_effort} illustrates the revision counts for each tool (\textbf{StepWrite}, \textbf{ChatGPT AVM}, \textbf{Dictation}) across both the \emph{write} and \emph{reply} tasks. A repeated-measures ANOVA revealed a significant main effect of tool on total revision count (\emph{F}(2,48)\,=\,41.67, \emph{p}\,<\,.001). \textbf{StepWrite} led to the fewest edits overall (\emph{M}\textsubscript{write}\,=\,1.52, \emph{SD}\,=\,2.33; \emph{M}\textsubscript{reply}\,=\,0.84, \emph{SD}\,=\,1.43), suggesting that its structured scaffolding helped produce near-complete drafts requiring minimal cleanup. \textbf{ChatGPT AVM} produced slightly higher but still relatively low revision counts (\emph{M}\textsubscript{write}\,=\,2.68, \emph{SD}\,=\,2.91; \emph{M}\textsubscript{reply}\,=\,1.56, \emph{SD}\,=\,2.00), reflecting variability in how participants polished the generative outputs. In contrast, \textbf{Dictation} required the most editing (\emph{M}\textsubscript{write}\,=\,7.60, \emph{SD}\,=\,3.37; \emph{M}\textsubscript{reply}\,=\,9.32, \emph{SD}\,=\,4.86), typically to fix recognition errors, insert punctuation, and restructure run-on sentences. Post-hoc comparisons (Bonferroni-corrected) showed that Dictation required significantly more edits than both StepWrite and ChatGPT (\emph{p}\,<\,.001), whereas the difference between StepWrite and ChatGPT was not significant. Overall, these results highlight StepWrite’s efficacy in guiding users to produce cleaner drafts from the outset.

\subsection{Text Quality and Structure}
We evaluated the linguistic quality of participants’ drafts (both original and revised) through standard readability metrics, sentence structure, and lexical diversity.

\subsubsection{Readability}

\begin{figure}[t]
    \centering
    \includegraphics[width=1\linewidth]{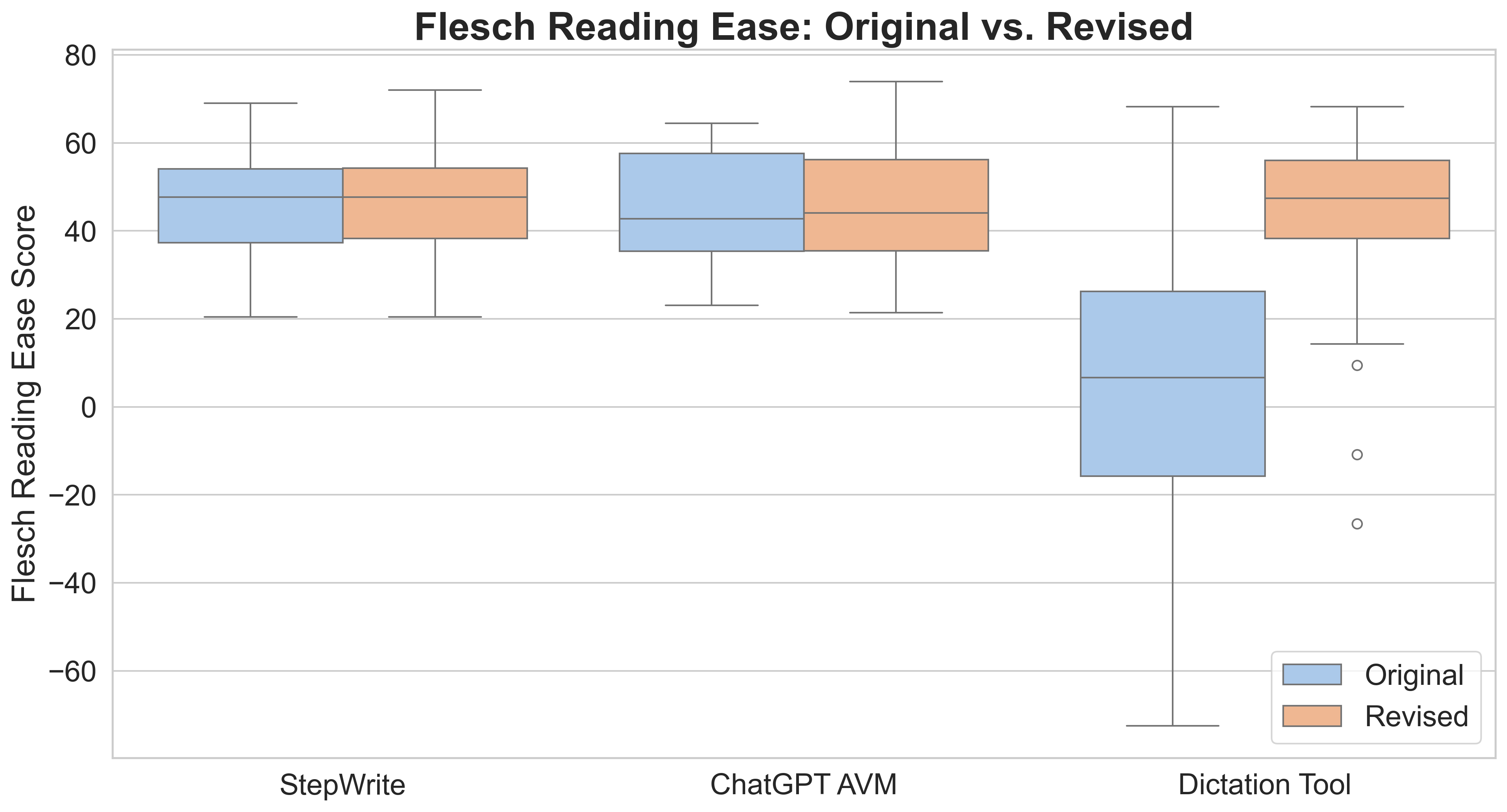}
    \caption{Flesch Reading Ease (FRE) across original and revised drafts. Higher scores mean more readable text. Dictation improved substantially after editing, while StepWrite and ChatGPT were already moderately readable from the start.}
    \label{fig:flesch_readability}
\end{figure}

\paragraph{Flesch Reading Ease (FRE)}
\autoref{fig:flesch_readability} shows FRE scores (range 0--100; higher indicates simpler, more readable text). A repeated-measures ANOVA revealed a significant effect of tool on FRE ($F(2,48) = 89.41$, $p < .001$) and an interaction with revision stage ($F(2,48) = 22.35$, $p < .001$). \textbf{StepWrite} and \textbf{ChatGPT} began with moderate readability (StepWrite $M = 46.13$; ChatGPT $M = 44.72$) that changed minimally after revision. In contrast, \textbf{Dictation} yielded very low initial readability ($M = 5.66$), primarily due to its unpunctuated, run-on outputs. After revisions, it rose dramatically ($M = 44.31$), matching the final readability levels of the AI tools.

\begin{figure}[t]
    \centering
    \includegraphics[width=1\linewidth]{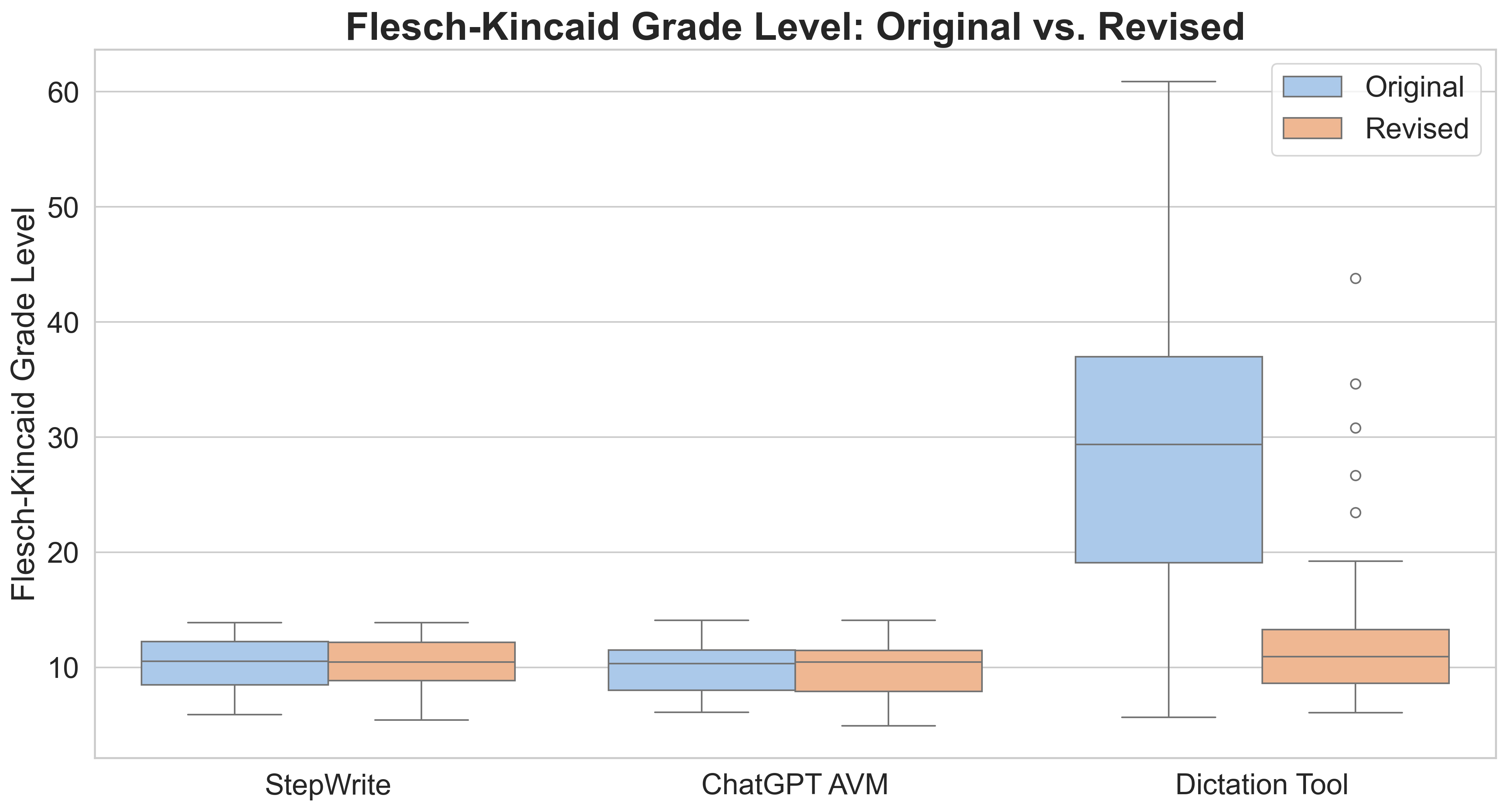}
    \caption{Flesch-Kincaid Grade Levels for original vs.\ revised texts. Dictation’s initial drafts were significantly more difficult to parse, while both AI tools were relatively stable and moderate in complexity.}
    \label{fig:grade_level_analysis}
\end{figure}

\paragraph{Flesch-Kincaid Grade Level.}
\autoref{fig:grade_level_analysis} presents Flesch-Kincaid Grade Levels (lower is simpler). Tool choice significantly affected text complexity (\emph{F}(2,48)\,=\,102.91, \emph{p}\,<\,.001). \textbf{StepWrite} (\emph{M}\,=\,10.41) and \textbf{ChatGPT} (\emph{M}\,=\,9.99) produced moderately complex yet coherent drafts, and their revised versions remained close to these initial values. \textbf{Dictation} started at a markedly higher complexity (\emph{M}\,=\,28.65), dropping to \emph{M}\,=\,12.84 after revision—demonstrating the substantial structural edits and punctuation insertions needed to transform raw speech outputs into more accessible text.

\subsubsection{Sentence Structure}

\begin{figure}[t]
    \centering
    \includegraphics[width=1\linewidth]{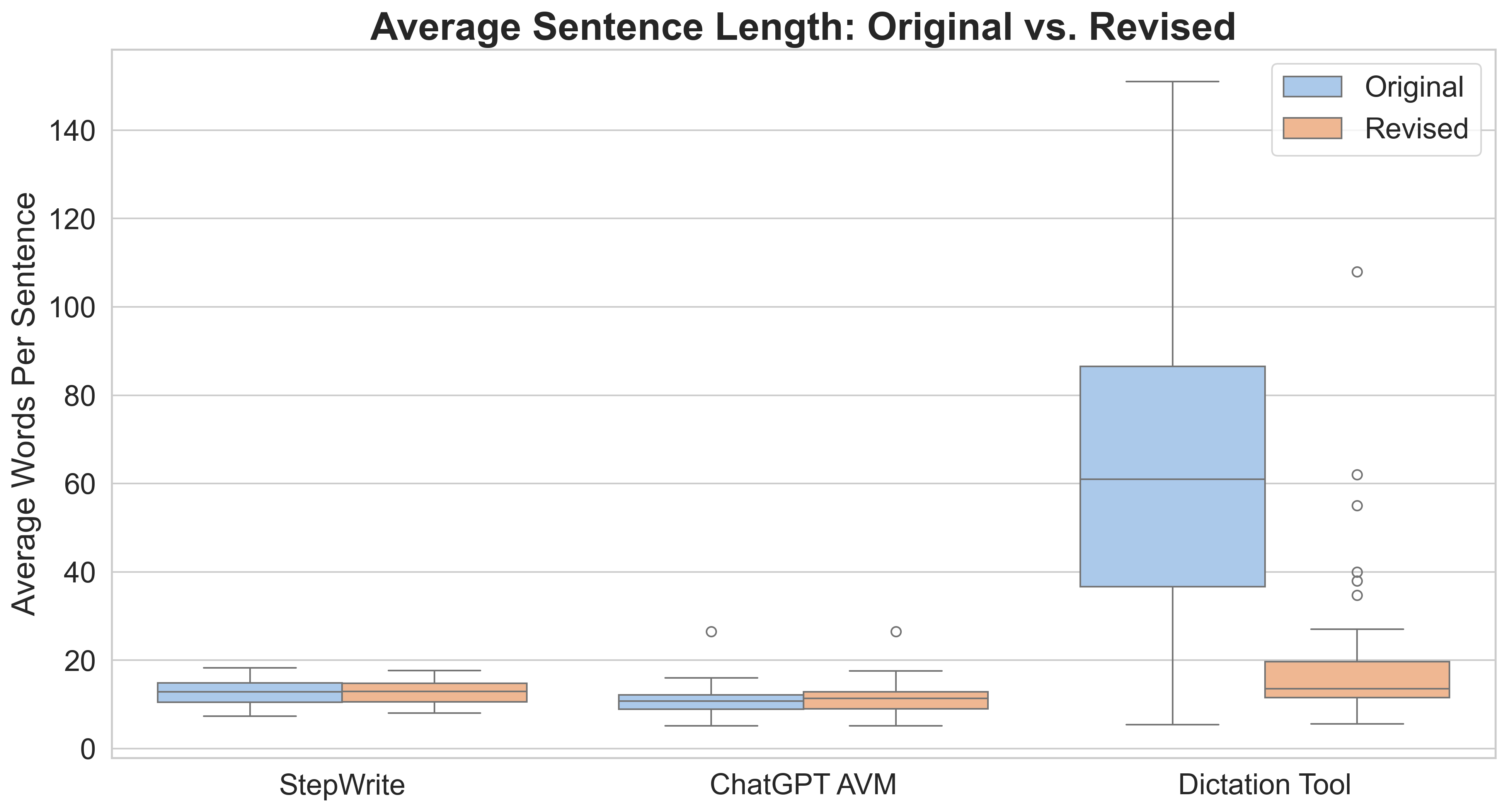}
    \caption{Average sentence length by tool. Dictation led to long, unsegmented sentences that were heavily shortened during revision, whereas StepWrite and ChatGPT produced cleanly segmented sentences from the outset.}
    \label{fig:sentence_length}
\end{figure}

\paragraph{Average Sentence Length.}
\autoref{fig:sentence_length} shows average sentence length for original and revised drafts. A repeated-measures ANOVA indicated that \textbf{Dictation} produced notably long, fragmented sentences before revision (\emph{M}\,=\,62.83 words), requiring extensive manual editing to shorten and punctuate (\emph{M}\,=\,19.28 words post-revision). By contrast, both \textbf{StepWrite} (\emph{M}\,=\,12.79 words) and \textbf{ChatGPT} (\emph{M}\,=\,10.81 words) generated relatively concise, well-segmented sentences from the outset, leading to minimal changes after participants’ edits.

\subsubsection{Lexical Diversity (Type-Token Ratio)}
\begin{figure}[t]
    \centering
    \includegraphics[width=1\linewidth]{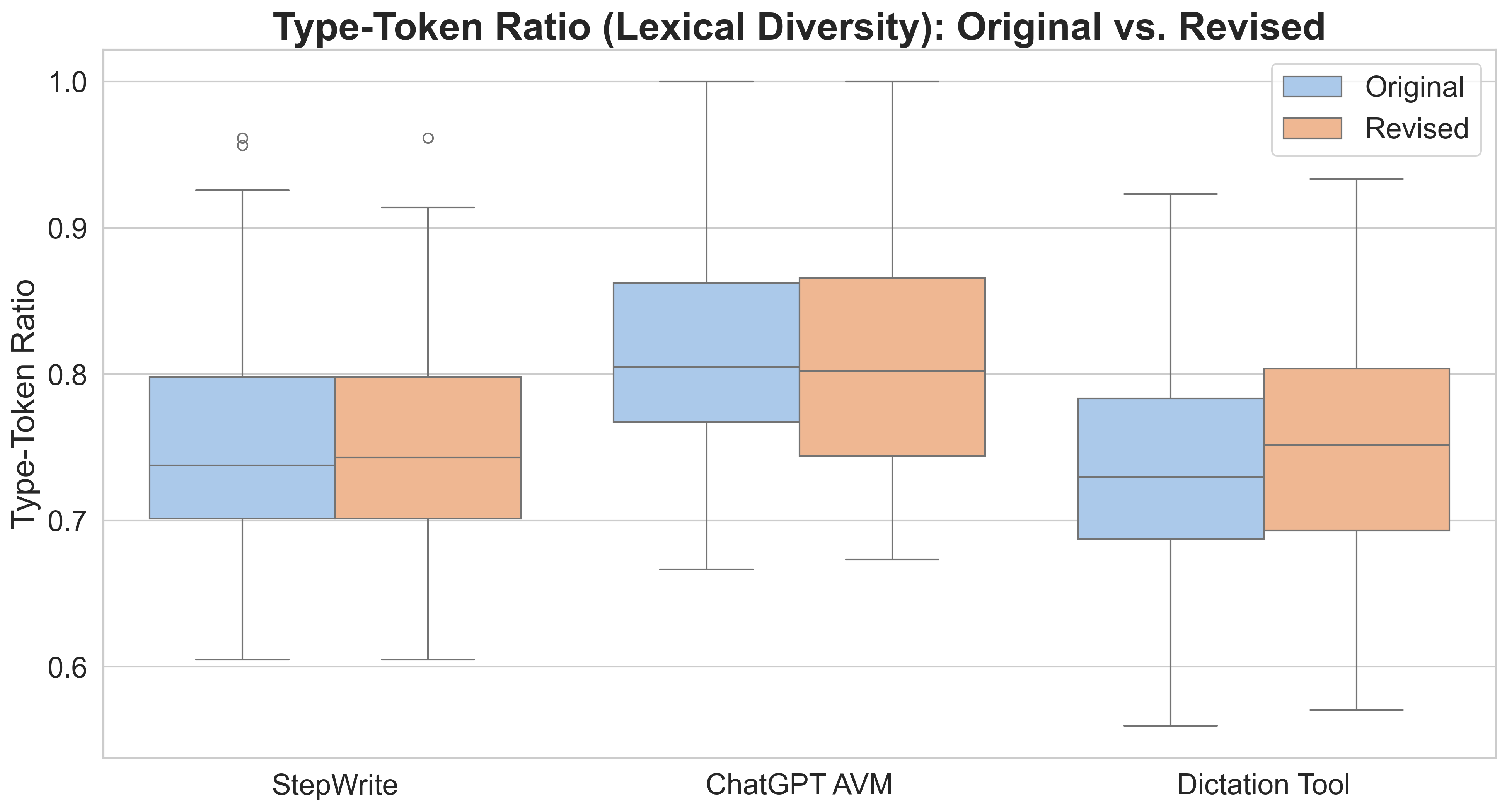}
    \caption{Type-token ratio (TTR) across original and revised drafts. ChatGPT outputs were most lexically diverse, while StepWrite was moderate. Dictation’s TTR was lowest initially but improved slightly post-edit.}
    \label{fig:ttr_comparison}
\end{figure}

\autoref{fig:ttr_comparison} displays the type-token ratio (TTR), where higher values indicate more varied vocabulary. \textbf{ChatGPT} attained the highest TTR (\emph{M}\,=\,0.8182), with \textbf{StepWrite} (\emph{M}\,=\,0.7529) next and \textbf{Dictation} (\emph{M}\,=\,0.7305) slightly lower. Revisions had minimal impact on TTR for all tools, suggesting that participants primarily focused on structural and grammatical refinements rather than vocabulary expansion.

\subsubsection{Semantic Diversity}

\begin{figure}[t]
  \centering
  \includegraphics[width=\linewidth]{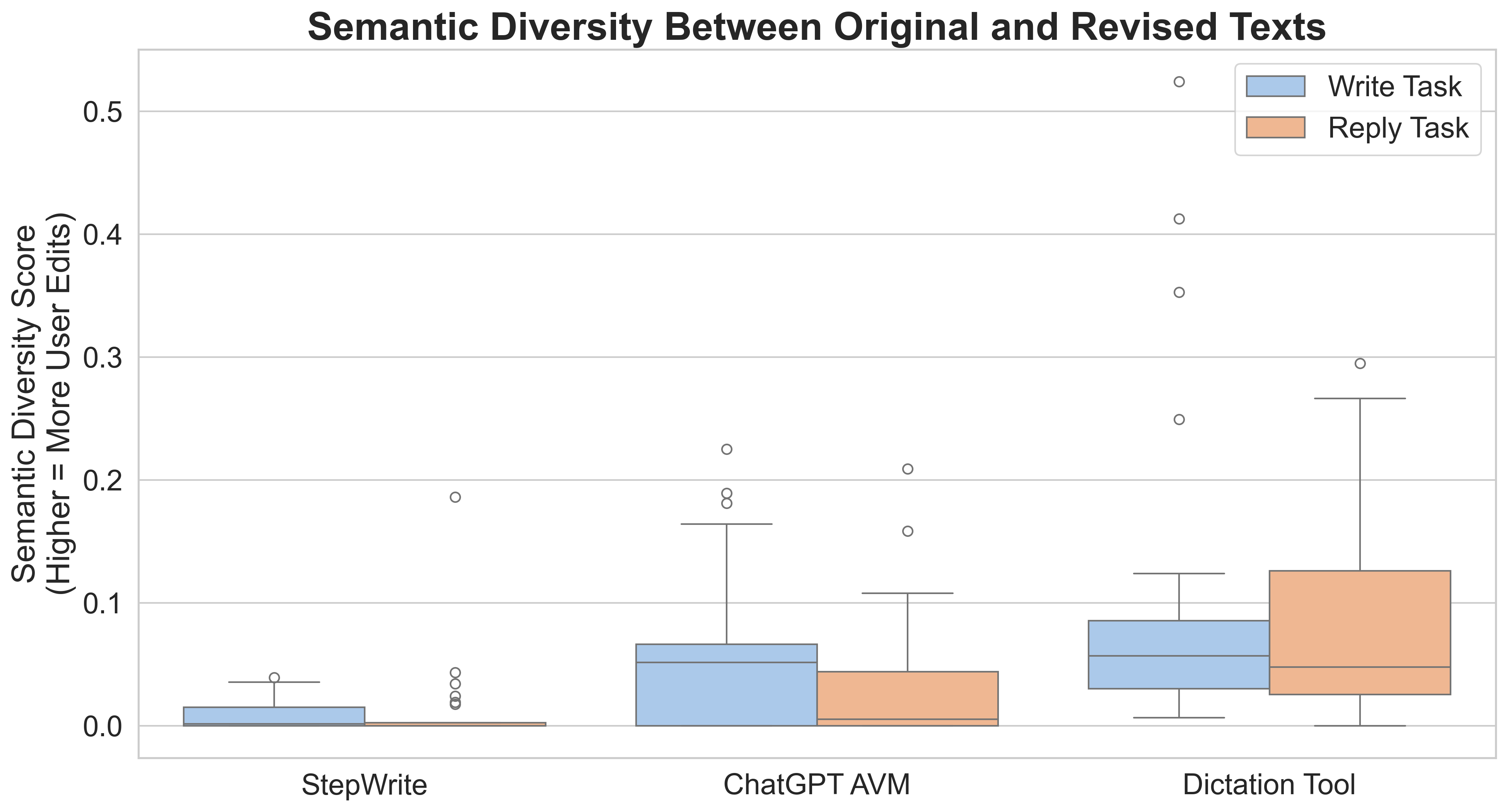}
  \caption{Semantic diversity (1 – similarity) between original and revised drafts.
  Lower scores indicate fewer meaning‑level edits.  Boxes show median and IQR;
  whiskers extend to 1.5 × IQR.}
  \label{fig:semantic_diversity}
\end{figure}

Semantic diversity captures meaning‑level edits.  
We embedded each draft’s original and revised versions with \texttt{gte-Qwen1.5-7B-instruct} \cite{li2023towards} and computed cosine similarity, then expressed diversity as $(1-\text{similarity})$. \autoref{fig:semantic_diversity} shows that \textbf{StepWrite} required virtually no semantic revision ($M=0.011,\; SD=0.028$), whereas \textbf{ChatGPT AVM} needed moderate adjustment ($M=0.049,\; SD=0.062$) and the \textbf{Dictation Tool} required the most ($M=0.095,\; SD=0.113$).

A two‑way repeated‑measures ANOVA confirmed a robust main effect of tool, $F(2,48)=15.45,\; p<.001$, and a smaller main effect of task, $F(1,24)=4.32,\; p=.048$; the interaction was not significant.  
Holm‑corrected pairwise comparisons showed StepWrite’s diversity was significantly lower than ChatGPT’s ($p=.004$) and Dictation’s ($p<.001$), and ChatGPT also outperformed Dictation ($p=.017$).

These results parallel the revision‑effort findings: StepWrite drafts are already aligned with user intent, ChatGPT drafts are “close enough” to tweak, and Dictation drafts often need substantial re‑wording.

\subsubsection{Final Draft Length}
\begin{figure}[t]
  \centering
  \includegraphics[width=\columnwidth]{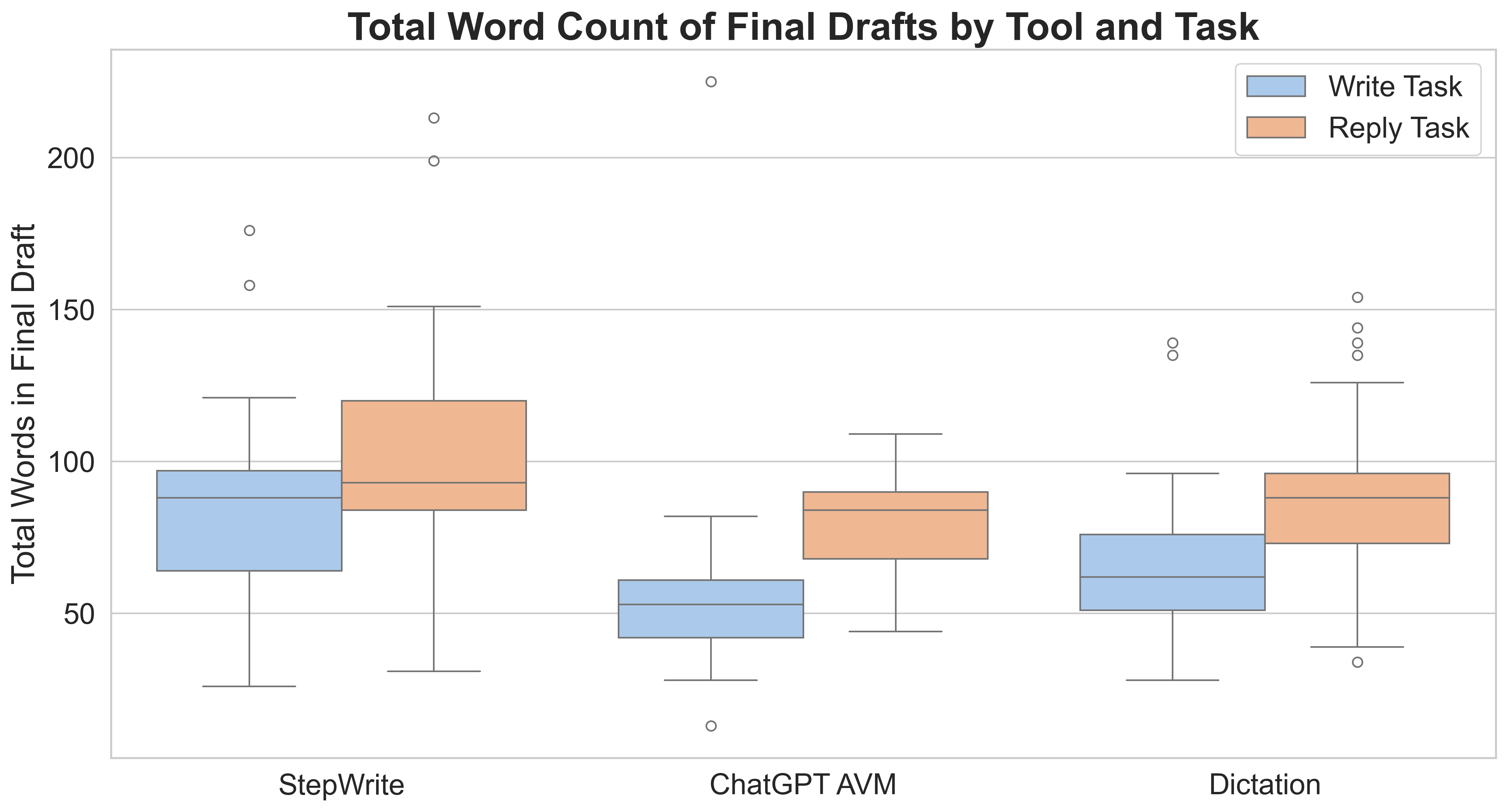}
  \caption{\hl{Total word count of final drafts across tool conditions and task types. StepWrite drafts were longer than the other tools, especially for replies.}}
  \label{fig:total_text_length_in_final_draft}
\end{figure}

\hl{We analyzed the total word count of final drafts to determine how tool and task influenced overall verbosity. As shown in Figure }
\autoref{fig:total_text_length_in_final_draft}, 
\hl{StepWrite produced longer drafts on average than both ChatGPT AVM and Dictation. In the write task, StepWrite drafts averaged 86.6 words (SD = 34.14), compared to 66.5 for Dictation and 58.5 for ChatGPT AVM. In the reply task, StepWrite averaged 104.3 words (SD = 40.56), while Dictation and ChatGPT AVM averaged 88.5 and 78.6 words, respectively. We attribute this increase to StepWrite’s step-by-step prompts, which encouraged users to include more detail and elaborate on their responses.}

\subsection{Temporal Efficiency}
\label{sec:temporal_analysis}
We measured how long participants spent composing drafts (\emph{drafting time}) and editing (\emph{revision time}) for each tool.

\subsubsection{Total Task Time}
\begin{figure}[t]
  \centering
  \includegraphics[width=1\linewidth]{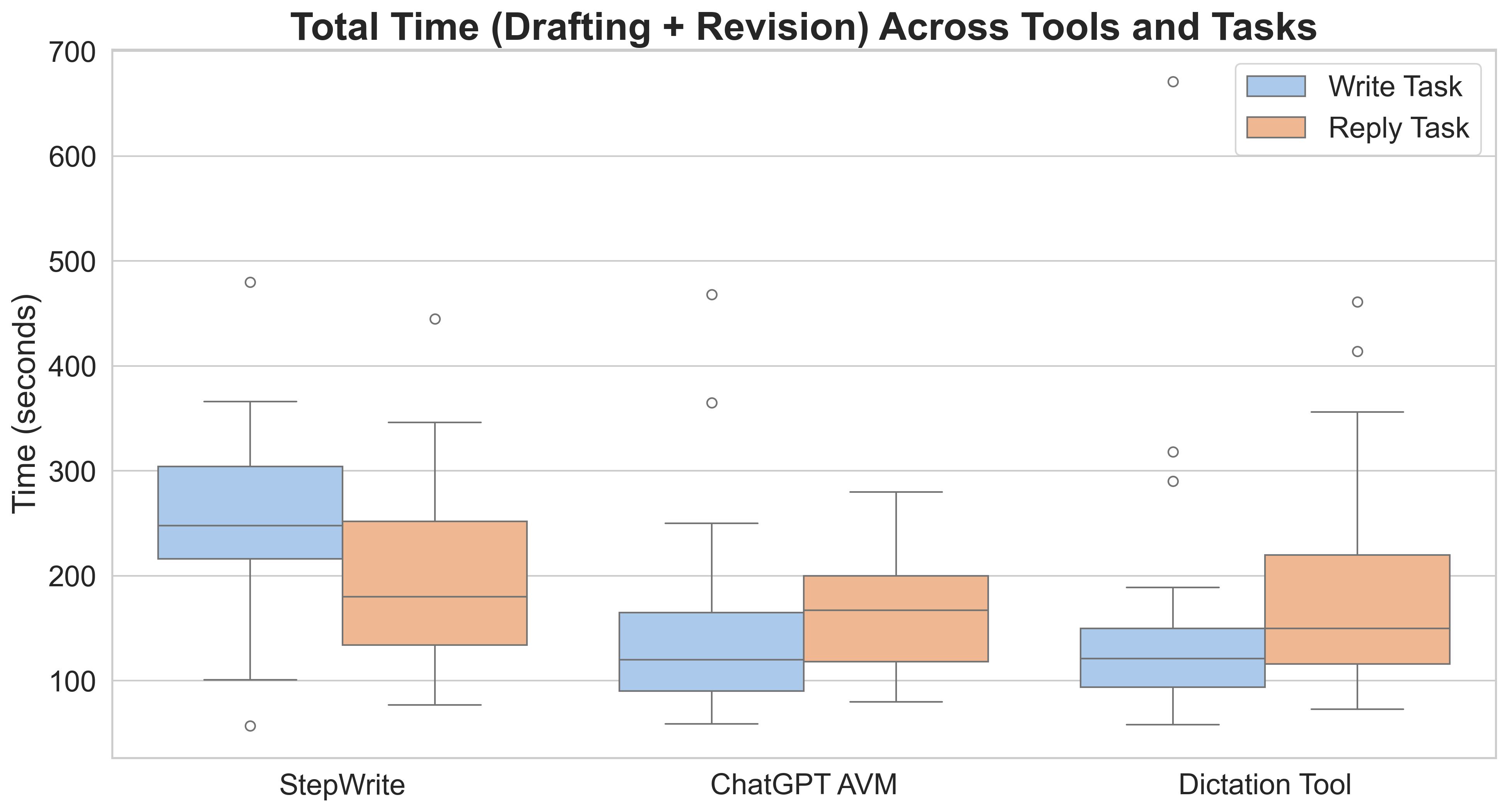}
  \caption{Total time (drafting + revision) for the \emph{write} and \emph{reply} tasks by tool. StepWrite’s prompt-by-prompt guidance added to overall time, while ChatGPT AVM was generally fastest.}
  \label{fig:total_time}
\end{figure}

\autoref{fig:total_time} shows total time (drafting + revision) across the \emph{write} and \emph{reply} tasks. A repeated-measures ANOVA revealed a significant effect of tool (\emph{F}(2,48)\,=\,17.02, \emph{p}\,<\,.001). \textbf{ChatGPT AVM} was fastest overall (Write: \emph{M}\,=\,150\,s; Reply: \emph{M}\,=\,168\,s), likely due to its rapid generative completion style and straightforward iterative correction. \textbf{Dictation} (\emph{M}\textsubscript{write}\,=\,154\,s; \emph{M}\textsubscript{reply}\,=\,185\,s) exhibited more variability because some participants quickly dictated usable drafts while others invested considerable time correcting recognition errors. \textbf{StepWrite} took the longest total time (Write: \emph{M}\,=\,248\,s; Reply: \emph{M}\,=\,200\,s), as its methodical stepwise prompts ensured coherent drafts but introduced additional latency in the process.

\subsubsection{Drafting vs.\ Revision Split}
\begin{figure}[h]
  \centering
  \includegraphics[width=\linewidth]{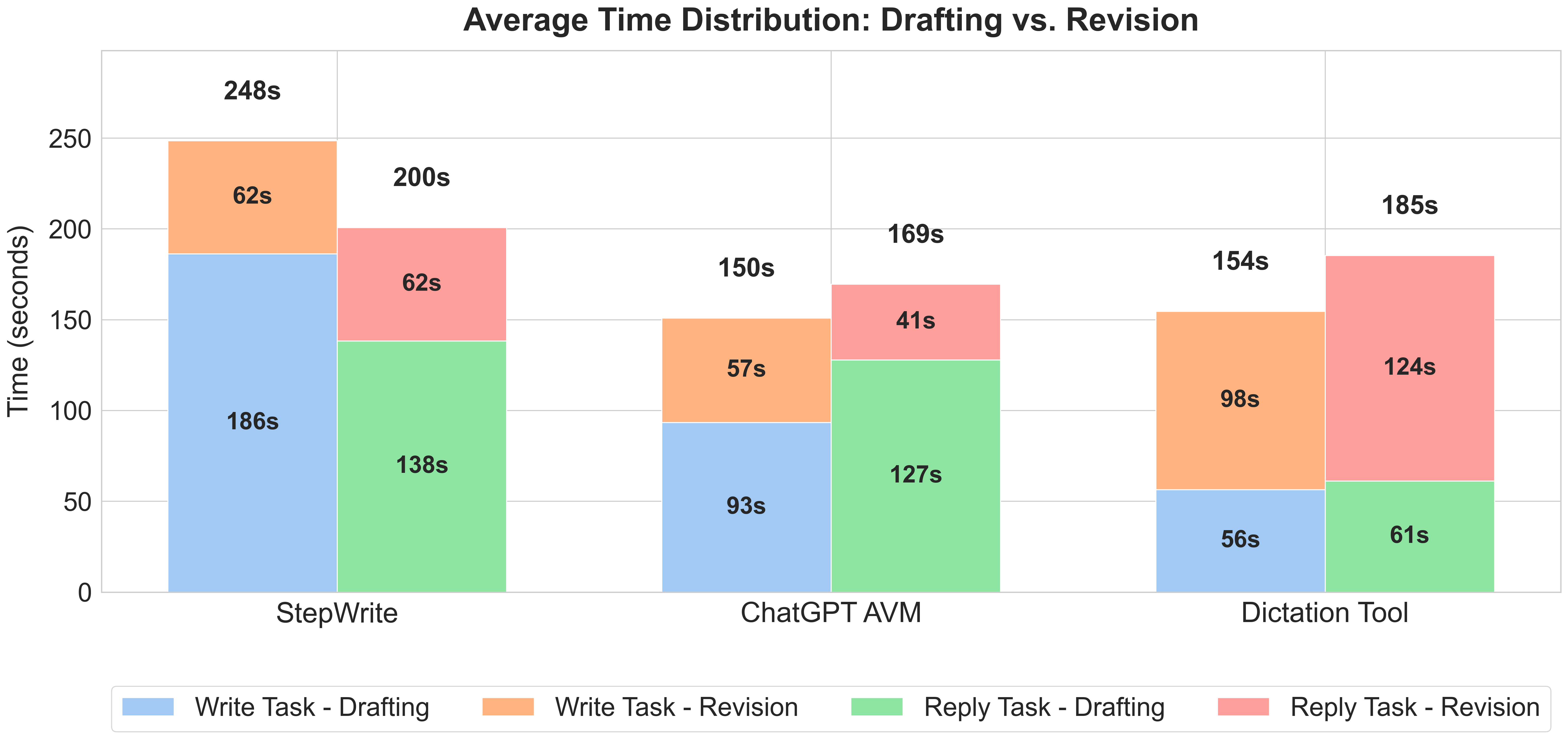}
  \caption{Average Time Distribution: Drafting vs. Revision. StepWrite required more initial input time but less revision, Dictation skewed toward revision time, and ChatGPT AVM demonstrated a balanced and efficient workflow.}
  \label{fig:time_distribution}
\end{figure}

\autoref{fig:time_distribution} illustrates how each tool’s total time was divided between \emph{drafting} and \emph{revision} phases. \textbf{StepWrite} front-loaded more effort into drafting (186\,s on Write; 138\,s on Reply), which helped minimize the subsequent revision phase to about 62\,s because the initial output was already well-structured. \textbf{Dictation} inverted this pattern by providing very quick drafting (\(\approx\)60\,s) at the cost of lengthy revision times (\(>\)100\,s), as participants fixed errors and reintroduced punctuation. \textbf{ChatGPT AVM} balanced drafting and revision (93--127\,s vs. 41--57\,s, respectively), leading to shorter overall durations.

\subsection{Necessity of StepWrite Questions}

Across all 25 participants, StepWrite issued 375 questions: 199 during the \textit{write} task and 176 during the \textit{reply} task. \hl{On average, participants received 7.96 questions in the write task (7.00 answered, 0.96 skipped; 87.9\% answer rate) and 7.04 in the reply task (6.24 answered, 0.80 skipped; 88.6\% answer rate). }

\hl {Questions were concise and consistent across tasks. The mean question length was 11.18 words (median = 11.00), and the mean answer length was 12.59 words (median = 11.00). Questions in the reply task were slightly longer (M = 13.01) than those in the write task (M = 9.79), though answer lengths remained stable across both.} Table~\ref{tab:necessity} summarizes annotation outcomes by task.

For reply e‑mails 81.8\% of questions were classified as \emph{necessary}, 11.4\% were skipped, and only 6.8\% were unnecessary, yielding an \emph{Essential Question Fraction} (EQF) of .818.  
In the write scenario the EQF was slightly lower at .744, with a larger share of unnecessary questions (13.6\%).  
Aggregated across both tasks, StepWrite achieved an overall EQF of .779—meaning roughly four of every five questions directly enabled content that survived all user revisions.

A chi‑square test of independence revealed a significant association between task type and question category ($\chi^2(2)=9.07,\; p=.011$), indicating that questions were more often indispensable when participants composed a reply e‑mail than when they drafted a new invitation.  
Despite this difference, the consistently high EQF across conditions confirms that StepWrite’s incremental questioning seldom digresses from information the user ultimately retains.

\hl{Upon deeper analysis, we found that questions marked as unnecessary or skipped often focused on highly specific logistical details—for example, “How many people are you inviting?” or “Who will cover transportation?”. These were occasionally perceived as too granular for the task. As one participant noted: “It asked me how many friends I was inviting to the museum, which didn't seem necessary because in an email context, it could potentially infer that from the number of people the email is being sent to” (P23). Other participants, however, appreciated the added structure: “The questions consistently exceeded my reasoning and were on point” (P16), and “All the questions were as expected” (P12). These diverging preferences indicate that the perceived utility of detailed prompts varies by user.}

\begin{table}[t]
  \centering
  \caption{Human coding of StepWrite questions. Percentages are in parentheses.}
  \label{tab:necessity}
  \begin{tabular}{lccc}
    \toprule
                 & \textbf{Necessary} & \textbf{Skipped} & \textbf{Unnecessary}\\
    \midrule
    Write (199)  & 148 (74.4) & 24 (12.1) & 27 (13.6)\\
    Reply (176)  & 144 (81.8) & 20 (11.4) & 12 (6.8)\\
    \midrule
    Total (375)  & 292 (77.9) & 44 (11.7) & 39 (10.4)\\
    \bottomrule
  \end{tabular}
\end{table}

\subsection{Evaluation of Tone Classification}

\hl{On the balanced evaluation set of 350 texts, our tone classifier achieved} \textbf{91.7\,\% overall accuracy}, a \textbf{macro‐F\textsubscript{1} of 0.912}, and a \textbf{weighted‐F\textsubscript{1} of 0.912}. Eleven of the fourteen tones registered F\textsubscript{1} scores of 0.86 or higher; \textit{apologetic}, \textit{encouraging}, \textit{surprised}, \textit{cooperative}, \textit{optimistic}, and \textit{empathetic} attained near-perfect performance. Results were lower for \textit{assertive} (F\textsubscript{1}\,=\,0.59) owing to modest recall, and for \textit{informal} and \textit{urgent} (F\textsubscript{1}\,$\approx$\,0.81–0.83). The relatively high support for \textit{formal} \hl{tone (83 instances) reflects the distribution of the original dataset, which consisted primarily of professional university correspondence.}

\hl{Importantly, these results were obtained using a zero-shot prompting approach. With few-shot examples, performance—particularly for lower-scoring categories—could likely be further improved. Overall, the classifier provides reliable tone labels for downstream alignment analysis, while also highlighting categories that may benefit from further prompt engineering}

\begin{table}[t]
\centering
\caption{\hl{Tone-classification results on the balanced evaluation set (350 messages).}}
    \label{tab:tone_results}
\begin{tabular}{lcccc}
\toprule
\textbf{Tone} & \textbf{Precision} & \textbf{Recall} & \textbf{F1} & \textbf{Support} \\
\midrule
Apologetic     & 1.00 & 1.00 & 1.00 & 22 \\
Encouraging    & 1.00 & 1.00 & 1.00 & 21 \\
Surprised      & 0.95 & 1.00 & 0.98 & 21 \\
Cooperative    & 0.95 & 1.00 & 0.98 & 21 \\
Optimistic     & 1.00 & 0.95 & 0.98 & 22 \\
Empathetic     & 1.00 & 0.95 & 0.97 & 20 \\
Concerned      & 1.00 & 0.94 & 0.97 & 16 \\
Friendly       & 0.93 & 1.00 & 0.96 & 25 \\
Diplomatic     & 0.94 & 0.94 & 0.94 & 16 \\
Formal         & 0.85 & 0.96 & 0.90 & 83 \\
Curious        & 1.00 & 0.75 & 0.86 & 24 \\
Urgent         & 0.74 & 0.95 & 0.83 & 21 \\
Informal       & 0.83 & 0.79 & 0.81 & 19 \\
Assertive      & 1.00 & 0.42 & 0.59 & 19 \\
\midrule
\textbf{Overall accuracy} & \multicolumn{4}{c}{\textbf{0.917}} \\
\textbf{Macro F1}         & \multicolumn{4}{c}{\textbf{0.912}} \\
\textbf{Weighted F1}      & \multicolumn{4}{c}{\textbf{0.912}} \\
\bottomrule
\end{tabular}
\end{table}

\subsection{User Experience}
\label{sec:user_experience}
We assessed user experience via \emph{perceived workload} (NASA TLX), \emph{usability} (SUS), \emph{emotional response} (EEQ), and our multidimensional Hands-Free Writing Tools Assessment (HFWTA).

\subsubsection{Workload: NASA TLX}
\begin{figure*}[t]
  \centering
  \includegraphics[width=\linewidth]{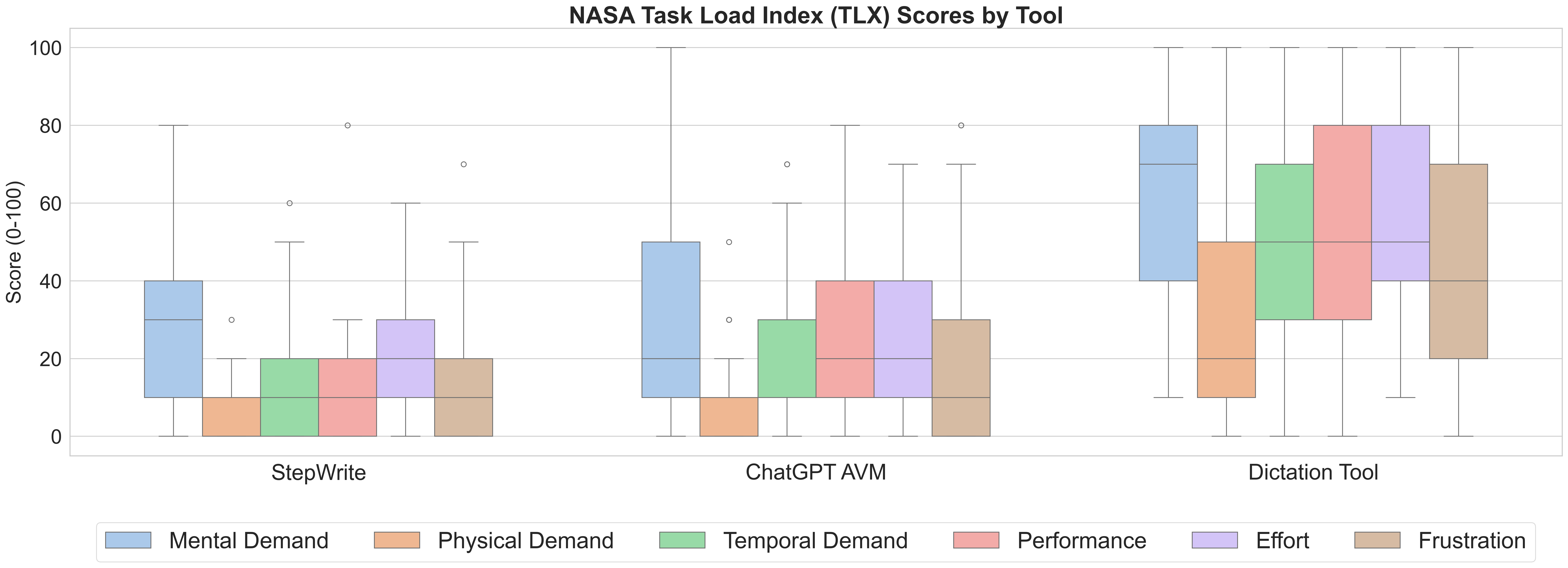}
  \caption{NASA TLX dimension scores (0--100). StepWrite was lowest on \emph{effort} and \emph{frustration}, whereas Dictation had substantially higher workload across all dimensions.}
  \label{fig:nasa_tlx_dimensions}
\end{figure*}

\autoref{fig:nasa_tlx_dimensions} summarizes NASA TLX dimension scores (0--100). A repeated-measures ANOVA indicated a significant main effect of tool on overall raw TLX (\emph{F}(2,48)\,=\,20.84, \emph{p}\,<\,.001). \textbf{StepWrite} registered the lowest overall workload (\emph{M}\,=\,16.8, \emph{SD}\,=\,11.4), with participants reporting minimal \emph{effort} or \emph{frustration}, even though the tool took more total time than others. \textbf{ChatGPT AVM} was moderately demanding (\emph{M}\,=\,22.5, \emph{SD}\,=\,17.1), owing to quick text generation alongside occasional user attention to steer the AI. \textbf{Dictation} was rated substantially more burdensome (\emph{M}\,=\,49.2, \emph{SD}\,=\,25.6), reflecting the frustration and effort required to correct recognition errors and reorganize unstructured output. Pairwise tests confirmed that both StepWrite and ChatGPT had significantly lower TLX than Dictation (\emph{p}\,<\,.001), while their difference was not significant (\emph{p}\,=\,.17).

\subsubsection{System Usability Scale (SUS)}
\begin{figure}[t]
  \centering
  \includegraphics[width=1\linewidth]{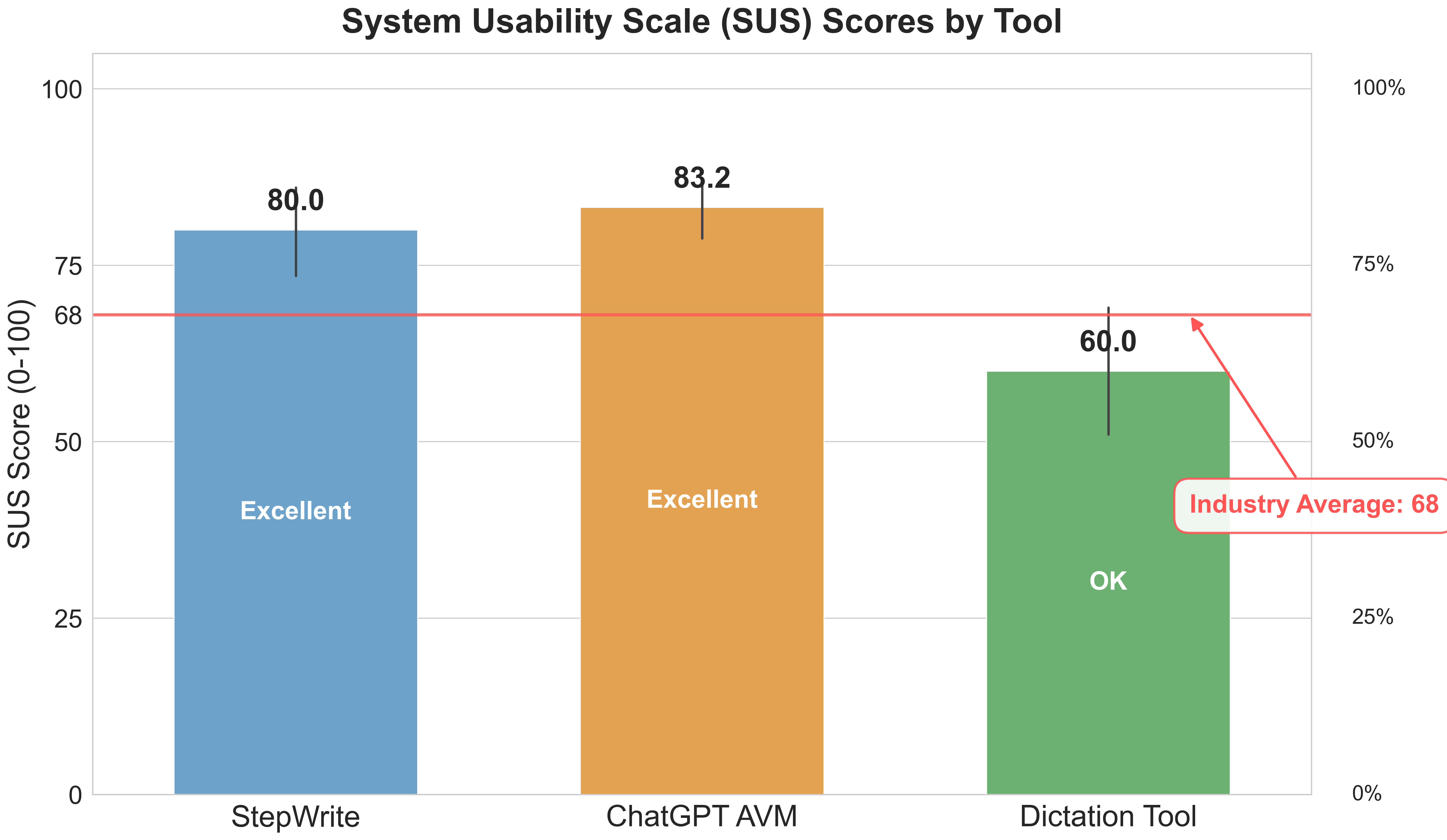}
  \caption{System Usability Scale (SUS) scores by tool. ChatGPT AVM and StepWrite scored well above the typical 68 benchmark; Dictation fell below it.}
  \label{fig:sus}
\end{figure}

\autoref{fig:sus} shows participants’ SUS scores (0--100). Scores above 68 are typically considered “usable.” \textbf{ChatGPT AVM} received the highest overall SUS score (\emph{M}\,=\,83.2, \emph{SD}\,=\,10.7), reflecting positive impressions of its conversational flexibility. \textbf{StepWrite} also surpassed the benchmark (\emph{M}\,=\,80.0, \emph{SD}\,=\,16.4). In contrast, \textbf{Dictation} (\emph{M}\,=\,60.0, \emph{SD}\,=\,23.9) was below 68, with users citing high error-correction overhead as a key usability barrier. An ANOVA confirmed significant differences (\emph{F}(2,48)\,=\,12.41, \emph{p}\,<\,.001), and Dictation was rated significantly lower than both StepWrite (\emph{p}\,=\,.001) and ChatGPT (\emph{p}\,<\,.001). There was no statistically significant difference between StepWrite and ChatGPT (\emph{p}\,=\,.42).

\subsubsection{Emotional Experience (EEQ)}
\begin{figure*}[t]
\centering
\includegraphics[width=\linewidth]{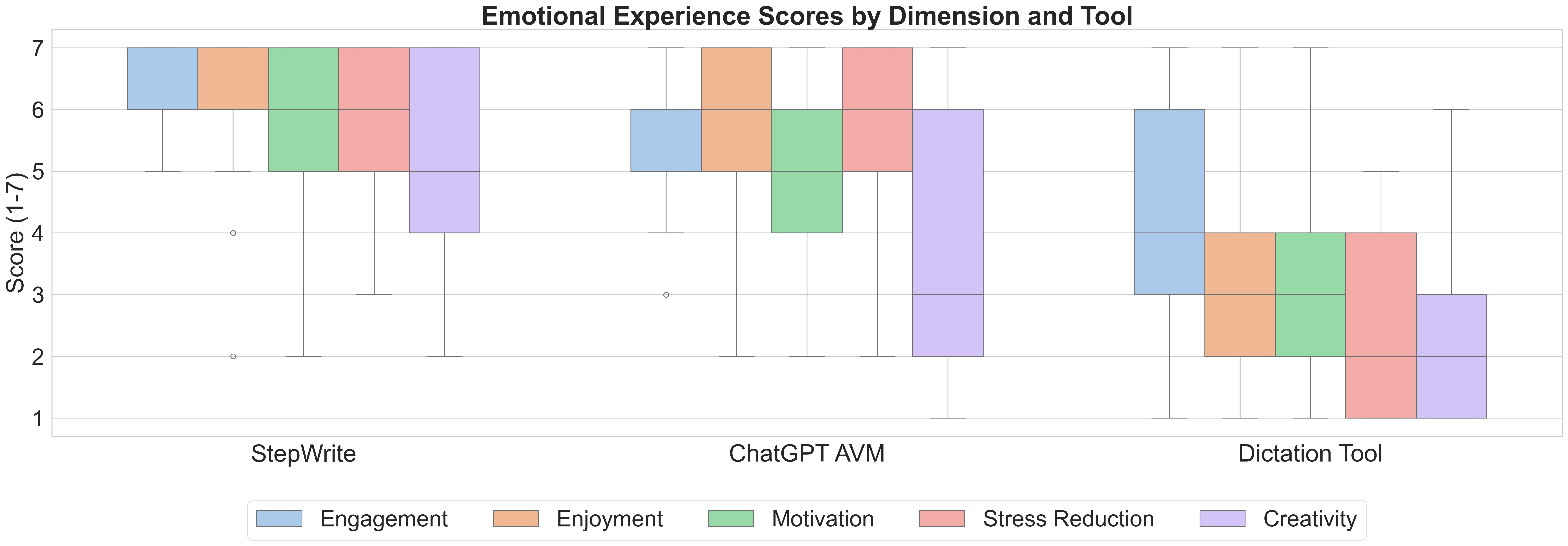}
\caption{Emotional Experience Questionnaire (EEQ) scores by dimension (1–7 scale). StepWrite elicited the highest overall emotional positivity; ChatGPT was close but especially good at reducing stress. Dictation scored significantly lower across all dimensions.}
\label{fig:eeq_dimensions}
\end{figure*}

Finally, the custom 5-item \emph{Emotional Experience Questionnaire} (EEQ) examined \emph{Engagement}, \emph{Enjoyment}, \emph{Motivation}, \emph{Stress Reduction}, and \emph{Creativity} using a 7-point Likert scale. \autoref{fig:eeq_dimensions} shows each dimension’s mean scores by tool. \textbf{StepWrite} received the highest overall emotional ratings (\emph{M}\,=\,5.76), especially in \emph{Engagement} (6.16) and \emph{Motivation} (5.96); participants consistently commented on feeling “guided” and “supported.” \textbf{ChatGPT AVM} followed (\emph{M}\,=\,5.10), often praised for its ability to reduce stress (5.64) through rapid text generation and prompt responsiveness. In contrast, \textbf{Dictation} scored significantly lower overall (\emph{M}\,=\,3.13), with heightened frustration, reduced creativity, and persistent stress due to transcription errors. A Friedman test ($\chi^2(2)\,=\,35.27$, \emph{p}\,<\,.001) and subsequent Wilcoxon signed-rank comparisons (Bonferroni-corrected) indicated that both StepWrite and ChatGPT outperformed Dictation (\emph{p}\,<\,.01), with no significant difference in overall EEQ between StepWrite and ChatGPT.

\subsubsection{Hands-Free Writing Tools Assessment (HFWTA)}

\begin{figure}[t]
  \centering
  \includegraphics[width=\linewidth]{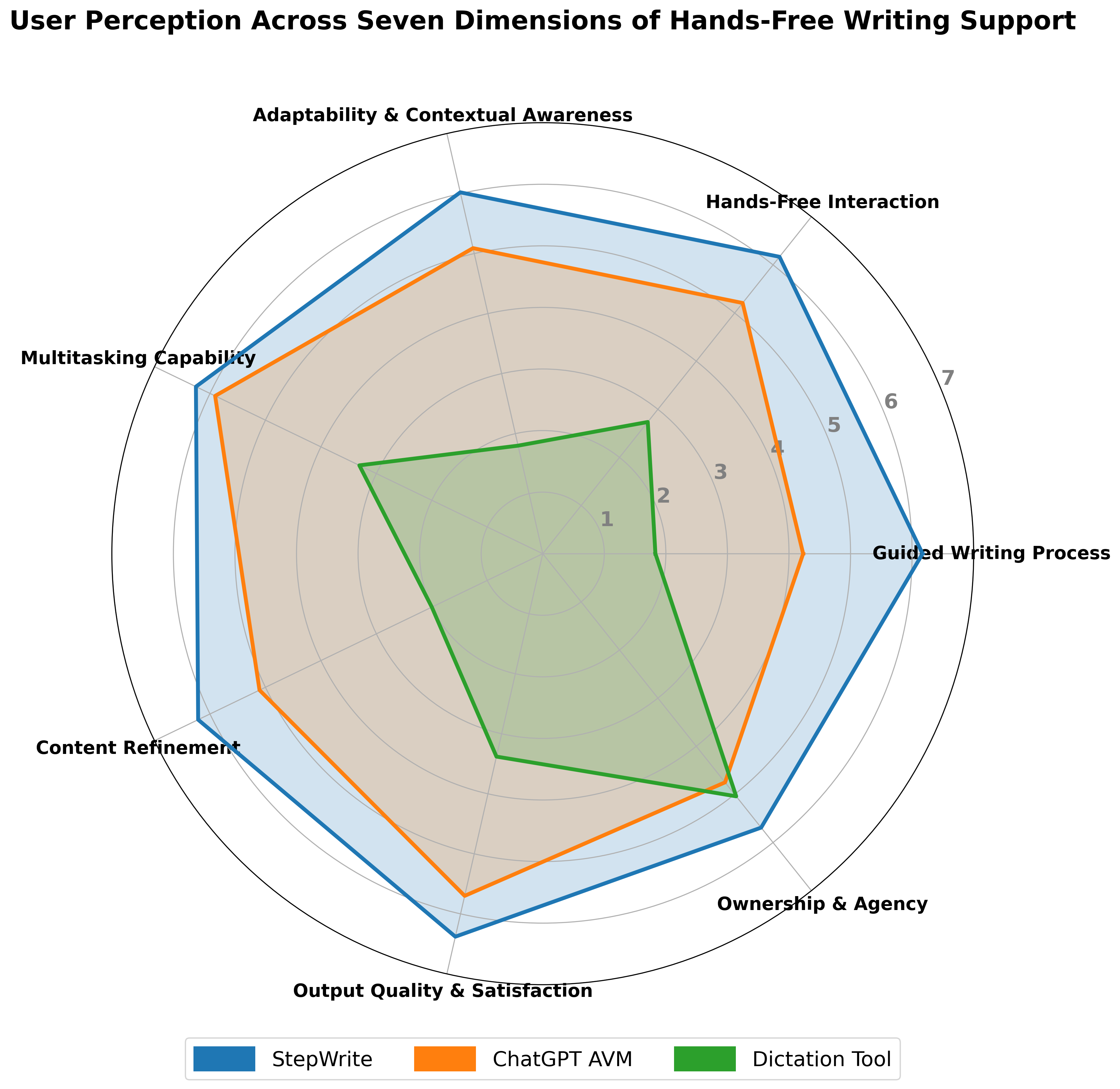}
  \caption{Radar chart comparison of the three writing tools across seven dimensions of hands-free writing support (scale: 1-7). StepWrite received the highest ratings across all measured dimensions, with the most pronounced advantages in Guided Writing Process and Adaptability.}
  \label{fig:radar_comparison}
\end{figure}

To gain deeper insights into tool-specific performance dimensions, we analyzed responses to our custom Hands-Free Writing Tools Assessment (HFWTA). \autoref{fig:radar_comparison} visualizes comparative performance across seven critical dimensions of hands-free writing support.

\textbf{StepWrite} received the highest overall rating (M = 6.13/7.00), with consistently high scores across all seven dimensions. It performed particularly well in \emph{Output Quality \& Satisfaction} (M = 6.38) and \emph{Multitasking Capability} (M = 6.25), indicating that participants valued its ability to produce quality content while allowing them to focus on secondary activities. \hl{While its score in \emph{Ownership \& Agency} (M = 5.69) was slightly lower than in other categories, it still outperformed all other tools in this dimension. This suggests that despite receiving strong guidance, users retained a high sense of authorship.}

\textbf{ChatGPT AVM} received moderate-to-high ratings (overall M = 5.14/7.00), with relative strengths in \emph{Multitasking Capability} (M = 5.91) and \emph{Output Quality \& Satisfaction} (M = 5.70). However, it scored notably lower in \emph{Guided Writing Process} (M = 4.23) and \emph{Ownership \& Agency} (M = 4.75), reflecting its conversational but less structured approach and participants' perception that its outputs sometimes superseded their own contributions.

\textbf{Dictation Tool} received significantly lower ratings across most dimensions (overall M = 2.87/7.00), with particular challenges in \emph{Adaptability \& Contextual Awareness} (M = 1.80) and \emph{Guided Writing Process} (M = 1.83). Its highest rating was in \emph{Ownership \& Agency} (M = 5.04)—exceeding ChatGPT AVM in this dimension—suggesting that despite usability challenges, participants still maintained a strong sense of authorship over dictated content.

The most substantial performance differences appeared in guidance-related dimensions, with a 4.34-point difference between StepWrite and Dictation Tool on \emph{Guided Writing Process} and a 4.22-point difference in \emph{Adaptability \& Contextual Awareness}. These results indicate that while all three tools enable hands-free composition, participants expressed a marked preference for approaches that offer structured scaffolding and contextual responsiveness over simple speech transcription. \hl{We also conducted a \textit{system use case} coding for StepWrite by qualitatively analyzing the HFWTA prompt “\emph{Describe a real-world scenario where you would find this tool particularly useful.}” (Section 9 of Appendix} \ref{sec:hfwta}). Two authors performed an inductive manifest content analysis. Working with the 25 participant responses, they first open-coded every scenario phrase, then iteratively merged related codes into a 15-category codebook, resolving all disagreements by discussion. The final pass yielded \textbf{81} coded scenario mentions, i.e.\ an average of 3.2 scenarios per participant. Table \ref{tab:stepwrite_scenarios} summarizes the frequencies. From this analysis, we identified three common themes:

\begin{itemize}
\item \textbf{Communication dominance.}  84\% of participants envisioned using StepWrite an email/text-composition aid. 
\item \textbf{Physically constrained contexts.}  A clear majority described situations where their hands or attention were occupied—gym sessions, cooking, commuting, laboratory work, and even cold weather. \item \textbf{Niche yet revealing scenarios.}  Specialized environments (e.g.\ taking notes during molten-steel experiments or soldering, wheelchair navigation) show how StepWrite can extend hands-free writing beyond communication and casual note-taking to safety-critical or accessibility-driven use cases.
\end{itemize}

\begin{table}[t]
  \caption{\hl{Self-reported StepWrite usage scenarios.}}
  \label{tab:stepwrite_scenarios}
  % \medium
  \begin{tabular}{lp{4.6cm}r}
    \toprule
    \textbf{Scenario} & \textbf{Activities} & \textbf{N} \\
    \midrule
     Email / messaging           & drafting or replying to emails, DMs                  & 21 \\
    Gym / workouts              & treadmill, weight-lifting, gym bike             &  9 \\
    Hands-full multitasking     & carrying items, “hands were full”                    &  8 \\
    Idea capture / notes        & brainstorming, jotting rough text                    &  7 \\
    Driving / commuting         & driving a car, riding a bus                          &  6 \\
    Walking / on-the-go         & campus walks, errands                                &  5 \\
    Cooking / food prep         & meal prep, stirring pots                             &  5 \\
    Urgent replies              & replying quickly, deadlines                          &  4 \\
    Academic writing            & class assignments, school papers                     &  4 \\
    Lab / technical work        & soldering, molten-steel experiments                  &  3 \\
    Running (outdoors)          & jogging, road running                                &  2 \\
    Accessibility needs         & wheelchair use, fatigued hands            &  2 \\
    Quiet desk environment      & low-noise desk work                                  &  2 \\
    Weather conditions     & winter weather, gloved hands                         &  2 \\
    Planning / to-do lists      & vacation planning, daily task lists                  &  1 \\
    \bottomrule
  \end{tabular}
\end{table}

\subsection{Order--Effect Analysis}

\hl{
After discarding every participant’s first-tool session, we compared the three sequence groups}—\emph{StepWrite First} ($n{=}9$), \emph{ChatGPT AVM First}
($n{=}8$), and \emph{Dictation Tool First} ($n{=}8$)—with Kruskal–Wallis tests \cite{kruskal1952use} on all performance and experience metrics.

\begin{description}[leftmargin=0pt]
\item[\textbf{Workload\,(NASA-TLX, 0–10 scale).}]
  StepWrite-First participants reported the highest subsequent workload
  ($M{=}4.24,\;SD{=}2.89$), ChatGPT-First were intermediate
  ($M{=}3.82,\;SD{=}2.56$), and Dictation-First the lowest
  ($M{=}1.17,\;SD{=}1.10$).  
  The difference was significant, $H(2)=17.91,\,p<.001$.

\item[\textbf{Revision effort\,(edit count).}]
  Means (SDs) were $4.89\;(4.41)$, $5.25\;(5.47)$, and $1.59\;(2.34)$ edits, respectively; $H(2)=11.31,\,p=.0035$.

\item[\textbf{Revision time.}]
  Subsequent editing took $79\,\text{s}\,(90.5)$,
  $87\,\text{s}\,(60.5)$, and $42\,\text{s}\,(34.2)$ on average;
  $H(2)=14.19,\,p<.001$.

\item[\textbf{Semantic diversity.}]
  Meaning-level change showed the same numerical trend
  ($M{=}0.054,\,0.059,\,0.025$) but was not reliable,
  $H(2)=3.74,\,p=.15$.
\end{description}

\hl{To check persistence, we repeated the test using only each participant’s
\emph{second} tool; all metrics converged (all $p>.06$), indicating that any priming or contrast from the first tool vanished after one additional session.

\textbf{Take-away.}  
Initial tool choice can momentarily colour users’ perception of effort, but the effect is short-lived. These transient differences do not alter the overall pattern reported earlier: when results are averaged across sequence, StepWrite still delivers the \textbf{lowest} sustained \textbf{cognitive load} and \textbf{revision burden}, independent of where it appears in the usage order. }

\subsection{Summary of Findings}
Across both writing and reply tasks, StepWrite’s structured Q\&A scaffold delivered the cleanest drafts with minimal edits, stable readability and complexity metrics, and the closest alignment to participants’ intent. Although StepWrite front-loaded more drafting time, it substantially reduced revision effort, achieved the lowest NASA-TLX workload scores, and earned the highest SUS and EEQ ratings for usability and emotional engagement.

ChatGPT AVM offered rapid generative drafts with moderate editing needs and the highest lexical diversity, striking a balance between speed and polish. It scored well on multitasking satisfaction but exhibited greater variability in how closely its outputs matched user intent.

By contrast, unstructured Dictation required minimal drafting time but imposed heavy editing overhead—long, run-on sentences, low initial readability, and substantial semantic revisions—resulting in the highest workload, lowest usability, and greatest stress.

These results highlight the value of incremental, context-aware questioning: adaptive planning (StepWrite) most effectively supports clean, intent-aligned, hands-free composition, while purely generative and speech-to-text approaches involve trade-offs among speed, flexibility, and user effort.

\section{Discussion}
\label{sec:discussion}

Our mixed-methods evaluation provides converging evidence that \textit{structured, context-aware scaffolding} meaningfully improves hands-free writing compared to both free-form dictation and open-ended conversational assistance. Below, we interpret the quantitative results in light of participants’ comments and observational notes, and we distill design implications for future voice-first authoring tools.

\subsection{Structured prompts reduce cognitive load and revision effort}

StepWrite’s question–answer workflow produced drafts that required, on average, 77\% fewer word-level edits compared to drafts produced using dictation and 40\% fewer compared to those produced using ChatGPT AVM (Section~\ref{sec:revision-effort}). Participants often attributed this efficiency to the system’s guided, incremental prompts. For instance, one participant noted, “the questions helped me remember what details to include” (P\#3), while another shared, “It was extremely helpful because it made me think of scenarios I probably would have forgotten” (P\#29). A third added, “I use a lot of filler words like ‘uhhh’ or ‘like,’ so I had to use more mental energy to not use those here because I knew they would be recorded in the output” (P\#22), suggesting that the structured format encouraged more deliberate, concise speech. Furthermore, StepWrite’s Essential Question Fraction (EQF) of 0.779 indicates that approximately 78\% of the questions asked directly contributed to the content in the final drafts, confirming the relevance and effectiveness of the guided prompts.

This structured support also lightened participants’ cognitive load. StepWrite had the lowest NASA TLX scores for \textit{mental demand, effort}, and \textit{frustration} (Figure~\ref{fig:nasa_tlx_dimensions}), and speech transcripts contained markedly fewer filler words compared to dictation. Participants described the experience as “refreshing and satisfying” (P\#1), citing reduced hesitation during speech and a clearer sense of direction.

\subsection{Guidance versus agency: finding the sweet spot}

Despite StepWrite’s overall popularity, some participants noted a trade-off between structured guidance and creative control. Several commented that the tool “asked more than I would normally include” (P\#10) or felt “a bit overkill” for users with strong writing confidence (P\#1). A few became fatigued by extended prompting: “After about 5 questions it became a little annoying, so I just issued \texttt{finish}” (P\#16). Others expressed the desire to “slip in extra context” or revisit earlier questions mid-flow (P\#4, P\#22). 
\hl{At the same time, participants highlighted StepWrite’s value for ideation and brainstorming: Pilot~P\#3 said, “I would use this to help me \emph{ideate and keep note of my ideas}. It lets me explore different directions just through voice,” P\#22 envisioned using it “\emph{for more brainstorming-type tasks where it would be helpful to go back and forth},” and P\#4 noted that the system “help[s] me take ideas I already have and decide where I want to provide specificity.”}
\hl{Conversely, some users felt the guidance could become cumbersome; and P\#22 felt the detailed dialogue “took longer than something more straightforward. As for ChatGPT AVM, two participants used it exclusively for grammar correction and tone refinement, describing it as a passive tool rather than a collaborator. This contrasted with StepWrite, which demanded and directed deeper engagement.
”}

The Hands-Free Writing Tools Assessment (HFWTA) captured this nuance: StepWrite earned the highest ratings in most design dimensions, and even though its score for \textit{Ownership \& Agency} (M = 5.69) was slightly lower than its own marks elsewhere, it still outperformed the two baseline tools on that dimension. Users consistently appreciated the tool’s quality and supportiveness, yet they also asked for lightweight “escape hatches”—for example, a free-response scratch pad or a real-time visual preview of the evolving text—to give them more flexible authorship. \hl{Taken together, these findings suggest that the sweet spot lies in preserving StepWrite’s structured scaffolding \emph{while} layering in low-friction avenues for divergent thinking and rapid idea capture.}

\subsection{Multitasking requires robust, forgiving speech interfaces}

Across all tools, the underlying speech recognition engine played a central role in shaping user experience. Dictation was especially fragile: users reported frustration with filler words, incorrect commands, and misinterpreted accents—“It recorded every ‘uh’ I made” (P\#12), “I couldn't modify anything hands-free” (P\#18), and “I had to stare at the screen to confirm what it wrote” (P\#29). Many participants abandoned inline punctuation entirely, leading to run-on sentences that required significant manual editing. This frustration was exacerbated for non-native speakers: several gave up on Word’s voice commands despite printed instruction lists, citing frequent misfires and lack of contextual correction. 

By contrast, StepWrite and ChatGPT partially mitigated STT issues by regenerating fluent outputs. However, participants still requested clearer cues for microphone state: “I wasn’t sure when ChatGPT was listening” (P\#7). Multimodal indicators (e.g., haptic feedback) could reduce this uncertainty, particularly for wearable displays.

\subsection{Order effects suggest transient scaffolding benefits}

Our counterbalanced design revealed short-term order effects. Participants who used StepWrite first often approached the second tool with greater clarity and efficiency, having already articulated key ideas. Several said that StepWrite “helped me plan better responses” (P\#2) or “structured my thinking so the rest was easier” (P\#16). Even those who preferred ChatGPT or Dictation noted that StepWrite primed their attention to tone, completeness, and clarity. However, quantitative analyses showed that these effects diminished by the third session, with all performance metrics converging across conditions. This suggests that while structured scaffolding may serve as a \textit{cognitive primer} for subsequent tools, its influence is likely ephemeral rather than enduring.

\subsection{Accessibility implications}

\hl{StepWrite’s voice-only workflow removes the need for precise finger input and sustained visual focus, making it a promising option for writers with limited upper-limb dexterity or learning disabilities that affect working memory. In our study, a manual-wheelchair user (P\#29) remarked that, with one hand always steering, “\emph{it’s hard to stop and type on my phone, so I’d use StepWrite for everyday emails or school papers},” while a participant with a learning disability (P\#16) said the system’s incremental questions were “\emph{superior and let me build a more complete answer—even in situations where I’d normally forget key details}.” Quantitatively, StepWrite yielded the lowest NASA-TLX mental-demand scores and required 86\,\% fewer word-level edits than Dictation and 45\,\% fewer than ChatGPT AVM} (Section~\ref{sec:revision-effort}) \hl{, demonstrating that its chunked, step-wise prompts lighten working-memory load—an approach consistent with structured-writing interventions shown to benefit people with intellectual and developmental disabilities (IDD)}~\cite{bakken2022single}. \hl{These observations point to adaptive, dialogue-based planning as a fruitful direction for future accessibility research—particularly for expanding writing autonomy among users with mobility impairments or IDD, and for anyone who benefits from structured, incremental guidance.}

\subsection{Toward Integrated Authoring Workflows}

As LLMs continue to improve in efficiency and can now run locally on-device—such as with Apple Intelligence, Gemini Nano, and open-source LLMs optimized for edge devices—we envision systems like StepWrite being embedded directly into mainstream authoring tools, including email clients, note-taking apps, and productivity suites. Rather than remaining standalone systems, structured voice-guided writing could become a native modality within everyday software, enabling fluid transitions between typing, dictation, and scaffolded composition. This shift would allow users to invoke guided writing in a lightweight, context-sensitive manner across a wide range of devices and environments.

\subsection{Design implications for future voice-first authoring tools} \label{design_implications}

\begin{enumerate}[leftmargin=*]
  \item \textbf{Adaptive planning beats static templates.} Our coding showed that 78\% of StepWrite prompts were essential to final drafts. Adaptive prompting—guided by user input or context—outperforms rigid forms, reducing unnecessary overhead while preserving relevance.
  
  \item \textbf{Expose process transparency.} ChatGPT AVM was occasionally distrusted for “just spitting out text without saying why” (P\#4). In contrast, StepWrite’s explicit questioning made its logic legible. Designing LLM interfaces to explain or expose intermediate steps could foster greater user trust.

   \item \hl{ \textbf{Scaffolding should adapt to user preferences.} 
   Some participants found highly specific prompts unnecessary, while others valued the added detail and structure. A future version of StepWrite could introduce a “detail specificity” setting that adapts to individual communication styles or allows users to control the level of detail. Alternatively, enabling memory could allow the system to learn preferences over time and automatically tailor scaffolding granularity to each user.}

\item \hl{\textbf{Provide lightweight escape hatches.} StepWrite’s speech detector employs a “thinking window” that ignores brief pauses, allowing authors to pause, reflect, and extend an utterance before the system responds. However, when revisiting an earlier prompt, revisions continue to rely on the \texttt{modify} command, which replaces the previous answer in full. Based on feedback from our study, we identified two complementary affordances to reduce this friction: (1) an always-listening micro-edit mode that supports incremental voice directives—e.g., “add the course name 'Linear Algebra' here” or “swap ‘Linear Algebra’ for ‘Data Structures’ ”—and integrates them into the existing response (planned for a future release), and (2) a continuously updated preview of the evolving draft to help users assess in real time whether their responses are sufficient. The latter was added to the StepWrite implementation following the study.}

\item \textbf{Minimize redundant effort.} \label{minimize_redundant_effort} \hl{Voice-first authoring tools should support user edits without requiring them to re-answer previously completed prompts. In the initial implementation of StepWrite, editing a prior response triggered the removal of all subsequent question--answer pairs. This approach was a deliberate tradeoff: by clearing potentially outdated content, the system ensured coherence and prevented irrelevant prompts from persisting. However, this strategy also introduced redundancy, with two participants expressing surprise that the system did not “remember” answers they had already provided. In response, we implemented a \textit{dependency analysis module} that evaluates the relationship between a modified response and downstream prompts. Only those that are logically, contextually, or semantically dependent on the edited answer are removed; unaffected prompts are retained. This selective regeneration was added to the StepWrite} (Appendix \ref{appendix:dependency-analysis}) following the study.

\item \textbf{Support modality switching.} Participants wanted to mute TTS when looking at a screen or speed up playback while walking (P\#4, P\#23). Systems should adapt feedback style (visual, auditory, haptic) to match situational attention and environment.

\end{enumerate}

\subsection{Limitations and Future Work}

While our controlled study demonstrates StepWrite’s promise, several limitations warrant further investigation. First, our participants were predominantly English‐speaking, university–educated adults comfortable with technology; results may not generalize to non–native speakers, older adults, or those with lower digital literacy. Second, we disabled personalization and long‐term memory, yet real‐world deployments could leverage these features to reduce redundant prompts—future work should evaluate how gradual adaptation affects both efficiency and user satisfaction. Third, we focused on short‐form emails and replies; scaling to longer documents (e.g., reports or essays) will likely require hierarchical outlines and section‐level navigation. Fourth, our simulated multitasking contexts (walking in a lab space, simple stationary tasks) do not capture the full variability of real environments—field studies in noisy, unpredictable settings (e.g., public transit, kitchens) are needed to assess robustness of speech recognition, latency, and user experience. Finally, we evaluated only U.S. English; extending StepWrite to other languages and cultural conventions may surface additional challenges. Addressing these limitations will be necessary for realizing broadly applicable, hands‐free and eye-free authoring tools.

\section{Conclusion}

Voice interfaces have long excelled at simple commands and short messages, but falter when users need to plan, structure, and revise longer texts while their hands or eyes are occupied. StepWrite bridges this gap by transforming composition into an adaptive, voice-driven dialogue: dynamic micro-prompts scaffold users’ intent step by step, offloading cognitive load and ensuring coherent, intent-aligned drafts.

In our within-subjects study (n=25), StepWrite outperformed both Microsoft Word’s dictation and ChatGPT Advanced Voice Mode. It yielded drafts requiring approximately 86 \% fewer word-level edits than dictation and 45 \% fewer than ChatGPT AVM, achieved the lowest NASA-TLX workload, and earned the highest SUS and EEQ scores for usability and engagement. Over 77 \% of StepWrite’s questions were essential to participants’ final texts, demonstrating the precision and relevance of its prompts.

By balancing structured guidance with user autonomy—supporting skips, edits, and a simple “finish” command—StepWrite preserves authorship while relieving users of planning and structural tracking. Its success in both stationary and mobile, hands-busy contexts showcase the power of context-aware scaffolding for wearable and multitasking scenarios.

Looking forward, we plan to personalize prompt granularity, integrate multimodal feedback (e.g., haptic cues), and extend hierarchical scaffolds for longer documents. Ultimately, we envision embedding adaptive voice scaffolding directly into everyday editors and wearable platforms, enabling users to compose complex texts—whether cooking dinner or walking between meetings—without pausing to type or look.

\medskip
\noindent\textit{“Because sometimes, the best way to write… is to speak your way there.”}


\begin{acks}
This work was supported by Apple Inc. and by compute resources provided by OpenAI. We thank the BIG Lab members for their thoughtful feedback and discussions, and the anonymous reviewers for their valuable suggestions. We also thank our study participants for generously sharing their time and perspectives.

Any views, opinions, findings, and conclusions or recommendations expressed in this material are those of the authors and should not be interpreted as reflecting the views, policies, or position, either expressed or implied, of Apple Inc.
\end{acks}

\bibliographystyle{ACM-Reference-Format}
\bibliography{references}


\begin{thebibliography}{61}

%%% the Web address.

\ifx \showCODEN    \undefined \def \showCODEN     #1{\unskip}     \fi
\ifx \showISBNx    \undefined \def \showISBNx     #1{\unskip}     \fi
\ifx \showISBNxiii \undefined \def \showISBNxiii  #1{\unskip}     \fi
\ifx \showISSN     \undefined \def \showISSN      #1{\unskip}     \fi
\ifx \showLCCN     \undefined \def \showLCCN      #1{\unskip}     \fi
\ifx \shownote     \undefined \def \shownote      #1{#1}          \fi
\ifx \showarticletitle \undefined \def \showarticletitle #1{#1}   \fi
\ifx \showURL      \undefined \def \showURL       {\relax}        \fi
% invisible to TeX
\providecommand\bibfield[2]{#2}
\providecommand\bibinfo[2]{#2}
\providecommand\natexlab[1]{#1}
\providecommand\showeprint[2][]{arXiv:#2}

\bibitem[Bakken et~al\mbox{.}(2022)]%
        {bakken2022single}
\bibfield{author}{\bibinfo{person}{Randi~Karine Bakken}, \bibinfo{person}{Kari-Anne~Botteg{\aa}rd N{\ae}ss}, \bibinfo{person}{Veerle Garrels}, {and} \bibinfo{person}{{\AA}ste~Mjelve Hagen}.} \bibinfo{year}{2022}\natexlab{}.
\newblock \showarticletitle{Single-case writing interventions for students with disorders of intellectual development: a systematic review and meta-analysis}.
\newblock \bibinfo{journal}{\emph{Education Sciences}} \bibinfo{volume}{12}, \bibinfo{number}{10} (\bibinfo{year}{2022}), \bibinfo{pages}{687}.
\newblock


\bibitem[Bereiter and Scardamalia(1982)]%
        {bereiter1982conversation}
\bibfield{author}{\bibinfo{person}{Carl Bereiter} {and} \bibinfo{person}{Marlene Scardamalia}.} \bibinfo{year}{1982}\natexlab{}.
\newblock \showarticletitle{From conversation to composition: The role of instruction in a developmental process}.
\newblock \bibinfo{journal}{\emph{Advances in instructional psychology}} \bibinfo{volume}{2}, \bibinfo{number}{1-64} (\bibinfo{year}{1982}).
\newblock


\bibitem[Brooke et~al\mbox{.}(1996)]%
        {SUS}
\bibfield{author}{\bibinfo{person}{John Brooke} {et~al\mbox{.}}} \bibinfo{year}{1996}\natexlab{}.
\newblock \showarticletitle{SUS-A quick and dirty usability scale}.
\newblock \bibinfo{journal}{\emph{Usability evaluation in industry}} \bibinfo{volume}{189}, \bibinfo{number}{194} (\bibinfo{year}{1996}), \bibinfo{pages}{4--7}.
\newblock


\bibitem[Burggr{\"a}f et~al\mbox{.}(2022)]%
        {burggraf2022preferences}
\bibfield{author}{\bibinfo{person}{Peter Burggr{\"a}f}, \bibinfo{person}{Moritz Beyer}, \bibinfo{person}{Jan-Philip Ganser}, \bibinfo{person}{Tobias Adlon}, \bibinfo{person}{Katharina M{\"u}ller}, \bibinfo{person}{Constantin Riess}, \bibinfo{person}{Kaspar Zollner}, \bibinfo{person}{Till Sa{\ss}mannshausen}, {and} \bibinfo{person}{Vincent Kammerer}.} \bibinfo{year}{2022}\natexlab{}.
\newblock \showarticletitle{Preferences for Single-Turn vs. Multiturn Voice Dialogs in Automotive Use Cases—Results of an Interactive User Survey in Germany}.
\newblock \bibinfo{journal}{\emph{IEEE Access}}  \bibinfo{volume}{10} (\bibinfo{year}{2022}), \bibinfo{pages}{55020--55033}.
\newblock


\bibitem[Ch et~al\mbox{.}(2022)]%
        {ch2022gesture}
\bibfield{author}{\bibinfo{person}{Nabil Al~Nahin Ch}, \bibinfo{person}{Diana Tosca}, \bibinfo{person}{Tyanna Crump}, \bibinfo{person}{Alberta Ansah}, \bibinfo{person}{Andrew Kun}, {and} \bibinfo{person}{Orit Shaer}.} \bibinfo{year}{2022}\natexlab{}.
\newblock \showarticletitle{Gesture and voice commands to interact with AR windshield display in automated vehicle: a remote elicitation study}. In \bibinfo{booktitle}{\emph{Proceedings of the 14th International Conference on Automotive User Interfaces and Interactive Vehicular Applications}}. \bibinfo{pages}{171--182}.
\newblock


\bibitem[Coenen et~al\mbox{.}(2021)]%
        {coenen2021wordcraft}
\bibfield{author}{\bibinfo{person}{A Coenen}, \bibinfo{person}{L Davis}, \bibinfo{person}{D Ippolito}, \bibinfo{person}{E Reif}, {and} \bibinfo{person}{A Yuan}.} \bibinfo{year}{2021}\natexlab{}.
\newblock \bibinfo{title}{Wordcraft: A human-AI collaborative editor for story writing. arXiv}.
\newblock


\bibitem[Dhillon et~al\mbox{.}(2024)]%
        {dhillon2024shaping}
\bibfield{author}{\bibinfo{person}{Paramveer~S Dhillon}, \bibinfo{person}{Somayeh Molaei}, \bibinfo{person}{Jiaqi Li}, \bibinfo{person}{Maximilian Golub}, \bibinfo{person}{Shaochun Zheng}, {and} \bibinfo{person}{Lionel~Peter Robert}.} \bibinfo{year}{2024}\natexlab{}.
\newblock \showarticletitle{Shaping human-ai collaboration: varied scaffolding levels in co-writing with language models}. In \bibinfo{booktitle}{\emph{Proceedings of the 2024 CHI Conference on Human Factors in Computing Systems}}. \bibinfo{pages}{1--18}.
\newblock


\bibitem[Edwards et~al\mbox{.}(2019)]%
        {edwards2019multitasking}
\bibfield{author}{\bibinfo{person}{Justin Edwards}, \bibinfo{person}{He Liu}, \bibinfo{person}{Tianyu Zhou}, \bibinfo{person}{Sandy~JJ Gould}, \bibinfo{person}{Leigh Clark}, \bibinfo{person}{Philip Doyle}, {and} \bibinfo{person}{Benjamin~R Cowan}.} \bibinfo{year}{2019}\natexlab{}.
\newblock \showarticletitle{Multitasking with Alexa: how using intelligent personal assistants impacts language-based primary task performance}. In \bibinfo{booktitle}{\emph{Proceedings of the 1st International Conference on Conversational User Interfaces}}. \bibinfo{pages}{1--7}.
\newblock


\bibitem[El~Alaoui et~al\mbox{.}(2023)]%
        {el2023building}
\bibfield{author}{\bibinfo{person}{Hamza El~Alaoui}, \bibinfo{person}{Zakaria El~Aouene}, {and} \bibinfo{person}{Violetta Cavalli-Sforza}.} \bibinfo{year}{2023}\natexlab{}.
\newblock \showarticletitle{Building intelligent chatbots: Tools, technologies, and approaches}. In \bibinfo{booktitle}{\emph{2023 3rd International Conference on Innovative Research in Applied Science, Engineering and Technology (IRASET)}}. IEEE, \bibinfo{pages}{1--12}.
\newblock


\bibitem[Esquivel et~al\mbox{.}(2024)]%
        {esquivel2024voice}
\bibfield{author}{\bibinfo{person}{Paola Esquivel}, \bibinfo{person}{Kayden Gill}, \bibinfo{person}{Mary Goldberg}, \bibinfo{person}{S~Andrea Sundaram}, \bibinfo{person}{Lindsey Morris}, {and} \bibinfo{person}{Dan Ding}.} \bibinfo{year}{2024}\natexlab{}.
\newblock \showarticletitle{Voice assistant utilization among the disability community for independent living: A rapid review of recent evidence}.
\newblock \bibinfo{journal}{\emph{Human Behavior and Emerging Technologies}} \bibinfo{volume}{2024}, \bibinfo{number}{1} (\bibinfo{year}{2024}), \bibinfo{pages}{6494944}.
\newblock


\bibitem[Feng et~al\mbox{.}(2025)]%
        {feng2025reflective}
\bibfield{author}{\bibinfo{person}{Yunhai Feng}, \bibinfo{person}{Jiaming Han}, \bibinfo{person}{Zhuoran Yang}, \bibinfo{person}{Xiangyu Yue}, \bibinfo{person}{Sergey Levine}, {and} \bibinfo{person}{Jianlan Luo}.} \bibinfo{year}{2025}\natexlab{}.
\newblock \showarticletitle{Reflective planning: Vision-Language Models for multi-stage long-horizon robotic manipulation}.
\newblock \bibinfo{journal}{\emph{arXiv preprint arXiv:2502.16707}} (\bibinfo{year}{2025}).
\newblock


\bibitem[Flesch(1948)]%
        {flesch1948new}
\bibfield{author}{\bibinfo{person}{Rudolph Flesch}.} \bibinfo{year}{1948}\natexlab{}.
\newblock \showarticletitle{A new readability yardstick}.
\newblock \bibinfo{journal}{\emph{Journal of Applied Psychology}} \bibinfo{volume}{32}, \bibinfo{number}{3} (\bibinfo{year}{1948}), \bibinfo{pages}{221}.
\newblock


\bibitem[Goodman et~al\mbox{.}(2022)]%
        {goodman2022lampost}
\bibfield{author}{\bibinfo{person}{Steven~M Goodman}, \bibinfo{person}{Erin Buehler}, \bibinfo{person}{Patrick Clary}, \bibinfo{person}{Andy Coenen}, \bibinfo{person}{Aaron Donsbach}, \bibinfo{person}{Tiffanie~N Horne}, \bibinfo{person}{Michal Lahav}, \bibinfo{person}{Robert MacDonald}, \bibinfo{person}{Rain~Breaw Michaels}, \bibinfo{person}{Ajit Narayanan}, {et~al\mbox{.}}} \bibinfo{year}{2022}\natexlab{}.
\newblock \showarticletitle{Lampost: Design and evaluation of an ai-assisted email writing prototype for adults with dyslexia}. In \bibinfo{booktitle}{\emph{Proceedings of the 24th International ACM SIGACCESS Conference on Computers and Accessibility}}. \bibinfo{pages}{1--18}.
\newblock


\bibitem[Graham and Perin(2007)]%
        {graham2007meta}
\bibfield{author}{\bibinfo{person}{Steve Graham} {and} \bibinfo{person}{Dolores Perin}.} \bibinfo{year}{2007}\natexlab{}.
\newblock \showarticletitle{A meta-analysis of writing instruction for adolescent students.}
\newblock \bibinfo{journal}{\emph{Journal of educational psychology}} \bibinfo{volume}{99}, \bibinfo{number}{3} (\bibinfo{year}{2007}), \bibinfo{pages}{445}.
\newblock


\bibitem[Haas et~al\mbox{.}(2020)]%
        {haas2020they}
\bibfield{author}{\bibinfo{person}{Gabriel Haas}, \bibinfo{person}{Jan Gugenheimer}, \bibinfo{person}{Jan~Ole Rixen}, \bibinfo{person}{Florian Schaub}, {and} \bibinfo{person}{Enrico Rukzio}.} \bibinfo{year}{2020}\natexlab{}.
\newblock \showarticletitle{“They Like to Hear My Voice”: Exploring Usage Behavior in Speech-Based Mobile Instant Messaging}. In \bibinfo{booktitle}{\emph{22nd International Conference on Human-Computer Interaction with Mobile Devices and Services}}. \bibinfo{pages}{1--10}.
\newblock


\bibitem[Harada et~al\mbox{.}(2007)]%
        {harada2007voicepen}
\bibfield{author}{\bibinfo{person}{Susumu Harada}, \bibinfo{person}{T~Scott Saponas}, {and} \bibinfo{person}{James~A Landay}.} \bibinfo{year}{2007}\natexlab{}.
\newblock \showarticletitle{VoicePen: Augmenting pen input with simultaneous non-linguisitic vocalization}. In \bibinfo{booktitle}{\emph{Proceedings of the 9th international conference on Multimodal interfaces}}. \bibinfo{pages}{178--185}.
\newblock


\bibitem[Hart(2006)]%
        {NASATLX}
\bibfield{author}{\bibinfo{person}{Sandra~G Hart}.} \bibinfo{year}{2006}\natexlab{}.
\newblock \showarticletitle{NASA-task load index (NASA-TLX); 20 years later}. In \bibinfo{booktitle}{\emph{Proceedings of the human factors and ergonomics society annual meeting}}, Vol.~\bibinfo{volume}{50}. Sage publications Sage CA: Los Angeles, CA, \bibinfo{pages}{904--908}.
\newblock


\bibitem[Hendrik et~al\mbox{.}(2006)]%
        {hendrik2006designing}
\bibfield{author}{\bibinfo{person}{Witt Hendrik}, \bibinfo{person}{Tom Nicolai}, {and} \bibinfo{person}{Holger Kenn}.} \bibinfo{year}{2006}\natexlab{}.
\newblock \showarticletitle{Designing a wearable user interface for hands-free interaction in maintenance applications}. In \bibinfo{booktitle}{\emph{PerCom Workshops}}.
\newblock


\bibitem[H{\"o}hne et~al\mbox{.}(2024)]%
        {hohne2024automatic}
\bibfield{author}{\bibinfo{person}{Jan~Karem H{\"o}hne}, \bibinfo{person}{Timo Lenzner}, {and} \bibinfo{person}{Joshua Claassen}.} \bibinfo{year}{2024}\natexlab{}.
\newblock \showarticletitle{Automatic speech-to-text transcription: evidence from a smartphone survey with voice answers}.
\newblock \bibinfo{journal}{\emph{International Journal of Social Research Methodology}} (\bibinfo{year}{2024}), \bibinfo{pages}{1--8}.
\newblock


\bibitem[Hu et~al\mbox{.}(2025)]%
        {hu2025enhancing}
\bibfield{author}{\bibinfo{person}{Wanqing Hu}, \bibinfo{person}{Jirong Tian}, {and} \bibinfo{person}{Yanyan Li}.} \bibinfo{year}{2025}\natexlab{}.
\newblock \showarticletitle{Enhancing student engagement in online collaborative writing through a generative AI-based conversational agent}.
\newblock \bibinfo{journal}{\emph{The Internet and Higher Education}}  \bibinfo{volume}{65} (\bibinfo{year}{2025}), \bibinfo{pages}{100979}.
\newblock


\bibitem[Huang et~al\mbox{.}(2023)]%
        {huang2023conveying}
\bibfield{author}{\bibinfo{person}{Chieh-Yang Huang}, \bibinfo{person}{Saniya Naphade}, \bibinfo{person}{Kavya~Laalasa Karanam}, {and} \bibinfo{person}{Ting-Hao'Kenneth' Huang}.} \bibinfo{year}{2023}\natexlab{}.
\newblock \showarticletitle{Conveying the Predicted Future to Users: A Case Study of Story Plot Prediction}.
\newblock \bibinfo{journal}{\emph{arXiv preprint arXiv:2302.09122}} (\bibinfo{year}{2023}).
\newblock


\bibitem[Ippolito et~al\mbox{.}(2022)]%
        {ippolito2022creative}
\bibfield{author}{\bibinfo{person}{Daphne Ippolito}, \bibinfo{person}{Ann Yuan}, \bibinfo{person}{Andy Coenen}, {and} \bibinfo{person}{Sehmon Burnam}.} \bibinfo{year}{2022}\natexlab{}.
\newblock \showarticletitle{Creative writing with an ai-powered writing assistant: Perspectives from professional writers}.
\newblock \bibinfo{journal}{\emph{arXiv preprint arXiv:2211.05030}} (\bibinfo{year}{2022}).
\newblock


\bibitem[Khurana et~al\mbox{.}(2024)]%
        {khurana2024just}
\bibfield{author}{\bibinfo{person}{Anjali Khurana}, \bibinfo{person}{Michael Glueck}, {and} \bibinfo{person}{Parmit~K Chilana}.} \bibinfo{year}{2024}\natexlab{}.
\newblock \showarticletitle{Do I Just Tap My Headset? How Novice Users Discover Gestural Interactions with Consumer Augmented Reality Applications}.
\newblock \bibinfo{journal}{\emph{Proceedings of the ACM on Interactive, Mobile, Wearable and Ubiquitous Technologies}} \bibinfo{volume}{7}, \bibinfo{number}{4} (\bibinfo{year}{2024}), \bibinfo{pages}{1--28}.
\newblock


\bibitem[Korkiakoski et~al\mbox{.}(2024)]%
        {korkiakoski2024preference}
\bibfield{author}{\bibinfo{person}{Mikko Korkiakoski}, \bibinfo{person}{Paula Alavesa}, {and} \bibinfo{person}{Panos Kostakos}.} \bibinfo{year}{2024}\natexlab{}.
\newblock \showarticletitle{Preference in voice commands and gesture controls with hands-free augmented reality with novel users}.
\newblock \bibinfo{journal}{\emph{IEEE Pervasive Computing}} (\bibinfo{year}{2024}).
\newblock


\bibitem[Kruskal and Wallis(1952)]%
        {kruskal1952use}
\bibfield{author}{\bibinfo{person}{William~H Kruskal} {and} \bibinfo{person}{W~Allen Wallis}.} \bibinfo{year}{1952}\natexlab{}.
\newblock \showarticletitle{Use of ranks in one-criterion variance analysis}.
\newblock \bibinfo{journal}{\emph{Journal of the American statistical Association}} \bibinfo{volume}{47}, \bibinfo{number}{260} (\bibinfo{year}{1952}), \bibinfo{pages}{583--621}.
\newblock


\bibitem[Larsen et~al\mbox{.}(2020)]%
        {larsen2020hands}
\bibfield{author}{\bibinfo{person}{Helene~H{\o}gh Larsen}, \bibinfo{person}{Alexander~Nuka Scheel}, \bibinfo{person}{Toine Bogers}, {and} \bibinfo{person}{Birger Larsen}.} \bibinfo{year}{2020}\natexlab{}.
\newblock \showarticletitle{Hands-free but not eyes-free: A usability evaluation of Siri while driving}. In \bibinfo{booktitle}{\emph{Proceedings of the 2020 conference on human information interaction and retrieval}}. \bibinfo{pages}{63--72}.
\newblock


\bibitem[Lee et~al\mbox{.}(2024)]%
        {lee2024gazepointar}
\bibfield{author}{\bibinfo{person}{Jaewook Lee}, \bibinfo{person}{Jun Wang}, \bibinfo{person}{Elizabeth Brown}, \bibinfo{person}{Liam Chu}, \bibinfo{person}{Sebastian~S Rodriguez}, {and} \bibinfo{person}{Jon~E Froehlich}.} \bibinfo{year}{2024}\natexlab{}.
\newblock \showarticletitle{GazePointAR: A context-aware multimodal voice assistant for pronoun disambiguation in wearable augmented reality}. In \bibinfo{booktitle}{\emph{Proceedings of the 2024 CHI Conference on Human Factors in Computing Systems}}. \bibinfo{pages}{1--20}.
\newblock


\bibitem[Lee et~al\mbox{.}(2022)]%
        {lee2022coauthor}
\bibfield{author}{\bibinfo{person}{Mina Lee}, \bibinfo{person}{Percy Liang}, {and} \bibinfo{person}{Qian Yang}.} \bibinfo{year}{2022}\natexlab{}.
\newblock \showarticletitle{Coauthor: Designing a human-ai collaborative writing dataset for exploring language model capabilities}. In \bibinfo{booktitle}{\emph{Proceedings of the 2022 CHI conference on human factors in computing systems}}. \bibinfo{pages}{1--19}.
\newblock


\bibitem[Li et~al\mbox{.}(2023)]%
        {li2023towards}
\bibfield{author}{\bibinfo{person}{Zehan Li}, \bibinfo{person}{Xin Zhang}, \bibinfo{person}{Yanzhao Zhang}, \bibinfo{person}{Dingkun Long}, \bibinfo{person}{Pengjun Xie}, {and} \bibinfo{person}{Meishan Zhang}.} \bibinfo{year}{2023}\natexlab{}.
\newblock \showarticletitle{Towards general text embeddings with multi-stage contrastive learning}.
\newblock \bibinfo{journal}{\emph{arXiv preprint arXiv:2308.03281}} (\bibinfo{year}{2023}).
\newblock


\bibitem[Lin et~al\mbox{.}(2024)]%
        {lin2024rambler}
\bibfield{author}{\bibinfo{person}{Susan Lin}, \bibinfo{person}{Jeremy Warner}, \bibinfo{person}{JD Zamfirescu-Pereira}, \bibinfo{person}{Matthew~G Lee}, \bibinfo{person}{Sauhard Jain}, \bibinfo{person}{Shanqing Cai}, \bibinfo{person}{Piyawat Lertvittayakumjorn}, \bibinfo{person}{Michael~Xuelin Huang}, \bibinfo{person}{Shumin Zhai}, \bibinfo{person}{Bj{\"o}rn Hartmann}, {et~al\mbox{.}}} \bibinfo{year}{2024}\natexlab{}.
\newblock \showarticletitle{Rambler: Supporting Writing With Speech via LLM-Assisted Gist Manipulation}. In \bibinfo{booktitle}{\emph{Proceedings of the 2024 CHI Conference on Human Factors in Computing Systems}}. \bibinfo{pages}{1--19}.
\newblock


\bibitem[Liu et~al\mbox{.}(2023)]%
        {liu2023older}
\bibfield{author}{\bibinfo{person}{Mingzhou Liu}, \bibinfo{person}{Caixia Wang}, {and} \bibinfo{person}{Jing Hu}.} \bibinfo{year}{2023}\natexlab{}.
\newblock \showarticletitle{Older adults’ intention to use voice assistants: Usability and emotional needs}.
\newblock \bibinfo{journal}{\emph{Heliyon}} \bibinfo{volume}{9}, \bibinfo{number}{11} (\bibinfo{year}{2023}).
\newblock


\bibitem[Mehra et~al\mbox{.}(2023)]%
        {mehra2023gist}
\bibfield{author}{\bibinfo{person}{Brinda Mehra}, \bibinfo{person}{Kejia Shen}, \bibinfo{person}{Hen~Chen Yen}, {and} \bibinfo{person}{Can Liu}.} \bibinfo{year}{2023}\natexlab{}.
\newblock \showarticletitle{Gist and Verbatim: Understanding Speech to Inform New Interfaces for Verbal Text Composition}. In \bibinfo{booktitle}{\emph{Proceedings of the 5th International Conference on Conversational User Interfaces}}. \bibinfo{pages}{1--11}.
\newblock


\bibitem[Mirowski et~al\mbox{.}(2023)]%
        {mirowski2023co}
\bibfield{author}{\bibinfo{person}{Piotr Mirowski}, \bibinfo{person}{Kory~W Mathewson}, \bibinfo{person}{Jaylen Pittman}, {and} \bibinfo{person}{Richard Evans}.} \bibinfo{year}{2023}\natexlab{}.
\newblock \showarticletitle{Co-writing screenplays and theatre scripts with language models: Evaluation by industry professionals}. In \bibinfo{booktitle}{\emph{Proceedings of the 2023 CHI conference on human factors in computing systems}}. \bibinfo{pages}{1--34}.
\newblock


\bibitem[Miura et~al\mbox{.}(2025)]%
        {miura2025understanding}
\bibfield{author}{\bibinfo{person}{Yusuke Miura}, \bibinfo{person}{Chi-Lan Yang}, \bibinfo{person}{Masaki Kuribayashi}, \bibinfo{person}{Keigo Matsumoto}, \bibinfo{person}{Hideaki Kuzuoka}, {and} \bibinfo{person}{Shigeo Morishima}.} \bibinfo{year}{2025}\natexlab{}.
\newblock \showarticletitle{Understanding and Supporting Formal Email Exchange by Answering AI-Generated Questions}.
\newblock \bibinfo{journal}{\emph{arXiv preprint arXiv:2502.03804}} (\bibinfo{year}{2025}).
\newblock


\bibitem[Nebeling et~al\mbox{.}(2016)]%
        {nebeling2016wearwrite}
\bibfield{author}{\bibinfo{person}{Michael Nebeling}, \bibinfo{person}{Alexandra To}, \bibinfo{person}{Anhong Guo}, \bibinfo{person}{Adrian~A de Freitas}, \bibinfo{person}{Jaime Teevan}, \bibinfo{person}{Steven~P Dow}, {and} \bibinfo{person}{Jeffrey~P Bigham}.} \bibinfo{year}{2016}\natexlab{}.
\newblock \showarticletitle{WearWrite: Crowd-assisted writing from smartwatches}. In \bibinfo{booktitle}{\emph{Proceedings of the 2016 CHI conference on human factors in computing systems}}. \bibinfo{pages}{3834--3846}.
\newblock


\bibitem[Nith et~al\mbox{.}(2024)]%
        {nith2024splitbody}
\bibfield{author}{\bibinfo{person}{Romain Nith}, \bibinfo{person}{Yun Ho}, {and} \bibinfo{person}{Pedro Lopes}.} \bibinfo{year}{2024}\natexlab{}.
\newblock \showarticletitle{SplitBody: Reducing Mental Workload while Multitasking via Muscle Stimulation}. In \bibinfo{booktitle}{\emph{Proceedings of the 2024 CHI Conference on Human Factors in Computing Systems}}. \bibinfo{pages}{1--11}.
\newblock


\bibitem[Oakley and Park(2007)]%
        {oakley2007designing}
\bibfield{author}{\bibinfo{person}{Ian Oakley} {and} \bibinfo{person}{Jun-Seok Park}.} \bibinfo{year}{2007}\natexlab{}.
\newblock \showarticletitle{Designing eyes-free interaction}. In \bibinfo{booktitle}{\emph{Haptic and Audio Interaction Design: Second International Workshop, HAID 2007 Seoul, South Korea, November 29-30, 2007 Proceedings 2}}. Springer, \bibinfo{pages}{121--132}.
\newblock


\bibitem[O'Brien and Toms(2010)]%
        {o2010development}
\bibfield{author}{\bibinfo{person}{Heather~L O'Brien} {and} \bibinfo{person}{Elaine~G Toms}.} \bibinfo{year}{2010}\natexlab{}.
\newblock \showarticletitle{The development and evaluation of a survey to measure user engagement}.
\newblock \bibinfo{journal}{\emph{Journal of the American Society for Information Science and Technology}} \bibinfo{volume}{61}, \bibinfo{number}{1} (\bibinfo{year}{2010}), \bibinfo{pages}{50--69}.
\newblock


\bibitem[Perrault et~al\mbox{.}(2013)]%
        {perrault2013watchit}
\bibfield{author}{\bibinfo{person}{Simon~T Perrault}, \bibinfo{person}{Eric Lecolinet}, \bibinfo{person}{James Eagan}, {and} \bibinfo{person}{Yves Guiard}.} \bibinfo{year}{2013}\natexlab{}.
\newblock \showarticletitle{Watchit: simple gestures and eyes-free interaction for wristwatches and bracelets}. In \bibinfo{booktitle}{\emph{Proceedings of the SIGCHI Conference on Human Factors in Computing Systems}}. \bibinfo{pages}{1451--1460}.
\newblock


\bibitem[Pradhan et~al\mbox{.}(2020)]%
        {pradhan2020use}
\bibfield{author}{\bibinfo{person}{Alisha Pradhan}, \bibinfo{person}{Amanda Lazar}, {and} \bibinfo{person}{Leah Findlater}.} \bibinfo{year}{2020}\natexlab{}.
\newblock \showarticletitle{Use of intelligent voice assistants by older adults with low technology use}.
\newblock \bibinfo{journal}{\emph{ACM Transactions on Computer-Human Interaction (TOCHI)}} \bibinfo{volume}{27}, \bibinfo{number}{4} (\bibinfo{year}{2020}), \bibinfo{pages}{1--27}.
\newblock


\bibitem[Ratcliff et~al\mbox{.}(1988)]%
        {ratcliff1988pattern}
\bibfield{author}{\bibinfo{person}{John~W Ratcliff}, \bibinfo{person}{David~E Metzener}, {et~al\mbox{.}}} \bibinfo{year}{1988}\natexlab{}.
\newblock \showarticletitle{Pattern matching: The gestalt approach}.
\newblock \bibinfo{journal}{\emph{Dr. Dobb’s Journal}} \bibinfo{volume}{13}, \bibinfo{number}{7} (\bibinfo{year}{1988}), \bibinfo{pages}{46}.
\newblock


\bibitem[Ruan et~al\mbox{.}(2016)]%
        {ruan2016speech}
\bibfield{author}{\bibinfo{person}{Sherry Ruan}, \bibinfo{person}{Jacob~O Wobbrock}, \bibinfo{person}{Kenny Liou}, \bibinfo{person}{Andrew Ng}, {and} \bibinfo{person}{James Landay}.} \bibinfo{year}{2016}\natexlab{}.
\newblock \showarticletitle{Speech is 3x faster than typing for english and mandarin text entry on mobile devices}.
\newblock \bibinfo{journal}{\emph{arXiv preprint arXiv:1608.07323}} (\bibinfo{year}{2016}).
\newblock


\bibitem[Ruan et~al\mbox{.}(2018)]%
        {ruan2018comparing}
\bibfield{author}{\bibinfo{person}{Sherry Ruan}, \bibinfo{person}{Jacob~O Wobbrock}, \bibinfo{person}{Kenny Liou}, \bibinfo{person}{Andrew Ng}, {and} \bibinfo{person}{James~A Landay}.} \bibinfo{year}{2018}\natexlab{}.
\newblock \showarticletitle{Comparing speech and keyboard text entry for short messages in two languages on touchscreen phones}.
\newblock \bibinfo{journal}{\emph{Proceedings of the ACM on Interactive, Mobile, Wearable and Ubiquitous Technologies}} \bibinfo{volume}{1}, \bibinfo{number}{4} (\bibinfo{year}{2018}), \bibinfo{pages}{1--23}.
\newblock


\bibitem[Schick et~al\mbox{.}(2023)]%
        {schick2023toolformer}
\bibfield{author}{\bibinfo{person}{Timo Schick}, \bibinfo{person}{Jane Dwivedi-Yu}, \bibinfo{person}{Roberto Dess{\`\i}}, \bibinfo{person}{Roberta Raileanu}, \bibinfo{person}{Maria Lomeli}, \bibinfo{person}{Eric Hambro}, \bibinfo{person}{Luke Zettlemoyer}, \bibinfo{person}{Nicola Cancedda}, {and} \bibinfo{person}{Thomas Scialom}.} \bibinfo{year}{2023}\natexlab{}.
\newblock \showarticletitle{Toolformer: Language models can teach themselves to use tools}.
\newblock \bibinfo{journal}{\emph{Advances in Neural Information Processing Systems}}  \bibinfo{volume}{36} (\bibinfo{year}{2023}), \bibinfo{pages}{68539--68551}.
\newblock


\bibitem[Schmidt et~al\mbox{.}(2020)]%
        {schmidt2020classifying}
\bibfield{author}{\bibinfo{person}{Maria Schmidt}, \bibinfo{person}{Ojashree Bhandare}, \bibinfo{person}{Ajinkya Prabhune}, \bibinfo{person}{Wolfgang Minker}, {and} \bibinfo{person}{Steffen Werner}.} \bibinfo{year}{2020}\natexlab{}.
\newblock \showarticletitle{Classifying cognitive load for a proactive in-car voice assistant}. In \bibinfo{booktitle}{\emph{2020 IEEE sixth international conference on big data computing service and applications (BigDataService)}}. IEEE, \bibinfo{pages}{9--16}.
\newblock


\bibitem[Shao et~al\mbox{.}(2024)]%
        {shao2024assisting}
\bibfield{author}{\bibinfo{person}{Yijia Shao}, \bibinfo{person}{Yucheng Jiang}, \bibinfo{person}{Theodore~A Kanell}, \bibinfo{person}{Peter Xu}, \bibinfo{person}{Omar Khattab}, {and} \bibinfo{person}{Monica~S Lam}.} \bibinfo{year}{2024}\natexlab{}.
\newblock \showarticletitle{Assisting in writing wikipedia-like articles from scratch with large language models}.
\newblock \bibinfo{journal}{\emph{arXiv preprint arXiv:2402.14207}} (\bibinfo{year}{2024}).
\newblock


\bibitem[Shen et~al\mbox{.}(2023)]%
        {shen2023beyond}
\bibfield{author}{\bibinfo{person}{Zejiang Shen}, \bibinfo{person}{Tal August}, \bibinfo{person}{Pao Siangliulue}, \bibinfo{person}{Kyle Lo}, \bibinfo{person}{Jonathan Bragg}, \bibinfo{person}{Jeff Hammerbacher}, \bibinfo{person}{Doug Downey}, \bibinfo{person}{Joseph~Chee Chang}, {and} \bibinfo{person}{David Sontag}.} \bibinfo{year}{2023}\natexlab{}.
\newblock \showarticletitle{Beyond summarization: Designing ai support for real-world expository writing tasks}.
\newblock \bibinfo{journal}{\emph{arXiv preprint arXiv:2304.02623}} (\bibinfo{year}{2023}).
\newblock


\bibitem[Shi et~al\mbox{.}(2023)]%
        {shi2023effidit}
\bibfield{author}{\bibinfo{person}{Shuming Shi}, \bibinfo{person}{Enbo Zhao}, \bibinfo{person}{Wei Bi}, \bibinfo{person}{Deng Cai}, \bibinfo{person}{Leyang Cui}, \bibinfo{person}{Xinting Huang}, \bibinfo{person}{Haiyun Jiang}, \bibinfo{person}{Duyu Tang}, \bibinfo{person}{Kaiqiang Song}, \bibinfo{person}{Longyue Wang}, {et~al\mbox{.}}} \bibinfo{year}{2023}\natexlab{}.
\newblock \showarticletitle{Effidit: An assistant for improving writing efficiency}. In \bibinfo{booktitle}{\emph{Proceedings of the 61st Annual Meeting of the Association for Computational Linguistics (Volume 3: System Demonstrations)}}. \bibinfo{pages}{508--515}.
\newblock


\bibitem[Singh et~al\mbox{.}(2018)]%
        {singh2018email}
\bibfield{author}{\bibinfo{person}{Aditya Singh}, \bibinfo{person}{Dibyendu Mishra}, \bibinfo{person}{Sanchit Bansal}, \bibinfo{person}{Vinayak Agarwal}, \bibinfo{person}{Anjali Goyal}, {and} \bibinfo{person}{Ashish Sureka}.} \bibinfo{year}{2018}\natexlab{}.
\newblock \showarticletitle{Email Dataset for Automatic Response Suggestion within a University"}.
\newblock  (\bibinfo{date}{2} \bibinfo{year}{2018}).
\newblock
\href{https://doi.org/10.6084/m9.figshare.5853057.v1}{doi:\nolinkurl{10.6084/m9.figshare.5853057.v1}}


\bibitem[Stoehr(1968)]%
        {stoehr1968tone}
\bibfield{author}{\bibinfo{person}{Taylor Stoehr}.} \bibinfo{year}{1968}\natexlab{}.
\newblock \showarticletitle{Tone and voice}.
\newblock \bibinfo{journal}{\emph{College English}} \bibinfo{volume}{30}, \bibinfo{number}{2} (\bibinfo{year}{1968}), \bibinfo{pages}{150--161}.
\newblock


\bibitem[Strayer et~al\mbox{.}(2016)]%
        {strayer2016talking}
\bibfield{author}{\bibinfo{person}{David~L Strayer}, \bibinfo{person}{Joel~M Cooper}, \bibinfo{person}{Jonna Turrill}, \bibinfo{person}{James~R Coleman}, {and} \bibinfo{person}{Rachel~J Hopman}.} \bibinfo{year}{2016}\natexlab{}.
\newblock \showarticletitle{Talking to your car can drive you to distraction}.
\newblock \bibinfo{journal}{\emph{Cognitive research: principles and implications}}  \bibinfo{volume}{1} (\bibinfo{year}{2016}), \bibinfo{pages}{1--17}.
\newblock


\bibitem[Taheri et~al\mbox{.}(2021)]%
        {taheri2021design}
\bibfield{author}{\bibinfo{person}{Atieh Taheri}, \bibinfo{person}{Ziv Weissman}, {and} \bibinfo{person}{Misha Sra}.} \bibinfo{year}{2021}\natexlab{}.
\newblock \showarticletitle{Design and evaluation of a hands-free video game controller for individuals with motor impairments}.
\newblock \bibinfo{journal}{\emph{Frontiers in Computer Science}}  \bibinfo{volume}{3} (\bibinfo{year}{2021}), \bibinfo{pages}{751455}.
\newblock


\bibitem[Thomas et~al\mbox{.}(2024)]%
        {thomas2024me}
\bibfield{author}{\bibinfo{person}{Justin Thomas}, \bibinfo{person}{Jigar Jogia}, \bibinfo{person}{Mariapaola Barbato}, {and} \bibinfo{person}{Richard Bentall}.} \bibinfo{year}{2024}\natexlab{}.
\newblock \showarticletitle{Me, not-me: Voice note use predicts self-voice recognition and liking}.
\newblock \bibinfo{journal}{\emph{Computers in Human Behavior Reports}}  \bibinfo{volume}{15} (\bibinfo{year}{2024}), \bibinfo{pages}{100446}.
\newblock


\bibitem[Tsimhoni et~al\mbox{.}(2004)]%
        {tsimhoni2004address}
\bibfield{author}{\bibinfo{person}{Omer Tsimhoni}, \bibinfo{person}{Daniel Smith}, {and} \bibinfo{person}{Paul Green}.} \bibinfo{year}{2004}\natexlab{}.
\newblock \showarticletitle{Address entry while driving: Speech recognition versus a touch-screen keyboard}.
\newblock \bibinfo{journal}{\emph{Human factors}} \bibinfo{volume}{46}, \bibinfo{number}{4} (\bibinfo{year}{2004}), \bibinfo{pages}{600--610}.
\newblock


\bibitem[V{\"o}lkel et~al\mbox{.}(2021)]%
        {volkel2021eliciting}
\bibfield{author}{\bibinfo{person}{Sarah~Theres V{\"o}lkel}, \bibinfo{person}{Daniel Buschek}, \bibinfo{person}{Malin Eiband}, \bibinfo{person}{Benjamin~R Cowan}, {and} \bibinfo{person}{Heinrich Hussmann}.} \bibinfo{year}{2021}\natexlab{}.
\newblock \showarticletitle{Eliciting and analysing users’ envisioned dialogues with perfect voice assistants}. In \bibinfo{booktitle}{\emph{Proceedings of the 2021 CHI conference on human factors in computing systems}}. \bibinfo{pages}{1--15}.
\newblock


\bibitem[Wan et~al\mbox{.}(2024)]%
        {wan2024felt}
\bibfield{author}{\bibinfo{person}{Qian Wan}, \bibinfo{person}{Siying Hu}, \bibinfo{person}{Yu Zhang}, \bibinfo{person}{Piaohong Wang}, \bibinfo{person}{Bo Wen}, {and} \bibinfo{person}{Zhicong Lu}.} \bibinfo{year}{2024}\natexlab{}.
\newblock \showarticletitle{" It Felt Like Having a Second Mind": Investigating Human-AI Co-creativity in Prewriting with Large Language Models}.
\newblock \bibinfo{journal}{\emph{Proceedings of the ACM on Human-Computer Interaction}} \bibinfo{volume}{8}, \bibinfo{number}{CSCW1} (\bibinfo{year}{2024}), \bibinfo{pages}{1--26}.
\newblock


\bibitem[Wei et~al\mbox{.}(2022)]%
        {wei2022chain}
\bibfield{author}{\bibinfo{person}{Jason Wei}, \bibinfo{person}{Xuezhi Wang}, \bibinfo{person}{Dale Schuurmans}, \bibinfo{person}{Maarten Bosma}, \bibinfo{person}{Fei Xia}, \bibinfo{person}{Ed Chi}, \bibinfo{person}{Quoc~V Le}, \bibinfo{person}{Denny Zhou}, {et~al\mbox{.}}} \bibinfo{year}{2022}\natexlab{}.
\newblock \showarticletitle{Chain-of-thought prompting elicits reasoning in large language models}.
\newblock \bibinfo{journal}{\emph{Advances in neural information processing systems}}  \bibinfo{volume}{35} (\bibinfo{year}{2022}), \bibinfo{pages}{24824--24837}.
\newblock


\bibitem[{Wikipedia contributors}(2024)]%
        {wikipediaGestaltPatternMatching}
\bibfield{author}{\bibinfo{person}{{Wikipedia contributors}}.} \bibinfo{year}{2024}\natexlab{}.
\newblock \bibinfo{title}{Gestalt Pattern Matching --- Wikipedia{,} The Free Encyclopedia}.
\newblock
\urldef\tempurl%
\url{https://de.wikipedia.org/wiki/Gestalt_Pattern_Matching}
\showURL{%
\tempurl}
\newblock
\shownote{Accessed: 2025-04-02}.


\bibitem[Xu et~al\mbox{.}(2023)]%
        {xu2023rewoo}
\bibfield{author}{\bibinfo{person}{Binfeng Xu}, \bibinfo{person}{Zhiyuan Peng}, \bibinfo{person}{Bowen Lei}, \bibinfo{person}{Subhabrata Mukherjee}, \bibinfo{person}{Yuchen Liu}, {and} \bibinfo{person}{Dongkuan Xu}.} \bibinfo{year}{2023}\natexlab{}.
\newblock \showarticletitle{Rewoo: Decoupling reasoning from observations for efficient augmented language models}.
\newblock \bibinfo{journal}{\emph{arXiv preprint arXiv:2305.18323}} (\bibinfo{year}{2023}).
\newblock


\bibitem[Xu et~al\mbox{.}(2007)]%
        {xu2007empirical}
\bibfield{author}{\bibinfo{person}{Shuang Xu}, \bibinfo{person}{Santosh Basapur}, \bibinfo{person}{Mark Ahlenius}, {and} \bibinfo{person}{Deborah Matteo}.} \bibinfo{year}{2007}\natexlab{}.
\newblock \showarticletitle{An empirical study on users’ acceptance of speech recognition errors in text-messaging}. In \bibinfo{booktitle}{\emph{International Conference on Human-Computer Interaction}}. Springer, \bibinfo{pages}{232--242}.
\newblock


\bibitem[Zhou et~al\mbox{.}(2024)]%
        {zhou2024glassmail}
\bibfield{author}{\bibinfo{person}{Chen Zhou}, \bibinfo{person}{Zihan Yan}, \bibinfo{person}{Ashwin Ram}, \bibinfo{person}{Yue Gu}, \bibinfo{person}{Yan Xiang}, \bibinfo{person}{Can Liu}, \bibinfo{person}{Yun Huang}, \bibinfo{person}{Wei~Tsang Ooi}, {and} \bibinfo{person}{Shengdong Zhao}.} \bibinfo{year}{2024}\natexlab{}.
\newblock \showarticletitle{GlassMail: Towards Personalised Wearable Assistant for On-the-Go Email Creation on Smart Glasses}. In \bibinfo{booktitle}{\emph{Proceedings of the 2024 ACM Designing Interactive Systems Conference}}. \bibinfo{pages}{372--390}.
\newblock


\end{thebibliography}


\appendix

\section{Hands-Free Writing Tools Assessment (HFWTA)}\label{sec:hfwta}

The following survey was used to evaluate participants' experiences with each writing tool across key dimensions of hands-free interaction. Participants rated each item on a 7-point Likert scale (1 = Strongly Disagree, 7 = Strongly Agree). This instrument was administered after completing tasks with each tool.

\subsection*{Section 1: Guided Writing Process}
\begin{itemize}
    \item The tool provided helpful structure and guidance for my writing process.
    \item The tool helped me include important details I might have otherwise forgotten.
    \item The tool's guidance felt tailored to the specific type of writing I was doing.
    \item The tool provided an appropriate amount of guidance (not too much, not too little).
\end{itemize}

\subsection*{Section 2: Hands-Free Interaction}
\begin{itemize}
    \item I could complete the entire writing task hands free.
    \item The tool allowed me to naturally pause and resume my workflow without losing context.
    \item I could easily make corrections or revisions entirely through voice commands.
    \item The tool rarely interrupted me while I was speaking.
    \item The tool provided clear audio feedback that kept me informed without requiring visual attention.
\end{itemize}

\subsection*{Section 3: Adaptability \& Contextual Awareness}
\begin{itemize}
    \item The tool adapted intelligently based on my previous input.
    \item When I provided vague information, the tool appropriately asked for clarification.
    \item The tool avoided asking for information I had already provided.
    \item The tool maintained appropriate context throughout the writing process.
\end{itemize}

\subsection*{Section 4: Multitasking Capability}
\begin{itemize}
    \item I could effectively use this tool while engaged in other activities (walking, eating, etc.).
    \item The tool required minimal visual attention to operate effectively.
    \item I felt confident that the tool was accurately capturing my input while I was multitasking.
\end{itemize}

\subsection*{Section 5: Content Refinement}
\begin{itemize}
    \item I could easily navigate to review or modify previous content.
    \item The modification process was intuitive and efficient.
    \item The tool effectively incorporated my changes into the final output.
\end{itemize}

\subsection*{Section 6: Output Quality \& Satisfaction}
\begin{itemize}
    \item The final output accurately reflected what I intended to communicate.
    \item The final output used the appropriate tone and style needed for the task.
    \item The final output correctly incorporated all the information I provided during the writing process.
    \item The tool helped me produce higher quality writing than I would have on my own.
\end{itemize}

\subsection*{Section 7: Overall Assessment}
\begin{itemize}
    \item This tool would be valuable for situations when I need to write while my hands are occupied.
    \item For hands-free writing, this tool was more efficient than the alternatives I've tried.
    \item I would choose this tool for future hands-free writing tasks.
\end{itemize}

\subsection*{Section 8: Ownership \& Agency}
\begin{itemize}
    \item I feel like the final output I created is truly \emph{my} writing.
    \item The final text sounds like me and reflects my voice.
    \item I had sufficient control over the process of writing.
    \item I had appropriate control over the final version of the text.
\end{itemize}

\subsection*{Section 9: Open-Ended Questions}

These open-ended questions were used to better understand participants’ subjective impressions and preferences. They were not included in the quantitative graphs but were used to qualitatively inform analysis and are cited in the paper where appropriate.

\begin{itemize}
    \item What specific features made this tool easier or harder to use in a hands-free context?
    \item How did you feel about the way the tool guided you through the writing process with questions? Was this helpful or limiting?
    \item How did the system's questions compare to how you would normally think through a writing task?
    \item Were there questions you expected the system to ask that it didn't? Or questions it asked that seemed unnecessary?
    \item What improvements would make this tool more effective for your specific writing needs?
    \item Were there any moments when you felt frustrated or limited by the interaction? Please describe.
    \item Describe a real-world scenario where you would find this tool particularly useful.
\end{itemize}

\section{Emotional Experience Questionnaire (EEQ)}\label{eeq}

EEQ was used to assess participants' affective responses to each tool. Participants rated their agreement with each statement using a 7-point Likert scale (1 = Strongly Disagree, 7 = Strongly Agree). This instrument was administered after participants completed all tasks with each tool.
\begin{itemize}
    \item I felt engaged while writing with this tool. \hfill \textit{(Engagement)}
    \item I enjoyed using this tool. \hfill \textit{(Enjoyment)}
    \item Using this tool made me feel motivated to complete my writing. \hfill \textit{(Motivation)}
    \item This tool reduced my stress while writing. \hfill \textit{(Stress Reduction)}
    \item This tool helped me feel more creative. \hfill \textit{(Creativity)}
\end{itemize}

\section{Tone Category Definitions}\label{tones}

We defined 14 tone categories to evaluate alignment between intended and generated tone. Each category reflects common communicative goals across formal, casual, and affective contexts.

\begin{itemize}
  \item \textbf{Formal}: Conforms to professional or institutional convention by opening and closing with courteous formulas, maintaining respectful distance throughout, and avoiding slang or overt emotion—even when the sentence structure or vocabulary is otherwise simple or includes contractions.

  \item \textbf{Informal}: Conversational and casual in both greeting and closing, omitting conventional formalities, favouring first-/second-person address and relaxed phrasing; it stays free of authoritative directives, which would shift the tone toward Assertive.

  \item \textbf{Friendly}: A friendly tone builds rapport through warm greetings, upbeat adjectives, polite assurances, and light humour; it stays steady and kind without diving into deep emotional validation (Empathetic) or making strong predictions about the future (Optimistic).

  \item \textbf{Diplomatic}: A diplomatic tone carefully navigates sensitive topics by using neutral vocabulary, gentle hedging (‘could’, ‘might’), and balanced phrasing that acknowledges multiple perspectives, explicitly avoiding blunt commands, deadlines, or one-sided judgments.

  \item \textbf{Urgent}: An urgent tone highlights immediate importance by pairing direct wording with explicit time cues such as ‘ASAP’, ‘by 2 PM today’, or references to events starting soon; its purpose is to trigger swift action, distinguishing it from mere assertiveness by its emphasis on speed.

  \item \textbf{Concerned}: A concerned tone expresses unease about potential problems; it employs conditional verbs (‘could’, ‘might’), tentative language, and references to possible negative outcomes to encourage caution, without the overt anxiety of Worried or the directive thrust of Urgent.

  \item \textbf{Optimistic}: An optimistic tone projects confidence in favourable future results; it relies on hopeful verbs (‘will’, ‘can’), visionary phrases (‘exciting opportunities ahead’), and uplifting language centred on forthcoming success rather than present rapport (Friendly) or sudden astonishment (Surprised).

  \item \textbf{Curious}: A curious tone signals genuine information-seeking; it is dominated by open-ended or clarifying questions and expressions of uncertainty, steering clear of imperatives, deadlines, or collaborative calls to action.

  \item \textbf{Encouraging}: An encouraging tone motivates and reassures; it supplies affirmations of ability (‘you’ve got this’), references to progress, and supportive language that boosts confidence without deep emotional empathy (Empathic) or explicit time pressure (Urgent).

  \item \textbf{Surprised}: A surprised tone communicates sudden astonishment at unexpected news; it features strong intensifiers, exclamatory punctuation, and short emotive bursts that foreground the shock itself rather than ongoing warmth, future optimism, or requests for action.

  \item \textbf{Cooperative}: A cooperative tone invites joint effort toward a shared goal; it consistently uses inclusive pronouns (‘we’, ‘our’), collaborative verbs (‘coordinate’, ‘work together’), and language that emphasises mutual benefit while avoiding solitary demands or purely polite formality.

  \item \textbf{Empathetic}: An empathetic tone shows understanding and compassion by explicitly naming or validating the other person’s feelings, offering support or flexibility, and using gentle, comforting phrasing distinct from simple friendliness or motivation.

  \item \textbf{Apologetic}: An apologetic tone takes responsibility for an error by stating an explicit apology (‘I’m sorry’), acknowledging fault, and describing corrective steps, thereby differentiating itself from neutral acknowledgements or defensive explanations.

  \item \textbf{Assertive}: An assertive tone delivers clear, confident instructions or expectations through imperative verbs or polite-imperative phrasing (‘please update’, ‘provide’), with minimal hedging and no necessary emphasis on tight timelines—distinguishing it from Urgent.
\end{itemize}

\section{Voice Command Reference} \label{voice_commands}

StepWrite supports a range of voice commands that enable hands-free navigation and control during the writing process. The system uses fuzzy matching to interpret input flexibly, so users do not need to speak the exact phrases listed below. Custom voice commands can also be defined by the user.

\begin{itemize}
  \item \textbf{Navigation} \\
  \textbf{Say:} \texttt{"next question"}, \texttt{"previous question"}, or \texttt{"skip question"} \\
  \textbf{Action:} Moves forward, backward, or skips the current question in the sequence.

  \item \textbf{Answer Editing} \\
  \textbf{Say:} \texttt{"modify answer"} \\
  \textbf{Action:} Revises the response to the current or a previous question.

  \item \textbf{Session Control} \\
  \textbf{Say:} \texttt{"pause writing"}, \texttt{"continue writing"}, or \texttt{"finish writing"} \\
  \textbf{Action:} Pauses, resumes, or ends the Q\&A session and generates the draft. Pausing disables input, useful if the user needs to talk to someone nearby without recording unintended speech. 

  \item \textbf{Interface Switching} \\
  \textbf{Say:} \texttt{"go to editor"}, \texttt{"return to questions"} \\
  \textbf{Action:} Switches between the editor view and the Q\&A interface.

  \item \textbf{Audio Playback} \\
  \textbf{Say:} \texttt{"play that again"}, \texttt{"stop speaking"} \\
  \textbf{Action:} Repeats or interrupts the most recent audio output.
\end{itemize}

More commands may be added in future versions.

\section{System Prompts}\label{appendix:prompts}

This appendix contains the prompts used in StepWrite, organized by functional modules. \footnote{Lengthy prompts are truncated for brevity; the complete versions are available in our GitHub repository.}
The prompting system consists of five main modules: \textit{question generation}, \textit{text generation}, \textit{fact checking}, \textit{tone classification}, and \textit{memory management}. Additionally, the \textit{dependency analysis} module referenced in Appendix \ref{design_implications} (\ref{minimize_redundant_effort}) is included as well.

\subsection{Question Generation Module}\label{appendix:question-generation}

\subsubsection{Write Question Prompt}\label{appendix:write-question}

This is the main investigative prompt that generates questions to gather minimal but essential information for text creation. It acts as an intelligent interviewer that helps users articulate their thoughts. The prompt is called iteratively to generate the next question in the conversation flow until sufficient information is gathered. Key features include extensive guidelines for what to ask and avoid, skip handling logic to prevent user frustration, logical question ordering and completion conditions, and context-aware detail gathering based on formality level.

\textbf{Prompt:}
\begin{lstlisting}[language=text, basicstyle=\small\ttfamily, breaklines=true, frame=single]
=== TASK ===
You are a system acting as an investigator & thinking partner to gather the minimal but essential information needed for another tool to craft a final text. Your role is to determine the single best next question to ask, always referencing the entire conversation to avoid repetition or irrelevant prompts. Your primary goal is to guide the user to provide enough information to fulfill their writing needs efficiently and without overwhelming them, while helping them discover and articulate their ideas through conversation.

=== Previous conversation ===
[Insert Q&A history]

=== CRITICAL REQUIREMENTS ===
1. NEVER ask about:
   - Titles, headings, or any formatting
   - Document structure or layout
   - Writing process or style preferences
   - Whether the user needs help (that's already implied)
   - How to phrase or word things
   - Font, spacing, or visual elements
   - Contact details unless explicitly needed for the task
   - "Anything else to add" type questions (use the dedicated optional prompt instead)
   - Whether to include standard sections (assume relevance based on context)
   - File formats or technical details
   - Whether to include references (unless the user has mentioned sources or research)
   - Whether to add appendices or attachments
   - Preferences about writing style or tone (assume a neutral, appropriate tone unless context suggests otherwise)
   - Language (always assume English)
   - Greetings, closings, or email formatting
   - Email addresses or contact details (unless the task explicitly requires gathering these)
   - Transportation or logistics (unless explicitly part of the core task)
   - Personal hobbies or interests from memory (unless directly and explicitly relevant to the immediate task)
   - Confirmations or verifications of previously given information
   - Any writing style elements (the writer LLM will handle this)
   - Goodbyes or closing remarks

2. ALWAYS ask about (if relevant, not already provided, and essential for the core message):
   - Core message or main points
   - Target audience (if not self)
   - Key objectives or goals
   - Important context or background
   - Deadlines or time-sensitive information
   - Budget/cost details if money is involved and relevant
   - Key stakeholders or decision makers
   - Specific details about the project or task
   - Any constraints or limitations
   - Required approvals or reviews

[Additional detailed guidelines continue...]

=== GUIDELINES ===
1. Review All Prior Context Before Proceeding
   - CRITICALLY IMPORTANT: Thoroughly analyze what information has ALREADY been provided through BOTH direct answers AND indirect mentions.
   - If a user mentions something even briefly, NEVER ask if they want to suggest specific details about that topic.
   - NEVER ask questions that can be logically inferred from previous answers.
   - Look for subtle implications and references in user responses that indirectly provide information.

2. Adopt an Investigator Mindset Focused on Essential Details and Idea Development
   - Focus on gathering ONLY the absolute minimum set of critical details required.
   - Frame questions to help users not just provide information but discover and refine their own thinking in the process.
   - Ask only one question at a time, seeking a specific piece of missing information that is clearly necessary.

[Additional guidelines continue...]

=== OUTPUT FORMAT ===
Return your result as valid JSON: 
{"question": "your question here","followup_needed": boolean}

- If "followup_needed" is false, return: {"question": "","followup_needed": false}
\end{lstlisting}

\subsubsection{Reply Question Prompts}\label{appendix:reply-question}

Reply prompts share the same structure as write prompts but include additional context from the original text being replied to.

\paragraph{Initial Reply Question Prompt}\label{appendix:initial-reply}

This prompt generates the first personalized question when replying to a message, ensuring it references specific content from the original message. It avoids generic questions and focuses on direct requests, time-sensitive issues, or the sender's main concerns. 

\textbf{Prompt:}
\begin{lstlisting}[language=text, basicstyle=\small\ttfamily, breaklines=true, frame=single]
=== TASK ===
Generate a SPECIFIC, PERSONALIZED first question to help someone reply to the message below.
This question should directly reference the content of the message in a way that feels personalized.

=== ORIGINAL MESSAGE ===
[Insert original text]

=== REQUIREMENTS ===
1. Create a SPECIFIC question that references the actual content of the message
2. The question must mention a key topic, request, or detail from the message
3. Avoid generic questions like:
   - "What main points do you want to address in your reply?"
   - "How would you like to respond to this message?"
   - "What do you want to say in your response?"
4. Instead, focus on a specific element in the message
5. Prioritize questions that address:
   - Direct requests in the message
   - Time-sensitive issues
   - The sender's main concern or point
   - Any decisions the recipient needs to make
6. Keep it concise (under 15 words)
7. Should be phrased as an open-ended question
8. Must sound natural and conversational

=== OUTPUT FORMAT ===
Return ONLY the question text, with no quotation marks, prefixes, or extra text.
\end{lstlisting}

\paragraph{Reply Question Prompt}\label{appendix:reply-continuation}

This prompt generates subsequent questions for the reply flow, incorporating the original message context and following the same guidelines as the write prompt. The key addition is that it includes the original message text as "extra context" and focuses on addressing points raised in the original communication while maintaining logical sequence and considering relationship context from the original sender.

\subsection{Text Generation Module}\label{appendix:text-generation}

\subsubsection{Write Output Prompt}\label{appendix:write-output}

This prompt generates the final text output based on the Q\&A conversation, incorporating user context and maintaining their voice. It is called after question generation is complete to produce the final text. The prompt ensures all user information is incorporated, including brief mentions, while maintaining the user's perspective and handling flexible preferences appropriately.

\textbf{Prompt:}
\begin{lstlisting}[language=text, basicstyle=\small\ttfamily, breaklines=true, frame=single]
=== TASK ===
Generate a coherent, concise response based on the conversation that captures the user's thinking process and ideas. Use this tone:
[Insert tone & its description]

=== Previous conversation ===
[Insert Q&A history]

=== CRITICAL OUTPUT FORMAT REQUIREMENTS ===
- Output ONLY the final text content itself
- DO NOT include any introductory text (like "Here's a draft:" or "Here's what I came up with:")
- DO NOT include any closing commentary (like "Let me know if you need any changes")
- DO NOT add dividers like "---" or "***" or similar formatting markers
- DO NOT include any meta-commentary about the text
- DO NOT wrap the output in quotes or code blocks
- Simply output the content directly, starting with the appropriate greeting if applicable

=== INTELLIGENCE REQUIREMENTS ===
- CRITICALLY IMPORTANT: Incorporate ALL information the user has provided, even if mentioned casually or briefly
- If the user mentioned something in passing, DEFINITELY include that perspective
- Pay special attention to brief mentions that might easily be missed but could be important to the user
- If the user indicates flexibility on a topic, reflect that flexibility rather than making up specific details
- If a user mentioned specific timing, location, people, or details, ensure they're accurately included
- Pay careful attention to the user's exact phrasing when they express preferences, concerns, or requests
- If the user skipped questions about a topic, do NOT include that topic in the response

=== Guidelines ===
- Use clear, straightforward language.
- Break down information into logical steps if needed.
- Keep sentences short and focused on the user's main points.
- Incorporate all essential details the user provided, no matter how briefly mentioned.
- Reflect the user's thought process, priorities, and reasoning as revealed through the conversation.
- Maintain the user's voice and perspective while providing structure and clarity.
- Emphasize topics where the user provided detailed responses or volunteered additional information.
- Follow the user's lead on what aspects of the content matter most to them.
- Never ask for additional details or clarification - use the information provided.
\end{lstlisting}

\subsubsection{Reply Output Prompt}\label{appendix:reply-output}

This prompt generates appropriate replies incorporating the original message context and user responses. The key addition compared to the write output prompt is that it includes the original message text and has specific greeting format requirements for replies, ensuring proper email etiquette and addressing all key points from the original message.

\textbf{Prompt:}
\begin{lstlisting}[language=text, basicstyle=\small\ttfamily, breaklines=true, frame=single]
=== TASK ===
Generate a clear and appropriate reply based on the user's responses to the questions that captures their thinking process and authentic voice. Use this tone:
[Insert tone & its description]

=== Original text they're replying to ===
[Insert original text]

=== Conversation with user's responses ===
[Insert Q&A history]

=== GREETING FORMAT REQUIREMENTS ===
- For emails: ALWAYS start with "Hello [Recipient's Name]," (extract the recipient's name from the original message)
- If recipient's name isn't clear from the original message, use an appropriate greeting like "Hello," or "Hi there,"
- NEVER start with the user's own name

[Same intelligence requirements and guidelines as write output prompt]
\end{lstlisting}

\subsection{Fact Checking Module}\label{appendix:fact-checking}

\subsubsection{Fact Check Prompt}\label{appendix:fact-check}

This prompt verifies that the generated output accurately represents the user's responses without changing the meaning or omitting important details. It is called after text generation to ensure accuracy and completeness. The prompt focuses on factual accuracy rather than appropriateness, allowing reasonable expansions and formatting improvements while flagging fundamental contradictions and omissions.

\textbf{Prompt:}
\begin{lstlisting}[language=text, basicstyle=\small\ttfamily, breaklines=true, frame=single]
=== TASK ===
You are a meticulous fact-checker responsible for ensuring the final output
faithfully represents the user's responses, exactly as they provided them.

=== ORIGINAL Q&A ===
[Insert Q&A history]

=== GENERATED OUTPUT ===
[Insert generated output]

=== GUIDELINES ===
1. Compare the user's responses in the Q&A with how they are represented in the generated output.
2. Your job is ONLY to verify that the output accurately reflects what the user said - NOT to judge if their answers were correct or appropriate for each question.
3. Be highly attentive to casual or brief mentions by the user that should be included in the output.

4. Verify that the user's responses appear accurately in the output, with these allowances:
   - Common spelling corrections are acceptable (e.g., "zoolm" -> "zoom", "tommorrow" -> "tomorrow")
   - Reasonable expansions of brief responses are fine (e.g., "ok" can be expanded into a proper response)
   - Standard formatting and professional conventions can be added
   - Grammar fixes and proper capitalization are allowed
   - Brand names can be properly capitalized/formatted (e.g., "facebook" -> "Facebook")
   - Common conversational elements and closings are acceptable (e.g., "Best regards", "Cheers", "Catch you later", "Take care")
   - Standard email/written communication elements can be added (e.g., greetings, sign-offs, well-wishes)
   - Technical terms can be properly formatted (e.g., "react" -> "React", "javascript" -> "JavaScript")
   - Partial or incomplete responses can be expanded with standard professional elements
   - Flexibility preferences (like "I'm flexible" or "any time works") can be appropriately reflected
   - Brief mentions can be contextually developed in a way consistent with the user's intent 
   - Standard conversational inferences are allowed (e.g., if user mentioned bringing something, output mentioning they'll bring it)
 

5. Only flag issues if:
   - The output fundamentally changes or contradicts the user's intended meaning
   - [Show if memory is on: The output adds major new claims or facts not implied by the memory]
   - The output completely ignores or omits the user's main point
   - The output misrepresents important details (beyond simple spelling/formatting fixes)
[Additional detailed guidelines continue...]

6. Do NOT flag issues for:
   - Whether the user's answer was appropriate for the question
   - Whether the user answered in the wrong section (ie: user answered "I want to travel" in the "What's your email subject?" question)
[Additional detailed guidelines continue...]

=== OUTPUT FORMAT ===
Return ONLY valid JSON (no extra text or backticks):
{
  "passed": boolean,
  "issues": [
    {
      "type": "missing" | "inconsistent" | "inaccurate" | "unsupported",
      "detail": "Description of the issue",
      "qa_reference": "Relevant Q&A excerpt or question"
    }
  ]
}
\end{lstlisting}

\subsubsection{Fact Correction Prompt}\label{appendix:fact-correction}

This prompt makes corrections to the generated text based on fact-checking results. It is called when fact-checking identifies issues that need correction.

\textbf{Prompt:}
\begin{lstlisting}[language=text, basicstyle=\small\ttfamily, breaklines=true, frame=single]
=== TASK ===
You are an AI assistant responsible for correcting content based on fact-checking results.

=== ORIGINAL Q&A ===
[Insert Q&A history]

=== CURRENT OUTPUT ===
[Insert generated output]

=== IDENTIFIED ISSUES ===
[Insert list of issues]

=== CRITICAL OUTPUT FORMAT REQUIREMENTS ===
- Output ONLY the final corrected text content itself
- DO NOT include any introductory text
- DO NOT include any closing commentary
- DO NOT add dividers or meta-commentary
- Simply output the corrected content directly

=== CRITICAL INTELLIGENCE REQUIREMENTS ===
- Make MINIMAL changes to fix ONLY the issues flagged
- DO NOT "over-correct" by changing things that weren't identified as issues
- IMPORTANT: Preserve brief mentions by the user - if the issue is that a brief mention was omitted, ensure it's included
- If a user expressed flexibility on a topic, maintain that flexibility in your correction
- Never add information about topics the user explicitly skipped questions about
- Pay close attention to specific facts, dates, times, and details mentioned by the user
- Maintain the exact meaning and intent of the user's responses
- ALWAYS prioritize what the user actually said over what sounds better or more complete

=== TASK ===
1. Review the original Q&A and the current output.
2. Address ONLY the issues flagged:
   - If a key fact or detail is missing (including brief mentions), insert it appropriately
   - If a fact is contradicted or misstated, correct it to match the user's Q&A
   - If the output introduces a major new claim that conflicts with the Q&A, remove or adjust it
   - If information appears about topics the user skipped, remove that information
3. Correction Strategy:
   - Make surgical, precise changes to fix only the specific issues
   - Preserve as much of the original output as possible
   - Only rewrite sections that directly contain issues
   - When adding missing information, place it in the most contextually appropriate location
4. Ensure the corrected version:
   - Stays concise
   - Preserves ALL of the user's key details, including those mentioned briefly
   - Maintains the user's unique voice and perspective
   - Captures their thinking process and reasoning
   - Respects this tone: [Insert tone & its description]
   - Does not remove benign expansions like greetings unless they cause a conflict
   - Maintains the same structure and organization unless the issues require changes
  - If the user said "no" to optional items, preferences, or arrangements that were asked as "Would you like/need/want X?", then completely omit mentioning X in the output
  - Rule of thumb: If saying "You don't need to X" or "No need to X" might come off as socially awkward or implies X was expected by default, omit mentioning X entirely.
\end{lstlisting}

\subsection{Tone Classification Module}\label{appendix:tone-classification}

\subsubsection{Tone Classification Prompt}\label{appendix:tone-prompt}

This prompt analyzes the conversation to determine the most appropriate tone for the generated text. 

\textbf{Prompt:}
\begin{lstlisting}[language=text, basicstyle=\small\ttfamily, breaklines=true, frame=single]
=== TASK ===
Analyze the conversation and determine the most appropriate tone for the response.

=== ORIGINAL TEXT ===
[OPTIONAL: insert additional context (reply)]

Pay special attention to the context of the original text, as it sets the expected formality and professionalism level of the conversation.


=== CONVERSATION Q&A ===
[Insert Q&A history]

=== TONE ANALYSIS GUIDELINES ===
1. Consider these factors:
   - The nature of the relationship (professional, personal, etc.)
   - The context and purpose of the communication
   - The level of formality in the user's responses
   - The emotional undertones in the conversation
   - The intended audience
   - The type of message (request, information, apology, etc.)

2. Classify into one of these categories:
[Insert tone categories referenced in the Tone Category Definitions Appendix]

3. Consider these aspects:
   - Word choice and vocabulary level
   - Sentence structure complexity
   - Use of contractions and idioms
   - Level of directness
   - Emotional expression
   - Cultural context

=== OUTPUT FORMAT ===
Return ONLY a JSON object with:
{
  "tone": "TONE_CATEGORY",
  "reasoning": "Brief explanation of classification"
}
\end{lstlisting}

\subsection{Memory Module}\label{appendix:memory}

\subsubsection{Memory Prompt}\label{appendix:memory-prompt}

This prompt provides user context from stored memories to personalize responses and avoid redundant questions. It is included in other prompts when memory is enabled to provide user context.

\textbf{Prompt:}
\begin{lstlisting}[language=text, basicstyle=\small\ttfamily, breaklines=true, frame=single]
=== USER CONTEXT ===
Consider the following information about the user when generating content:
[Insert object of memories]

Additional Guidelines:
- Use this information to avoid asking questions about things we already know
- For texts that require a name signature, use the user's full name
- For emails, always start with "Hello [Recipient's Name]" (never with the user's name)
- For replies, use the user's information to personalize the response
- Only reference this information when relevant to the current task
- Do not expose or directly mention that you have this stored information
- Use this context to make suggestions more personalized when appropriate
- If a memory detail conflicts with what the user explicitly says in the current conversation, always prioritize the user's current input
- Don't suggest memory details unless they're directly relevant to the current request
\end{lstlisting}

\subsubsection{Memory Fact Check Prompt}\label{appendix:memory-fact-check}

This prompt prevents the fact-checking system from flagging information derived from user memories as incorrect. It is included in fact-checking prompts when memory is enabled, formatting memory context for fact-checking and treating memory details as verified facts. 

\textbf{Prompt:}
\begin{lstlisting}[language=text, basicstyle=\small\ttfamily, breaklines=true, frame=single]
=== MEMORY-AWARE FACT CHECKING ===
The following information exists in the user's memory context and should NOT be flagged as inconsistencies or unsupported claims:
[Insert object of memories]

Important:
- Do not flag information as missing, inconsistent, or unsupported if it matches or is derived from these memory items
- Personal details, preferences, or context from these memories are considered valid even if not explicitly mentioned in the Q&A
- Names, locations, or other specific details from memories should be treated as verified facts
- Any reasonable expansion or natural use of this contextual information is acceptable
\end{lstlisting}

\subsection{Dependency Analysis Module}\label{appendix:dependency-analysis}

\subsubsection{Dependency Analysis Prompt}\label{appendix:dependency-prompt}

This prompt determines which subsequent questions should be invalidated when a user changes an earlier answer. It is called when a user modifies a previous response to maintain conversation consistency. The prompt analyzes questions based on semantic dependency, logical flow, context dependency, and specificity matching to determine which questions are affected by the change and should be invalidated.

\textbf{Prompt:}
\begin{lstlisting}[language=text, basicstyle=\small\ttfamily, breaklines=true, frame=single]
=== TASK ===
You are a dependency analysis system for a conversational writing assistant. A user has changed their answer to an earlier question, and you need to determine which subsequent questions are affected by this change and should be invalidated.

=== CONTEXT ===
The user changed their answer to Question: [Insert ID of changed question]
- Original answer: [Insert original answer]
- New answer: [Insert new answer]

=== ALL QUESTIONS AND CURRENT ANSWERS ===
[Insert Q&A history]

=== ANALYSIS CRITERIA ===
A question should be marked as AFFECTED if:
1. Semantic Dependency: The question's meaning, relevance, or appropriateness changes based on the new answer
2. Logical Flow: The question no longer makes logical sense in the conversation flow
3. Context Dependency: The question assumes information that is no longer valid
4. Specificity Mismatch: The question is too specific for a now-broader answer, or too broad for a now-specific answer

A question should be marked as UNAFFECTED if:
1. Topic Independence: The question addresses a completely different aspect that remains relevant
2. Generic Applicability: The question is general enough to apply regardless of the change
3. Orthogonal Concerns: The question deals with separate concerns (e.g., tone, audience, timing) that aren't impacted

=== INSTRUCTIONS ===
1. Analyze each question that comes AFTER the changed question (Q${changedQuestionId})
2. For each subsequent question, determine if it's AFFECTED or UNAFFECTED
3. Provide clear reasoning for your decision
4. Focus on logical dependencies, not just keyword matching
5. Consider the conversation flow and context

=== OUTPUT FORMAT ===
Return a JSON object with this exact structure:
{
  "affectedQuestions": [
    {
      "questionId": number,
      "question": "exact question text",
      "status": "AFFECTED" | "UNAFFECTED",
      "reasoning": "clear explanation of why this question is/isn't affected"
    }
  ],
  "summary": "brief summary of the overall impact"
}
\end{lstlisting}

\appendix
\section{System Architecture}
\label{appendix:diagrams}

\subsection{System Diagram}
\label{appendix:system_diagram}

\begin{figure*}[h]
    \centering
    \includegraphics[width=\textwidth]{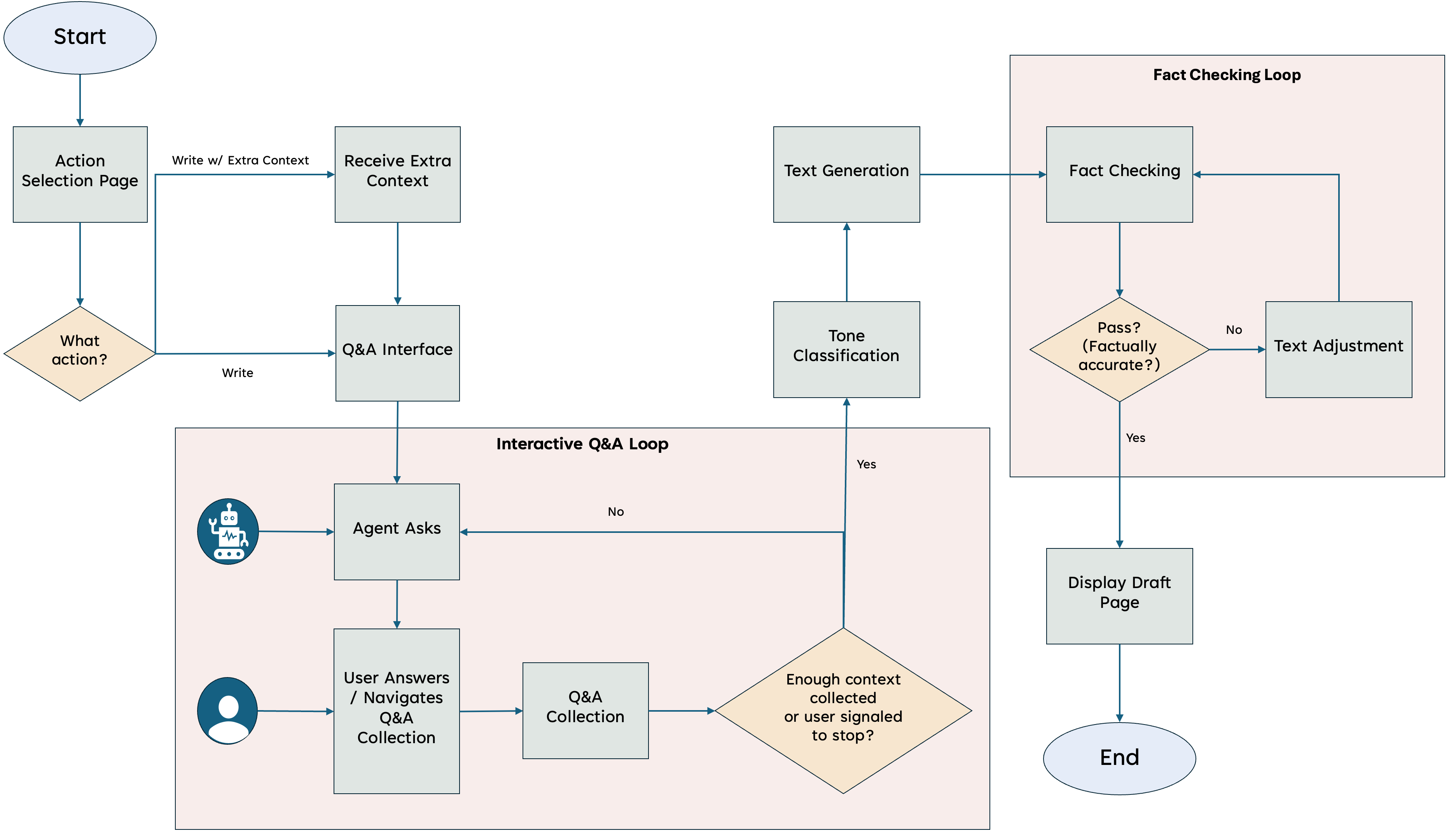}
    \caption{The StepWrite system diagram, illustrating the full end-to-end workflow. This includes the interactive Q\&A loop, tone classification, text generation, and the fact-checking loop.}
    \label{fig:stepwrite_system_diagram}
\end{figure*}

This diagram shows the architecture and flow of StepWrite. After selecting an action (e.g., write or reply), the user enters an interactive Q\&A loop, where the system adaptively collects information through voice prompts. Once sufficient context is gathered or the user signals completion, the system classifies tone and generates a draft. The draft then enters a fact-checking loop for verification and refinement. Only after passing all checks is the final output displayed.

\subsection{Speech-to-Text (STT) Pipeline}
\label{appendix:stt_pipeline}

\begin{figure*}[h]
    \centering
    \includegraphics[width=\textwidth]{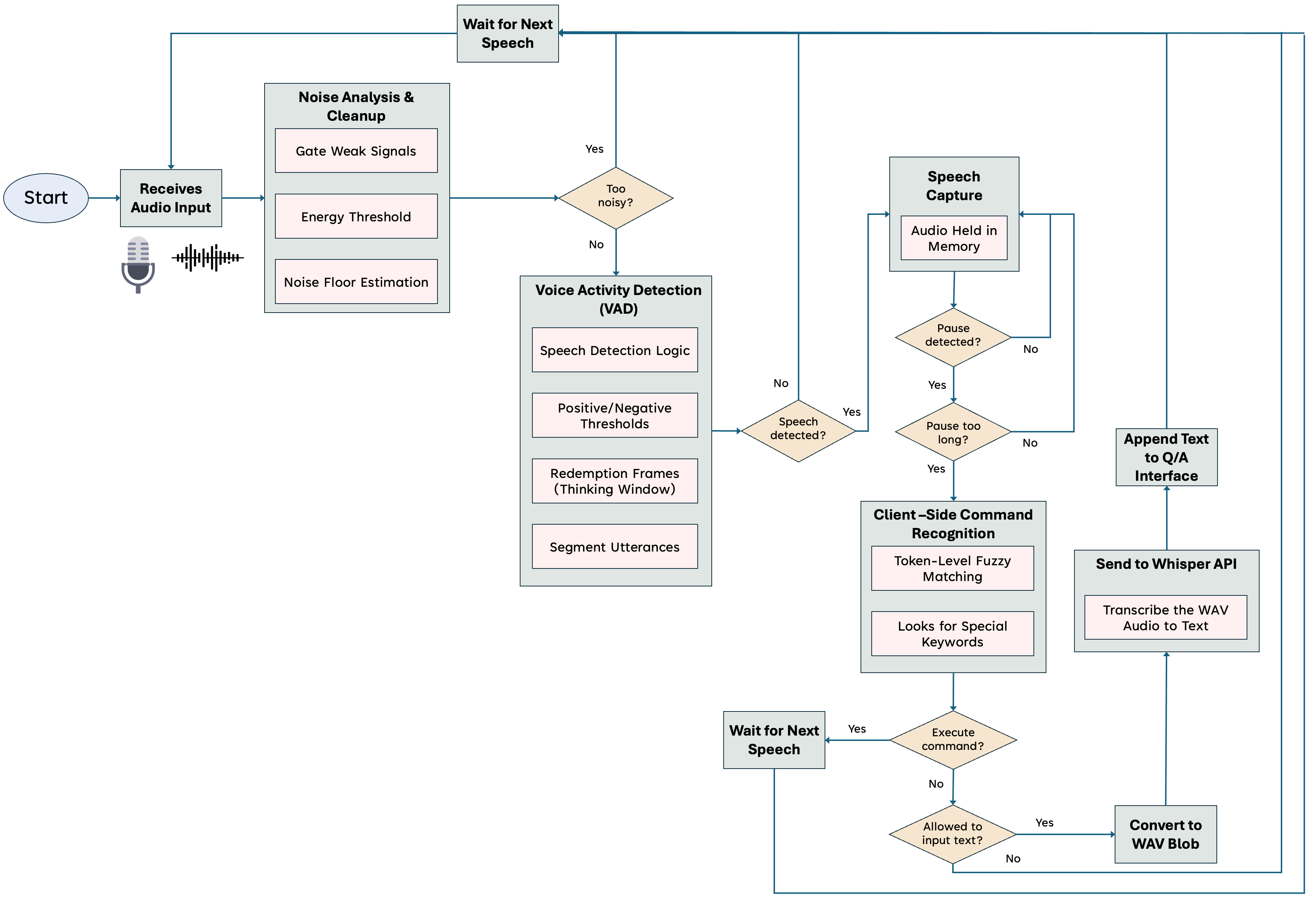}
    \caption{StepWrite’s speech-to-text (STT) pipeline. The system performs noise filtering, voice activity detection (VAD), client-side command recognition, and server-side transcription. Only clean, non-command utterances are forwarded to the Q\&A module.}
    \label{fig:stt_pipeline}
\end{figure*}

This diagram shows the STT pipeline used by StepWrite. Audio input from the microphone is first passed through a noise analysis and cleanup module. Voice activity detection (VAD) then segments valid utterances using thresholds and pause timing logic. Client-side command recognition identifies and executes supported macros before transcription. If the utterance is not a command and input is allowed, the audio is sent to the Whisper API for transcription. The resulting text is then appended to the interactive Q\&A interface for further processing.

\subsection{Text-to-Speech (TTS) Pipeline}
\label{appendix:tts_pipeline}

\begin{figure*}[h]
    \centering
    \includegraphics[width=\textwidth]{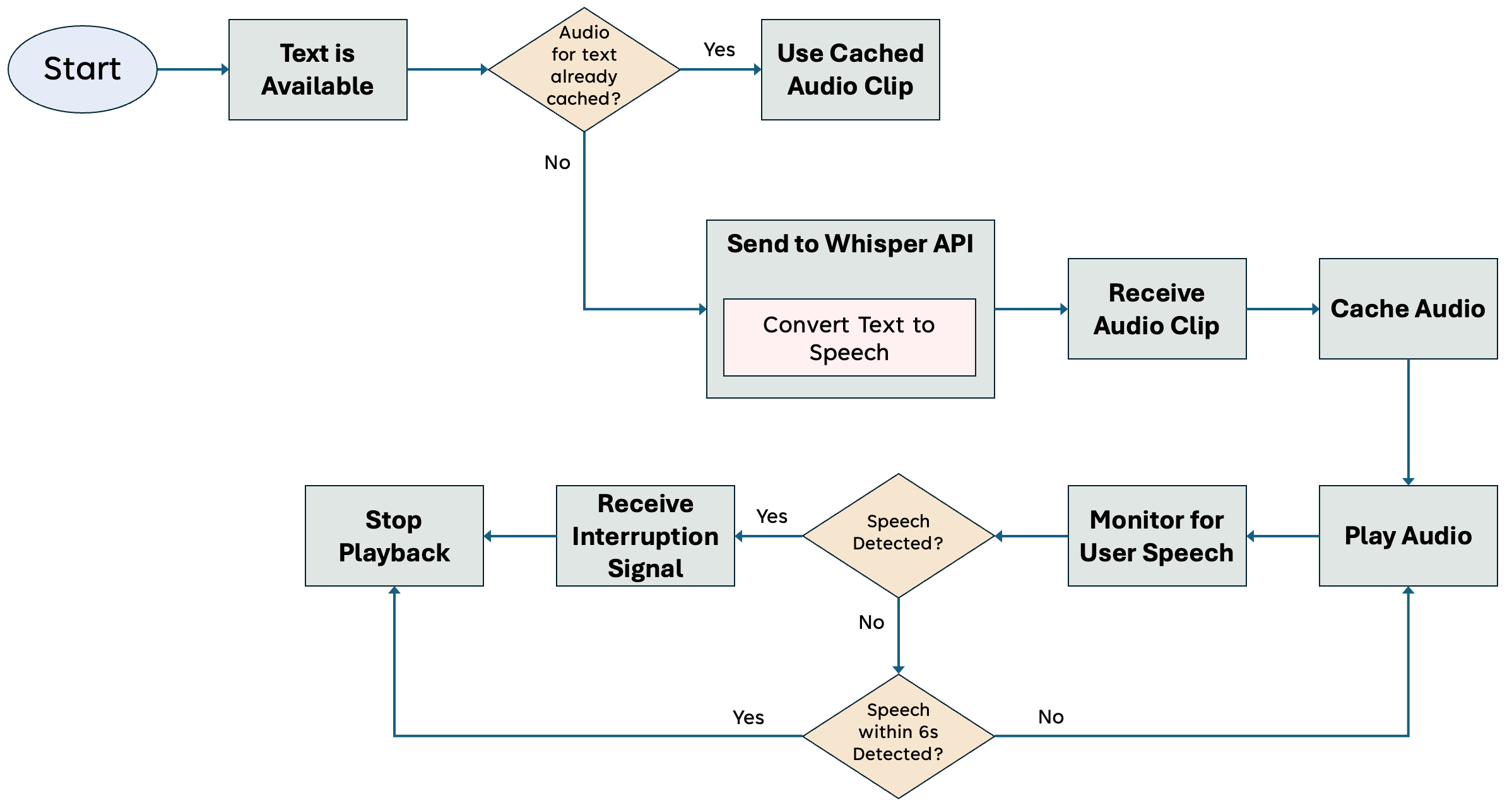}
    \caption{StepWrite’s text-to-speech (TTS) pipeline. The system converts response text into audio, caches it for reuse, and monitors for user interruptions during playback to support real-time interaction.}
    \label{fig:tts_pipeline}
\end{figure*}

This diagram shows the TTS pipeline used by StepWrite. When text is available for playback, the system checks whether a corresponding audio clip has already been cached. If not, it sends the text to a server-side API for synthesis. The resulting audio is cached and played back to the user. During playback, the system continuously monitors for speech-based interruptions, including immediate responses and voice activity within a 6-second window. If an interruption is detected, playback is stopped and the system resumes listening.

\end{document}